\documentclass{aa}  

\usepackage{graphicx, color}
\usepackage{txfonts}
\usepackage{comment}
\usepackage{booktabs}
\usepackage[allcolors=blue]{hyperref}
\newcommand{\orcidlink}[1]{\protect\href{https://orcid.org/#1}{\protect\includegraphics[width=8pt]{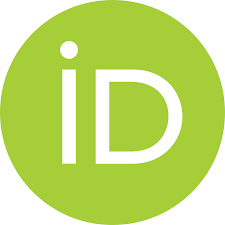}}}
\begin{document}

   \title{A BCool survey of stellar magnetic cycles}
%\subtitle{Subtitle}
   \titlerunning{A BCool survey of stellar magnetic cycles}

   \author{S. Bellotti \inst{1,2}\orcidlink{0000-0002-2558-6920}
          \and
          P. Petit \inst{2}\orcidlink{0000-0001-7624-9222}
          \and
          S. V. Jeffers\inst{3}\orcidlink{}
          \and
          S. C. Marsden\inst{4}\orcidlink{0000-0001-5522-8887}
          \and
          J. Morin \inst{5}\orcidlink{0000-0002-4996-6901}
          \and 
          A. A. Vidotto \inst{1}\orcidlink{0000-0001-5371-2675}
          \and 
          C. P. Folsom\inst{6}\orcidlink{0000-0002-9023-7890}
          \and
          V. See \inst{7,8}\orcidlink{0000-0001-5986-3423}
          \and
          J.-D. do Nascimento, Jr.\inst{9,10}\orcidlink{0000-0001-7804-2145}
          }
   \authorrunning{Bellotti et al.}
    
   \institute{
            Leiden Observatory, Leiden University,
            PO Box 9513, 2300 RA Leiden, The Netherlands\\
            \email{bellotti@strw.leidenuniv.nl}
        \and
            Institut de Recherche en Astrophysique et Plan\'etologie,
            Universit\'e de Toulouse, CNRS, IRAP/UMR 5277,
            14 avenue Edouard Belin, F-31400, Toulouse, France
        \and
            Th\"uringer Landessternwarte Tautenburg, Sternwarte 5, D-07778 Tautenburg, Germany
        \and 
            Centre for Astrophysics, University of Southern Queensland, 
            Toowoomba, QLD 4350, Australia
        \and 
             Laboratoire Univers et Particules de Montpellier,
             Universit\'e de Montpellier, CNRS,
             F-34095, Montpellier, France
        \and
             Tartu Observatory, University of Tartu, Observatooriumi 1, T\~oravere, 61602, Estonia
        \and 
             Science Division, Directorate of Science, 
             European Space Research and Technology Centre (ESA/ESTEC),
             Keplerlaan 1, 2201 AZ, Noordwijk, The Netherlands
        \and 
            School of Physics \& Astronomy, University of Birmingham, Edgbaston, Birmingham B15 2TT, UK
        \and 
            Center for Astrophysics, Harvard \& Smithsonian, 60 Garden Street, Cambridge, MA 02138, USA
        \and
            Dep. de F\'isica, Univ. Federal do Rio Grande do Norte-UFRN, Natal, RN, 59078-970, Brazil
             }
   \date{Received ; accepted }

% \abstract{}{}{}{}{} 
% 5 {} token are mandatory
 
  \abstract
   {The magnetic cycle on the Sun consists of two consecutive 11-yr sunspot cycles and exhibits a polarity reversal around sunspot maximum. Although solar dynamo theories have progressively become more sophisticated, the details as to how the dynamo sustains magnetic fields are still subject of research. Observing the magnetic fields of Sun-like stars can bring useful insights to contextualise the solar dynamo.}   
   {With a long-term spectropolarimetric monitoring of stars, the BCool survey studies the evolution of surface magnetic fields to understand how dynamo-generated processes are influenced by key ingredients, like mass and rotation. Here, we focus on six Sun-like stars with mass between 1.02 and 1.06 M$_\odot$ and with 3.5-21~d rotation period (or 0.3-1.8 in Rossby number), thus it is a practical sample to study magnetic cycles across distinct activity levels.}
   {We analysed high-resolution spectropolarimetric data collected with ESPaDOnS, Narval and Neo-Narval between 2007 and 2024 within the BCool programme. We measured longitudinal magnetic field from least-squares deconvolution line profiles and we inspected its long-term behaviour with both a Lomb-Scargle periodogram and a Gaussian process. We then applied Zeeman-Doppler imaging to reconstruct the large-scale magnetic field geometry at the stellar surface for different epochs.}
   {Two of our slow rotators, namely HD\,9986 and HD\,56124 (P$_\mathrm{rot}\sim$20\,d) exhibit repeating polarity reversals of the radial or toroidal field component on shorter time scales than the Sun (5 to 6\,yr). HD\,73350 (P$_\mathrm{rot}\sim$12\,d) has one polarity reversal of the toroidal component and HD\,76151 (P$_\mathrm{rot}$=17\,d) may have short-term evolution (2.5\,yr) modulated by the long-term (16\,yr) chromospheric cycle. Our two fast rotators, HD~166435 and HD~175726 (P$_\mathrm{rot}$=3-5\,d), manifest complex magnetic fields without an evident cyclic evolution.}
   {Our findings indicate the potential dependence of the magnetic cycles nature with stellar rotation period. For the two stars with likely cycles, the polarity reversal time scale seems to decrease with decreasing rotation period or Rossby number. These results represent important observational constraints for dynamo models of solar-like stars.} 

   \keywords{Stars: magnetic field --
                Stars: activity --
                Techniques: polarimetric
               }

   \maketitle

%
%-------------------------------------------------------------------

\section{Introduction}

The activity cycle of the Sun is characterised by the quasi-periodic evolution of the surface sunspot distribution. Such variation of sunspot number, size, and latitude over a timescale of 11\,yr was noticed early by \citet{Schwabe1844} and \citet{Maunder1904}. This is accompanied by a polarity reversal of the magnetic field as expressed by Hale's laws \citep{Hale1919}, revealing the underlying magnetic cycle of 22\,yr. The 11-yr long variation is also known as the Shwabe cycle and the 22-yr long evolution as the Hale cycle. The magnetic cycle is thus formed by two consecutive sunspot cycles, with the polarity reversal of the poloidal and toroidal field occurring around sunspot maximum \citep[see the reviews of][]{Hathaway2010,Hathaway2015}. During the magnetic cycle, the amount of magnetic energy in the poloidal and toroidal large-scale field components varies, and the obliquity of the poloidal-dipolar component oscillates between axisymmetric and non-axisymmetric configurations \citep{Sanderson2003,DeRosa2012,Vidotto2018,Finley2023}.

Understanding the solar magnetic cycle and the dynamo loop, that is the alternating generation of poloidal and toroidal field components, is an active field of research \citep[][for a recent review]{Charbonneau2020}. It is generally accepted that the transformation of a poloidal field into a toroidal one occurs via differential rotation with anisotropic turbulence \citep[$\Omega$ effect;][]{Parker1955}, while the reverse process is debated and can be described by cyclonic turbulence \citep[$\alpha$ effect;][]{Parker1955} or by the dispersal of magnetic flux by the poleward migration of decaying bipolar magnetic regions \citep{Babcock1961,Leighton1969}, or by magnetohydrodynamical instabilities at the level of the tachocline \citep[e.g.][]{Schussler2003,Dikpati2009,Chatterjee2011}. All these models use mean-field approximation, in which convection is not included, as opposed to global magneto-convection models, in which convection and its effects are included self-consistently \citep[see e.g.][and references therein]{Charbonneau2020}. The tachocline is the thin interface between the solidly rotating radiative core and the differentially rotating convective envelope in the solar interior \citep{Spiegel1992}. Moreover, numerical simulations of dynamo models have become increasingly sophisticated, but a number of difficulties remain, such as reproducing the solar convection and differential rotation \citep{Kapyla2023}. Although the Sun is an important benchmark to studies of activity of solar-like stars, solar dynamo models have also been unable to reproduce the saturation of activity seen with different proxies \citep[e.g.][]{Wright2018,See2019,Reiners2022}. 

In this context, observations of magnetic cycles in other stars provide key information to understand how stellar parameters, such as mass and rotation period, impact the internal dynamo processes \citep[][for a recent review]{Jeffers2023,Charbonneau2023}. Investigating the existence of cycles on other stars is performed via distinct techniques. Monitoring the fluctuation in atmospheric heating as conveyed by the emission reversal in the cores of chromospheric lines \citep[e.g. Ca~\textsc{ii} H\&K][]{Leighton1959,Hall2008} is a primary approach, which was used extensively for solar-like stars during the Mt. Wilson project \citep{Wilson1968,Baliunas1995} and beyond \citep{BoroSaikia2018a,Baum2022,Isaacson2024}. Likewise, long-term photometric time series can reveal the periodic variation of stellar brightness associated to the evolving distribution of surface inhomogeneities like spots and faculae \citep{Olah2009,Strassmeier2009,Ozdarcan2010,Ferreira-Lopes2015,SuarezMascareno2016,Lehtinen2016,Clements2017,Reinhold2017}. For the Sun, \citet{White1981} showed that the brightness variations are correlated to the evolution of chromospheric emission lines. Furthermore, stellar cycles can be identified by the variability of the coronal X-ray emission \citep[e.g.][]{Gudel2004,Hempelmann2006,DeWarf2010,Robrade2012,SanzForcada2013,Coffaro2020}, by the reversals or evolution of polarised radio emission \citep{Route2016,Bloot2024}, and by the influence of the magnetic field on acoustic mode properties \citep{Garcia2010,Mathur2013,Regulo2016}. Recently, studies have shown the potential of using flare statistics as probes for stellar cycles \citep{Feinstein2024,Wainer2024}.

The long-term spectropolarimetric monitoring of a star is also a powerful technique, because it allows one to trace the secular evolution of the large-scale magnetic field geometry as reconstructed with Zeeman-Doppler imaging \citep[ZDI][]{Semel1989,DonatiBrown1997}. For the Sun, \citet{Vidotto2018} and \citet{Lehmann2021} investigated the evolution of the large-scale magnetic field during a Schwabe cycle as it would be seen by ZDI, that is analysing the observables that are recovered reliably by ZDI. They showed that the axisymmetric and poloidal energy fractions of the large-scale magnetic field peak around solar cycle minimum, while the toroidal component increases during solar cycle maximum. Such evolution of the axisymmetry and toroidal component is correlated to the varying latitude of emergence of sunspots during the cycle \citep[as displayed by the butterfly diagram;][]{Maunder1904,Charbonneau2020}, making them suitable diagnostics to search for solar-like cycles on other stars \citep{Lehmann2021}. More generally, the aim of long-term spectropolarimetric monitoring is to discern similar or contrasting trends relative to the solar magnetic cycle, in the form of polarity reversals and/or varying complexity of the field geometry.

The BCool program \footnote{\href{https://bcool.irap.omp.eu/}{https://bcool.irap.omp.eu/}} \citep{Marsden2014} has now reached a baseline of 15-20\,yr, which is suitable for inspecting the secular evolution of stellar magnetic topologies with spectropolarimetry. Previous studies within BCool have explored different spectral types ranging between F and K~type \citep[see][for a review]{Jeffers2023}. Clear examples of magnetic cycles are $\tau$~Boo \citep[F7 type, $P_\mathrm{cyc}=$120\,d][]{Donati2008b,Fares2009,Fares2013,Mengel2016,Jeffers2018}, $\kappa$~Cet \citep[G5 type, $P_\mathrm{cyc}=$10\,yr][]{doNascimento2016,BoroSaikia2022}, 61~Cyg~A \citep[K5 type, $P_\mathrm{cyc}=$7.3\,yr][]{BoroSaikia2016,BoroSaikia2018}, $\varepsilon$~Eri \citep[K2 type, $P_\mathrm{cyc}=$3\,yr modulated by a longer cycle of 13\,yr][]{Jeffers2022}. Of these, only $\tau$~Boo and 61~Cyg~A manifest large-scale polarity reversals in phase with chromospheric activity cycles \citep[see e.g.][]{Jeffers2023}. Stars with putative magnetic cycles were also found in the same spectral range, such as HD~75332 \citep[F7 type][]{Brown2021}, HD~78366 \citep[G0 type][]{Morgenthaler2011}, HD~19077 \citep[K1 type][]{Petit2009,Morgenthaler2011}, while others exhibit fast evolution of the topology without evident polarity reversals such as HN~Peg \citep[G0 type][]{BoroSaikia2015}, HD~171488 \citep[G2 type][]{Marsden2006,JeffersDonati2008,Jeffers2011}, and EK~Dra \citep[G5 type][]{Waite2017}, or a stable behaviour like $\chi$~Dra \citep[F7 type][]{Marsden2023}. Finally, evidence for magnetic cycles on M dwarfs was found more recently \citep{Bellotti2023b,Lehmann2024,Bellotti2024}, although not as part of the BCool program.

In this paper, we present the long-term spectropolarimetric monitoring of six solar-like stars that was carried out as part of the BCool program. The observations were collected with the twin optical spectropolarimeters ESPaDOnS\footnote{\href{https://www.cfht.hawaii.edu/Instruments/Spectroscopy/Espadons/}{https://www.cfht.hawaii.edu/Instruments/Spectroscopy/Espadons/}} and Narval, and its recent upgrade Neo-Narval\footnote{\href{https://www.news.obs-mip.fr/neo-narval-pic-du-midi/}{https://www.news.obs-mip.fr/neo-narval-pic-du-midi/}}, with a time span of $\sim17$\,yr, from 2007 to 2024. Such a baseline is suitable to start to inspect the long-term temporal variation of the longitudinal magnetic field via periodograms and Gaussian Processes, and to examine the yearly evolution of the large-scale topology of the stellar magnetic field with ZDI. 

The paper is structured as follows. In Sect.~\ref{sec:observations}, we describe the ESPaDOnS, Narval, and Neo-Narval observations and in Sect.~\ref{sec:Bl} the computation of longitudinal magnetic field from circularly polarised spectra. The tools and assumptions used to perform temporal analyses and Gaussian process regression are outlined in Sect.~\ref{sec:time_analysis}, and the principles of Zeeman-Doppler imaging in Sect.~\ref{sec:zdi}. We present our results in Sect.~\ref{sec:results} for each star, and we discuss our findings in Sect.~\ref{sec:discussion}. Finally, we draw our conclusions in Sect.~\ref{sec:conclusions}.

\section{Observations}\label{sec:observations}

Our study focuses on six solar-like stars that were observed as part of the BCool program \citep{Marsden2014}: HD\,9986, HD\,56124, HD\,73350, HD\,76151, HD\,166435, and HD\,175726. The properties are listed in Table~\ref{tab:stars_properties}. The effective temperature of our sample stars ranges from 5790 to 5998\,K and the mass between 1.022 and 1.058\,M$_\odot$. HD\,9986 and HD\,56124 are the most similar to the Sun in terms of rotation and age, with a rotation rate that is at most 1.3 faster the solar value. HD\,166435, and HD\,175726 are the fastest rotators among our stars, with rotation rates 7.8 and 6.6 times solar, and correspondingly they are the most magnetically active. Finally, HD\,73350 and HD\,76151 show an intermediate rotation, with a rotation rate of 2.2 and 1.5 times faster than solar, respectively.  Although small, our sample of stars is representative of Sun-like stars with different activity levels, and it is thus suitable to investigate the presence and shape of magnetic cycles depending on stellar rotation. Ultimately, this helps us putting the solar Hale cycle into a broader context.

\subsection{ESPaDOnS, Narval, and Neo-Narval}

We analysed optical spectropolarimetric observations collected with Narval between 2007 and 2019. Narval is the spectropolarimeter on the 2~m T\'elescope Bernard Lyot (TBL) at the Pic du Midi Observatory in France \citep{Donati2003}, which operates between 370 and 1050\,nm at high resolution ($R\sim 65,000$). As of September 2019, Narval was upgraded to Neo-Narval, with the installation of a new detector and improved velocimetric capabilities \citep{LopezAriste2022}. The instrument maintains the main performances of Narval: a spectral coverage from 380 to 1050 nm, and a median spectral resolving power of $65,000$ after data reduction. From 2019 to 2024, our observations were performed with Neo-Narval. We also included ESPaDOnS observations in our analyses, which is the twin spectropolarimeter on the 3.6 m Canada-France-Hawaii-Telescope (CFHT) located atop Mauna Kea in Hawaii \citep{Donati2003}. Combining observations of these instruments improves the temporal sampling of our time series, considering that ESPaDOnS is mounted at CFHT for a small fraction of time and (Neo-)Narval suffers from poorer weather conditions at TBL.

A polarimetric sequence is obtained from four consecutive sub-exposures. Each sub-exposure is taken with a different rotation of the retarder waveplate of the polarimeter relative to the optical axis. The observations were carried out in circular polarisation mode, hence they provide unpolarised (Stokes~$I$), circularly polarised (Stokes~$V$) and null (Stokes~$N$) high-resolution spectra. The Stokes~$I$ spectrum is computed by summing the four sub-exposures, the Stokes~$V$ spectrum from the ratio of sub-exposures with orthogonal polarisation states, while the Stokes~$N$ from the ratio of sub-exposures with the same polarisation states. The Stokes~$N$ spectrum is a useful check for the presence of spurious polarisation signatures \citep[see][for more details]{Donati1997,Bagnulo2009,Tessore2017}. The data were reduced with the \texttt{LIBRE-ESPRIT} pipeline \citep{Donati1997}, and the continuum-normalised spectra were retrieved from PolarBase \citep{Petit2014}. For Neo-Narval observations, a different reduction pipeline was used \citep{LopezAriste2022}.

We used least-squares deconvolution \citep[LSD;][]{Donati1997} to compute average line profiles from the unpolarised, circularly polarised and null spectra. In practice, we adopted the python implementation \textsc{lsdpy} \footnote{Available at \url{https://github.com/folsomcp/LSDpy}}. This numerical technique combines the information of thousands of photospheric spectral lines included in a synthetic line list, which is a series of Dirac delta functions located at each absorption line in the stellar spectrum and with the associated line features such as depth, and Land\'e factor (encapsulating the line sensitivity to Zeeman effect and indicated as $g_\mathrm{eff}$). To respect the requirement of self-similarity \citep[e.g.][]{Kochukhov2010a}, the spectral lines contained in the list are only metal lines (hydrogen and helium lines are excluded). The line lists were produced using the Vienna Atomic Line Database\footnote{\url{http://vald.astro.uu.se/}} \citep[VALD,][]{Ryabchikova2015}. The effective temperature and the surface gravity of the model were selected to be close to the value reported in the literature. They contain information of atomic lines with known Land\'e factor and with depth larger than 40\% the level of the unpolarised continuum. 

\begin{figure}
    \includegraphics[width=\columnwidth]{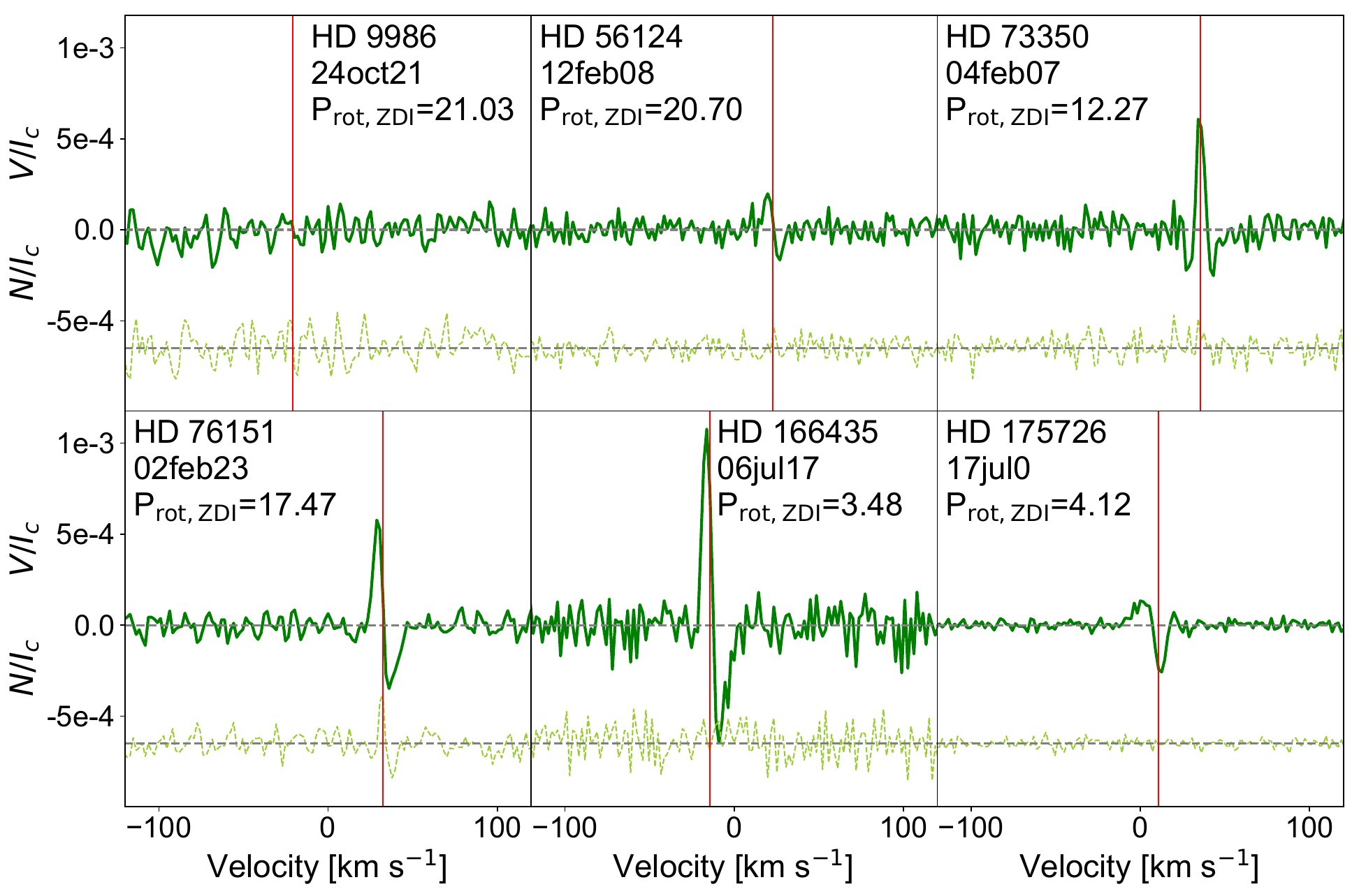}
    \caption{Least-squares deconvolution profiles for the six solar-like stars examined in this work. Each panel corresponds to a different star and contains one typical example of the Stokes~$V$ (solid green line) and Stokes~$N$ (dashed green line) profiles. The vertical red dotted line indicates the radial velocity of the star, and the stellar rotation period obtained with ZDI and date of observations are included.
	}\label{fig:stokes_VN}
\end{figure}

The full list of observations is provided in Table~\ref{tab:log} and examples of Stokes~$V$ profiles are shown in Fig.~\ref{fig:stokes_VN}. The vertical dotted line in the plots indicates the radial velocity of the star. The latter is computed as the centroid of the Stokes~$I$ profile, which is modelled with a Voigt kernel and a linear component to account for residuals of continuum normalisation. We recorded substantially lower S/N in Stokes~$V$ LSD profiles for six observations of HD\,9986 on 11 October 2011, 17 November 2020, 07 September 2021, 26 September 2021, 24 October 2021, and 06 Feb 2023, and a double-peaked Stokes~$I$ profile on 28 October 2012, clearly outlying with respect to all other Stokes~$I$ profiles. These seven observations were therefore not used for the analyses. We did not detect a clear Zeeman signature in circularly polarised light for the 2020 and 2021 Neo-Narval time series, hence they were not used in the analyses outlined below. In addition, we removed two low-S/N observations for HD\,56124 on 2 November 2017 and 19 November 2021, eight observations for HD\,76151 on 25 February 2021, 25 March 2021, 11 May 2021, 17 May 2021, 29 January 2022, 17 January 2022, 28 February 2023, 25 January 2024, one observation for HD\,166435 on 31 August 2020, and one observation for HD\,175726 on 10 July 2008.

Using the Stokes~$N$ LSD profiles to check for spurious signals, we noticed that some of the Narval observations exhibited a signature with positive sign centred at the radial velocity of the star. In previous studies \citep{Folsom2016,Bellotti2023a}, such signal was attributed to an imperfect background subtraction during data reduction, and it was removed by computing the LSD profiles using only the red region of the spectra ($\lambda>500$\,nm), but such mitigation was not effective in our case. We noticed that the Stokes~$N$ signal was not present for all stars, and in most cases it only manifested for a limited number of observations within an epoch. Furthermore, when the signal was present, its shape appeared to be systematically the same, but without affecting the Stokes~$V$ profile in an evident way. Indeed, the Stokes~$V$ profile shape and amplitude is the same between two observations close in time, whether the Stokes~$N$ profile is present or not. The Stokes~$N$ signal is likely stemming from an instrumental effect because, following the same reasoning as \citet{Mathias2018}, we did not find this signal in the ESPaDOnS observations of HD\,76151 on 7 and 9 January 2018, whereas it was present in the Narval observation on 24 January 2018. Furthermore, there are observations where the Stokes~$N$ signature is present, while there is no detected Stokes~$V$ signature, which suggests that this Stokes~$N$ signature does not leak into Stokes~$V$. In conclusion, despite the presence of a Stokes~$N$ signal in some observations, the spectra can be used for reliable spectropolarimetric characterisation of the stellar magnetic field. 

In the next sections, the observations will be phased with the following ephemeris
\begin{align}
    \mathrm{HJD} = \mathrm{HJD}_0 + \mathrm{P}_\mathrm{rot}\cdot n_\mathrm{cyc},
    \label{eq:ephemeris}
\end{align}
where HJD$_\mathrm{0}$ is the heliocentric Julian Date reference (the first one of the time series for each star, see Table~\ref{tab:log}), P$_\mathrm{rot}$ is the stellar rotation period of the star (see Table~\ref{tab:stars_properties}), and $n_\mathrm{cyc}$ represents the number of the rotation cycle. In Table~\ref{tab:stars_properties}, we also list the Rossby number, which is the rotation period normalised by the convective turnover time ($Ro=\mathrm{P}_\mathrm{rot}/\tau_\mathrm{cyc}$), and encapsulates the interplay between convection and rotation, two main ingredients for stellar dynamo. The values were computed by \citet{See2019}.

\setlength{\tabcolsep}{4.5pt}
\begin{table*}[!t]
\caption{Properties of our sample stars in comparison to the Sun.} 
\label{tab:stars_properties}     
\centering                       
\begin{tabular}{l c c c c c c c c c c c c c}    
\toprule
Name & N$_\mathrm{obs}$ & V  & Dist  & T$_\mathrm{eff}$  & $\log g$ & Mass & Radius & Age & $Ro$ & P$_\mathrm{rot}$ & $v_\mathrm{eq}\sin i$ & $i$ & d$\Omega$ \\
& & [mag] & [pc] & [K] & & [M$_\odot$] & [R$_\odot$] & [Gyr] &  & [d] & [km\,s$^{-1}$] & [$^\circ$] & [rad\,d$^{-1}$]\\
\midrule
Sun        & $\ldots$ & $-$26.7 & \ldots& 5772 & 4.44 & 1.000 & 1.00 & 4.50 & 2.19 & 25.4 & 2.0 & \ldots & 0.07\\
HD\,9986   & 120 & 6.76 & 25.44 & 5805 & 4.43 & 1.022 & 1.04 & 3.74 & 1.80 & $21.03\pm0.44$ & 2.6  & 60 & \ldots\\ 
HD\,56124  & 74 & 6.94 & 27.25 & 5848 & 4.45 & 1.029 & 1.01 & 3.88 & 1.50 & $20.70\pm0.32$ & 1.5  & 40 & \ldots\\ 
HD\,73350  & 33 & 6.73 & 24.35 & 5802 & 4.49 & 1.038 & 0.98 & 1.43 & 0.93 & $12.27\pm0.13$ & 4.0  & 70 & \ldots\\ 
HD\,76151  & 149 & 6.00 & 16.85 & 5790 & 4.49 & 1.056 & 1.00 & 2.09 & 1.50 & $17.47\pm0.81$ & 1.2  & 30 & \ldots\\ 
HD\,166435 & 82 & 6.83 & 24.41 & 5843 & 4.47 & 1.039 & 0.99 & 0.27 & 0.48  & $3.48\pm0.01$  & 7.9  & 40 & $0.14\pm0.01$\\ 
HD\,175726 & 89 & 6.71 & 26.59 & 5998 & 4.43 & 1.058 & 1.06 & 0.38 & 0.31  & $4.12\pm0.03$  & 12.3 & 70 & $0.15\pm0.03$\\ 
\bottomrule 
\end{tabular}
\tablefoot{The columns are: identifier of the star, total number of spectropolarimetric observations, V band apparent magnitude, distance, effective temperature, surface gravity, mass, radius, age, Rossby number, rotation period computed with ZDI, projected equatorial velocity, stellar inclination, and latitudinal differential rotation rate computed with ZDI. Visual magnitudes were extracted from SIMBAD \citep{Wenger2000} and the distances were computed from $Gaia$ parallaxes \citep{GaiaCollaboration2020}. The stellar inclination was inferred from geometrical considerations (see Sect.~\ref{sec:zdi}). The Rossby number is taken from \citet{See2019} and the remaining parameters from \citet{Marsden2014} and references therein. The solar parameters were extracted or derived from the values in the NASA Sun fact sheet available at \url{https://nssdc.gsfc.nasa.gov/planetary/factsheet/sunfact.html}. The age of the Sun was taken from \citet{Guenter1989}.}
\end{table*}
\setlength{\tabcolsep}{6pt}

\subsection{TESS}\label{sec:tess}

All our stars except HD\,175726 were observed by the Transiting Exoplanet Survey Satellite \citep[$TESS$;][]{Ricker2014}. Considering that the typical time span of $TESS$ light curves is 20-30\,d, and that our primary usage is to infer stellar rotation periods, we decided to use photometric data only for our fast rotator HD\,166435. This way, the light curves are representative of multiple stellar rotations and can be used efficiently for temporal analyses (see Sect.\ref{sec:periodogram}). For the remaining stars, their rotation period is of the same order of magnitude as the light curve time span, therefore an extraction of the stellar rotation period is not reliable. In addition, for quiet stars like these, the photometric amplitude can become very small, which makes the extraction of rotational modulation from $TESS$ light curves even more challenging.

HD\,166435 was observed by $TESS$ in June/July 2020, 2021, and 2022 as part of sector 26, 40, and 53, respectively. We analysed the Pre-search Data Conditioning Single Aperture Photometry (PDC-SAP) light curves publicly available at the Mikulski Archive for Space Telescope (MAST)\footnote{\url{https://archive.stsci.edu/}}, in which the reduction pipeline has already corrected the photometric flux for instrumental systematics. We further removed data points whose quality flag was different than zero, symbolising data conditions outside nominal values (e.g. flares).

Each light curve of HD\,166435 shows a smooth modulation of the photometric flux, as shown in Fig.~\ref{fig:LSP_tess}. Following \citet{Petit2021}, we binned the data using a window of 0.2\,d in order to reduce the number of data points while preserving the light curve modulation (we also used a window of 0.05\,d but the results did not change). The error bar of each bin was computed using either the median error of the bin or an inverse-variance weighting scheme \citep{Petit2021}. The results of the temporal analysis (see Sec~\ref{sec:results}) are robust with respect to the choice of error bar formalism.

\section{Longitudinal magnetic field}\label{sec:Bl}

The longitudinal magnetic field (B$_l$) is the line-of-sight component of the magnetic field integrated over the stellar disk. We use the centre-of-gravity prescription of \citet{Rees1979} to compute B$_l$. Formally, it is the first-order moment of the Stokes~$V$ LSD profile
\begin{equation}
\mathrm{B}_l\;[G] = \frac{-2.14\cdot10^{11}}{\lambda_0 \mathrm{g}_{\mathrm{eff}}c}\frac{\int vV(v)dv}{\int(I_c-I)dv} \,,
\label{eq:Bl}
\end{equation}
where $\lambda_0$ and $\mathrm{g}_\mathrm{eff}$ are the normalisation wavelength (in nm) and Land\'e factor of the LSD profiles, $I_c$ is the continuum level, $v$ is the radial velocity associated to a point in the spectral line profile in the star's rest frame (in km\,s$^{-1}$) and $c$ the speed of light in vacuum (in km\,s$^{-1}$). 

For all our stars, we set the normalisation parameters to $\lambda_0=700$\,nm and $\mathrm{g}_\mathrm{eff}=1.2$. The velocity range over which the integration is carried out should encompass the width of both Stokes~$I$ and $V$ LSD profiles. One way to determine the velocity interval is to visually inspect the median Stokes~$V$ profile and identify its lobes. Another way consists of computing the standard deviation per velocity bin of the Stokes~$V$ profile, across the observations. This procedure allows one to easily locate regions of large dispersion, which correspond to the lobes of the Stokes~$V$ profile. We set the velocity interval to 20\,km\,s$^{-1}$ for all our stars except HD\,76151 and HD\,175726 for which we set 15\,km\,s$^{-1}$ and 25\,km\,s$^{-1}$, respectively. The same ranges are used for the ZDI reconstructions (see Sect.~\ref{sec:zdi}).

The longitudinal magnetic field is a practical magnetic activity diagnostics because of its sensitivity to magnetic regions on the visible stellar hemisphere. The surface distribution of the magnetic regions may not be axisymmetric, making the variations of B$_l$ modulated to the stellar rotation period. For this reason, the stellar rotation period can be inferred via periodograms \citep{Hebrard2016,Folsom2018,Petit2021,Klein2021,Carmona2023} or Gaussian process regression \citep[e.g.][]{Yu2019,Fouque2023,Donati2023,Bellotti2023a,Rescigno2024}. Moreover, the direct link with the Stokes~$V$ Zeeman signatures makes B$_l$ a useful tool for a preliminary assessment of large-scale magnetic field topologies \citep[e.g.][]{Bellotti2023b,Lehmann2024,Bellotti2024}. This quantity represents an average over the stellar disk, while tomographic inversion (see Sect.\ref{sec:zdi}) provides more details of the magnetic geometry.

\section{Temporal analysis}\label{sec:time_analysis}

\subsection{Periodogram}\label{sec:periodogram}

We applied a generalised Lomb-Scargle periodogram \citep[][and references therein]{Zechmeister2009} to the full B$_l$ time series, in order to search for the main periodicities in the time series. The algorithm proceeds by fitting sinusoidal models at distinct period values (or equivalently, frequency) over a selected grid \citep[for more details see e.g.][]{VanderPlas2018}. This way it is possible to characterise the periodic content for a time series with uneven cadence. The metric for the significance of a periodicity is the False Alarm probability (FAP), which measures how likely it is that random noise can generate a signal with the same periodicity. 

In this work, we considered a grid of periodicities between 1 and $10^4$\,d, to investigate both short (i.e. rotation) and long (i.e. cycles) time scales. We also computed the window function, which is a good indicator of spurious signals and aliases due to the observing cadence in the data sets \citep{VanderPlas2018}.

\subsection{Gaussian Process regression}\label{sec:gp}

We employed Gaussian Processes (GP) to characterise the long-term evolution of the longitudinal magnetic field. They are a statistical tool to define a probability distribution over functions, which is especially practical to find a functional form that describes the variations of a time series \citep[for more details see for instance][]{Haywood2014,Angus2018,Aigrain2022}. Compared to a standard Lomb-Scargle periodogram, the GP model allows more flexibility by including additional evolution time scales that make the variations deviate from strictly periodic, which is the case also for the Sun \citep{Usoskin2008,Charbonneau2010}. Moreover, \citet{Olspert2018} applied a quasi-periodic GP on chromospheric $S$-index data of solar-like stars to search for cycles, and found that such statistical tool performs better than a periodogram.

We adopted the quasi-periodic covariance kernel
\begin{equation}\label{eq:qp_kernel}
    k(t,t') = \theta_1^2\exp\left[-\frac{(t-t')^2}{\theta_2^2}-\frac{1}{\theta_4^2}\sin^2\left(\frac{\pi(t-t')}{\theta_3}\right) \right] + S^2\delta_{t,t'},
\end{equation}
where $\delta_{t,t'}$ is a Kronecker delta, and $\theta_i$ are the hyperparameters of the model: $\theta_1$ is the amplitude of the curve in G, $\theta_2$ is the evolution timescale in d expressing how rapidly the modulation of B$_l$ evolves, $\theta_3$ is the recurrence timescale (i.e. the rotation period P$_\mathrm{rot}$) in d, and $\theta_4$ is the smoothness factor which determines the harmonic structure of the curve (dimensionless). We added an additional hyperparameter ($S$, in G) to account for the excess of uncorrelated noise, which acts only on the diagonal of the covariance matrix. The log likelihood function to maximise is the following
\begin{equation}\label{eq:gp_like}
    \log \mathcal{L} = -\frac{1}{2}\left( n\log(2\pi) + \log|K+\Sigma| + y^T(K+\Sigma)^{-1}y \right),
\end{equation}
where $y$ is the array containing the $n$ values of B$_l$ we measured, $K$ is the covariance matrix built with the kernel in Eq.~\ref{eq:qp_kernel}, and $\Sigma$ is the diagonal variance matrix of our measured B$_l$.

A nested sampling algorithm \citep{Skilling2004} is used to explore the posterior distribution of the five hyperparameters ($\theta_i$ and $S$) by means of the python package \textsc{cpnest} \citep{DelPozzo2022}. Nested sampling was applied with 2,000 live points and using uniform priors for all the hyperparameters. The details of the adopted prior distributions are given in Table~\ref{tab:gp}. The error bars are the 16th and 84th percentiles of the posterior distribution, with which it is possible to capture asymmetries of the distribution and potential harmonic (multi-peak) structures, as described in Sect.\ref{sec:results}.

\renewcommand{\arraystretch}{1.5}
\begin{table*}[!t]
\caption{Results of the GP fit carried out on the B$_l$ time series for all our stars.}
\label{tab:gp}
\centering
\begin{tabular}{l l l l l l l l} 
\hline
Hyperparameter & Prior & HD\,9986 & HD\,56124 & HD\,73350 & HD\,76151 & HD\,166435 & HD\,175726\\
\hline
B$_l$ amplitude [G] ($\theta_1$) & $\mathcal{U}(0,100)$ & $0.8^{+0.5}_{-0.3}$ & $1.8^{+0.7}_{-0.5}$ & $3.2^{+1.6}_{-0.9}$ & $2.8^{+0.4}_{-0.4}$ & $5.2^{+1.7}_{-1.2}$ & $4.7^{+1.0}_{-0.9}$\\
Evolution time [d] ($\theta_2$) & $\mathcal{U}(1,3000)$ & $852^{+497}_{-375}$ & $511^{+390}_{-275}$ & $1497^{+1002}_{-931}$ & $232^{+40}_{-41}$ & $652^{+541}_{-293}$ & $148^{+1954}_{-140}$\\
P$_\mathrm{rot}$ [d] ($\theta_3$) & $\mathcal{U}(1,50)^*$ & $22.76^{+2.36}_{-2.36}$ & $21.32^{+1.96}_{-2.01}$ & $14.20^{+13.06}_{-1.79}$ & $16.70^{+0.18}_{-0.16}$ & $3.52^{+0.01}_{-0.03}$ & $4.04^{+0.11}_{-0.11}$\\
Smoothness ($\theta_4$) & $\mathcal{U}(0.1,1.2)$ & $0.4^{+0.3}_{-0.2}$ & $0.9^{+0.3}_{-0.3}$ & $0.2^{+0.4}_{-0.3}$ & $1.2^{+0.1}_{-0.1}$ & $1.0^{+0.2}_{-0.2}$ & $0.1^{+0.2}_{-0.1}$\\
Uncorrelated noise [G] ($S$) & $\mathcal{U}(0,100)$ & $0.03^{+0.17}_{-0.11}$ & $0.93^{+0.26}_{-0.24}$ & $1.28^{+0.56}_{-0.61}$ & $0.71^{+0.11}_{-0.11}$ & $3.91^{+0.44}_{-0.39}$ & $1.45^{+0.73}_{-0.85}$\\
$\chi^2_r$ & & 0.63 & 2.2 & 1.50 & 1.74 & 18.9 & 0.85\\
Residuals (RMS, [G]) & & 0.83 & 1.7 & 2.13 & 1.06 & 4.26 & 1.27\\
\hline
\end{tabular}
\tablefoot{The columns are: hyperparameter name, uniform prior distribution of the form $\mathcal{U}(\mathrm{min},\mathrm{max})$, and mode of the posterior distribution for each star. The error bars are the 16th and 84th percentiles of the posterior distribution. The rows list the five hyperparameters of the GP along with the $\chi^2$ of the model and the RMS scatter of the residuals. $^*$ the uniform prior was restricted to 1-10\,d for HD\,175726 and was changed to a Gaussian prior $\mathcal{G}(3.47\mbox{ d},0.10\mbox{ d})$ for HD\,166435 (see Sect.~\ref{sec:results}).}
\end{table*}
\renewcommand{\arraystretch}{1.}

\section{Zeeman-Doppler imaging}\label{sec:zdi}

Zeeman-Doppler imaging was applied to reconstruct the large-scale magnetic field topology for the stars in our study. One map was obtained for each epoch in which a star was observed, provided a sufficient number of observations were collected or a sufficient number of circularly polarised Zeeman signatures were detected. The ZDI algorithm inverts a time series of Stokes~$V$ LSD profiles into a magnetic field map in an iterative fashion \citep[for more information see][]{Skilling1984,DonatiBrown1997}. More precisely, synthetic Stokes~$V$ profiles are compared and updated with respect to the observed ones at each iteration, until convergence at a specific target $\chi^2_r$ is reached. Such a problem is ill-posed, meaning that infinite solutions could fit the observed data equally well, thus ZDI employs a regularisation scheme based on maximum entropy to choose a solution \citep{Skilling1984}. The algorithm searches for the maximum-entropy solution at a given $\chi^2$ level, that is, the magnetic field configuration compatible with the data and with the lowest information content. 

The magnetic field vector is expressed as the sum of poloidal and toroidal components, each described via a spherical harmonics formalism. Specifically, we employed the decomposition described in \citet{Lehmann2022}. The simulated spherical surface of the star is divided into 1000 cells of approximately equal area and the local Stokes~$I$ and $V$ profiles for each cell are calculated assuming the weak-field approximation. Stokes~$I$ LSD profiles are modelled with a Voigt kernel, and the weak-field approximation allows us to describe Stokes~$V$ as proportional to the first derivative of $I$ with respect to velocity
\begin{equation}\label{eq:wfa}
    V(v) = -\Delta\lambda_B\cos\gamma\frac{\mathrm{d}I}{\mathrm{d}v},
\end{equation}
where $\Delta\lambda_B$ is the Zeeman splitting in wavelength units and $\gamma$ is the angle between the magnetic field vector and the line of sight \citep[see][for more details]{Landi1992}. The choice of weak-field approximation is typically valid until the field strength reaches 1\,kG \citep{Kochukhov2010a}, and it is justified in our work because local field strengths do not exceed 70\,G for any of our stars (see Sect.~\ref{sec:results}). Note that magnetic fields at unresolved spatial levels likely exceed 1\,kG, as demonstrated by Zeeman broadening measurements \citep[e.g.][]{Robinson1980,Kochukhov2020,Hahlin2023}.

Our model further assumes that there are no large-scale brightness inhomogeneities over the stellar surface, so that all synthetic Stokes~$I$ profiles do not vary over the photosphere. This assumption is probably well verified for low-activity stars for which, by analogy with the Sun, most brightness inhomogeneities (e.g. starspots) are expected to
be restricted to spatial scales much smaller than the typical extent of magnetic regions resolved here. 

We employed the \texttt{zdipy} code described in \citet{Folsom2018}. We set the linear limb darkening coefficient to 0.7 \citep{Claret2011} and the maximum degree of spherical harmonic coefficients to $\ell_\mathrm{max}=8$, except for the fast rotators, for which we use $\ell_\mathrm{max}=15$. This choice is dictated by the projected equatorial velocity ($v_\mathrm{eq}\sin i$) of our stars. Note however, that most of the magnetic energy is stored in the $\ell\leq5$ modes as explained in Sect.~\ref{sec:results} and listed in Table~\ref{tab:zdi_output} \citep[see also][for more details]{Lehmann2019}.

The \textsc{zdipy} code includes solar-like latitudinal differential rotation as a function of colatitude ($\theta$), expressed in the form
\begin{equation}\label{eq:diff_rot}
\Omega(\theta) = \Omega_\mathrm{eq} - d\Omega\sin^2(\theta),
\end{equation}
where $\Omega_\mathrm{eq}=2\pi/\mathrm{P}_\mathrm{rot}$ is the rotational frequency at equator and $d\Omega$ is the differential rotation rate in rad\,d$^{-1}$. For all epochs of each star, we jointly searched for the optimised value of equatorial projected rotation period and $d\Omega$ following \citet{Donati2000} and \citet{Petit2002}. We generated a grid of (P$_\mathrm{rot}$, $d\Omega$) pairs and searched for the pair that minimised the $\chi^2$ distribution between observations and synthetic LSD profiles, at a fixed entropy level. The best parameters are measured by fitting a 2D paraboloid to the $\chi^2$ distribution, and the error bars are obtained from a variation of $\Delta\chi^2 = 1$ away from the minimum \citep{Press1992,Petit2002}. The latitudinal differential rotation search was performed for the epochs whose time span is between two and five weeks, allowing the latitudinal surface shear to distort the magnetic features and be possibly detected. If an epoch spanned more than five weeks, we performed the search on both the full epoch and subsets of it, provided that the number of observations examined is at least ten and with reasonable longitudinal coverage of the stellar rotation. We proceeded this way since it is known that the magnetic field topology of Sun-like stars may change rapidly on time scales of months \citep[e.g.][]{Morgenthaler2011,Jeffers2018}.

All the stars in our sample have rotation period estimates, computed from chromospheric activity indicators in \citet{Marsden2014}. When applying ZDI, we decided to optimise the stellar rotation period for each star. Unless this is performed in conjunction with the differential rotation search, the P$_\mathrm{rot}$ optimisation proceeds in a similar manner, but it generates a $\chi^2_r$ distribution in 1D instead of 2D. The final value and error bars are obtained by fitting a parabola to the minimum of the $\chi^2_r$ curve. For each star, we optimised P$_\mathrm{rot}$ for every epoch in which ZDI is applicable. We then computed the median P$_\mathrm{rot}$ and its error bar as the standard deviation of the measurements. The median value, which is reported in Table~\ref{tab:stars_properties}, is assumed for ZDI reconstructions of all epochs for a specific star (see Sect.~\ref{sec:results} for more details). The colour of the maps encodes the polarity and strength (in G) of the magnetic field, therefore it highlights whether a polarity reversal has occurred.

The stellar inclination was estimated comparing the stellar radius provided in the literature with the projected radius $R\sin i=\mathrm{P}_\mathrm{rot}v_\mathrm{eq}\sin i/50.59$, where $R\sin(i)$ is measured in solar radii, P$_\mathrm{rot}$ in days, and $v_\mathrm{eq}\sin i$ in km\,s$^{-1}$. If the estimated inclination was larger than $80^{\circ}$, we adopted a value of $70^{\circ}$ to conservatively prevent mirroring effects between the stellar north and south pole. Indeed, for a high inclination value, an ambiguity between north and south hemisphere would appear, and the spherical harmonics modes with odd $\ell$ and $m = 0$ would cancel out. The properties of the ZDI maps and the results of the differential rotation search are summarised in Table~\ref{tab:zdi_output}. 

\begin{table*}[!t]
\caption{Properties of the magnetic maps.} 
\label{tab:zdi_output}     
\centering                       
\begin{tabular}{l c c c c c c c c c c c c}      
\toprule
Star & Epoch & $\chi^2_r$ & B$_V$   & B$_\mathrm{max}$ & $f_\mathrm{pol}$ & $f_\mathrm{tor}$  & $f_\mathrm{dip}$   & $f_\mathrm{quad}$  & $f_\mathrm{oct}$   & $f_\mathrm{axisym}$ & $f_\mathrm{axisym,pol}$ & $f_\mathrm{axisym,tor}$ \\
& & & [G] & [G] & [\%] & [\%] & [\%] & [\%] & [\%] & [\%] & [\%] & [\%] \\
\midrule
HD\,9986  & 2008.08 & 1.00 & 1.5 & 4.1 & 74.8 & 25.2 & 66.7 & 21.0 & 11.1 & 37.7 & 18.1 & 95.7\\
          & 2010.76 & 1.20 & 1.3 & 4.1 & 87.5 & 12.5 & 58.5 & 25.4 & 11.8 & 27.1 & 18.7 & 86.2\\
          & 2011.78 & 1.12 & 1.2 & 2.7 & 87.7 & 12.3 & 61.0 & 24.4 & 11.2 & 24.2 & 14.7 & 92.2\\
          & 2012.85 & 1.01 & 1.6 & 3.2 & 98.6 & 1.4  & 79.9 & 10.1 & 7.3  & 16.2 & 16.2 & 21.9\\
          & 2017.76 & 1.20 & 1.9 & 3.6 & 77.1 & 22.9 & 67.1 & 20.6 & 10.6 & 55.3 & 43.6 & 95.1\\
          & 2018.74 & 1.07 & 2.6 & 5.0 & 58.4 & 41.6 & 88.1 & 8.4  & 2.7  & 60.1 & 32.1 & 99.3\\
          & 2023.09 & 0.97 & 1.9 & 4.5 & 79.0 & 21.0 & 83.9 & 10.9 & 3.4  & 19.4 & 0.1 & 92.0\\
\midrule
HD\,56124 & 2008.08 & 1.10 & 3.3 & 6.7 & 94.8 & 5.2 & 97.3 & 2.4 & 0.2 & 90.7 & 90.4 & 96.2\\
          & 2011.90 & 1.15 & 2.3 & 4.6 & 99.6 & 0.4 & 92.2 & 5.0 & 2.4 & 80.3 & 80.3 & 63.2\\
          & 2017.88 & 1.14 & 0.7 & 1.4 & 95.6 & 4.4 & 94.0 & 4.3 & 1.6 & 85.7 & 85.5 & 90.1\\
          & 2021.29 & 0.97 & 2.5 & 5.4 & 98.0 & 2.0 & 87.8 & 6.8 & 4.4 & 68.8 & 69.5 & 37.4\\
\midrule
HD\,73350 & 2007.09 & 1.80 & 10.2 & 31.4 & 54.2 & 45.8 & 37.4 & 26.6 & 23.4 & 43.7 & 0.4  & 94.2\\
          & 2011.06 & 1.40 & 11.3 & 20.9 & 47.4 & 52.6 & 68.2 & 19.4 & 9.5 & 79.8 & 58.9 & 98.7\\
          & 2012.04 & 1.25 & 6.1  & 12.6 & 99.1 & 0.9  & 83.4 & 9.8 & 5.1  & 46.8 & 46.6 & 76.3\\
\midrule
HD\,76151 & 2007.09 & 1.30 & 3.7 & 7.6  & 97.7 & 2.3 & 93.0 & 5.2  & 1.6 & 74.5 & 74.2 & 85.6\\
          & 2009.95 & 1.25 & 2.7 & 5.4  & 99.0 & 1.0 & 92.5 & 5.6  & 1.8 & 86.0 & 85.9 & 96.6\\
          & 2012.05 & 1.37 & 1.0 & 2.1  & 98.0 & 2.0 & 82.4 & 11.4 & 5.9 & 44.8 & 43.8 & 94.8\\
          & 2015.95 & 2.65 & 6.2 & 12.4 & 95.9 & 4.1 & 93.8 & 4.8  & 1.2 & 93.0 & 92.7 & 98.8\\
          & 2017.02 & 1.48 & 2.1 & 4.6  & 98.4 & 1.6 & 86.1 & 8.5  & 4.9 & 45.2 & 44.6 & 82.1\\
          & 2019.02 & 1.60 & 2.1 & 4.8  & 96.9 & 3.1 & 80.1 & 11.8 & 7.5 & 5.2  & 2.5  & 89.7\\
          & 2021.25 & 1.90 & 4.7 & 9.7  & 97.7 & 2.3 & 95.0 & 4.4  & 0.6 & 95.2 & 95.3 & 91.0\\
          & 2022.07 & 1.55 & 3.6 & 8.2  & 83.7 & 16.3 & 91.6 & 6.9  & 1.2 & 72.6 & 69.6 & 88.0\\
          & 2022.28 & 1.22 & 3.3 & 8.2  & 97.1 & 2.9 & 83.9 & 9.4 & 5.4 & 5.0 & 3.8 & 42.8\\
          & 2023.10 & 1.77 & 8.5 & 17.5 & 95.6 & 4.4 & 93.6 & 5.0 & 1.2 & 87.8 & 87.5 & 94.4\\
          & 2024.06 & 1.24 & 4.0 & 9.0 & 97.5 & 2.5 & 85.6 & 8.7 & 5.0 & 3.9 & 3.8 & 10.6\\
\midrule
HD\,166435 & 2010.51 & 2.00 & 12.7 & 45.6 & 66.6 & 33.4 & 22.2 & 26.9 & 22.9 & 32.2 & 11.2 & 74.3\\
           & 2010.60 & 2.00 & 15.5 & 35.8 & 61.5 & 38.5 & 33.2 & 32.1 & 21.2 & 56.6 & 37.3 & 87.4\\
           & 2011.52 & 2.00 & 23.4 & 62.7 & 68.2 & 31.7 & 34.4 & 29.1 & 20.6 & 49.7 & 37.7 & 75.5\\
           & 2016.49 & 4.00 & 8.6  & 23.3 & 87.4 & 12.6 & 32.1 & 16.1 & 20.3 & 18.9 & 14.2 & 52.1\\
           & 2017.35 & 2.50 & 18.3 & 53.8 & 62.3 & 37.7 & 40.9 & 20.7 & 16.5 & 51.3 & 27.1 & 91.2\\
           & 2020.59 & 1.50 & 19.2 & 43.1 & 65.1 & 34.9 & 67.5 & 9.5  & 8.9  & 66.1 & 56.2 & 84.5\\
\midrule
HD\,175726 & 2008.55 & 1.60 & 10.4 & 24.3 & 70.0 & 30.0 & 37.4 & 26.6 & 8.3  & 45.9 & 28.9 & 85.8\\
           & 2008.63 & 1.70 & 2.9  & 7.4  & 78.9 & 21.1 & 34.5 & 35.9 & 20.2 & 35.4 & 21.8 & 86.2\\
           & 2016.53 & 1.60 & 8.0  & 20.6 & 89.2 & 10.8 & 31.6 & 22.2 & 13.6 & 47.1 & 48.5 & 36.1\\
           & 2024.53 & 0.84 & 6.2  & 14.1 & 86.8 & 13.2 & 20.1 & 30.3 & 30.9 & 22.9 & 15.8 & 69.1\\
           & 2024.63 & 1.00 & 11.1 & 32.8 & 94.6 & 5.4 & 17.5 & 41.6 & 24.5 & 27.0 & 27.8 & 12.9\\
\bottomrule                                
\end{tabular}
\tablefoot{The following quantities are listed: star's name, median epoch of observations in decimal units, target $\chi^2_r$ of the ZDI reconstruction, mean unsigned magnetic strength, maximum unsigned magnetic strength, poloidal and toroidal magnetic energies as a fraction of the total energy, dipolar, quadrupolar, and octupolar magnetic energy as a fraction of the poloidal energy, axisymmetric magnetic energy as a fraction of the total energy, poloidal axisymmetric energy as a fraction of the poloidal energy, toroidal axisymmetric energy as a fraction of the toroidal energy.}
\end{table*}

\section{Results}\label{sec:results}

\subsection{HD\,9986 (HIP\,7585)}

HD\,9986 is a solar analog \citep{PortodeMello2014,Datson2015} and the star in our sample with properties most similar to the Sun (see Table~\ref{tab:stars_properties}). It is a G5~dwarf with an age of 3.7\,Gyr and a rotation period of 22.4\,d \citep{Marsden2014}. Previous studies have reported measurements of the chromospheric activity index $\log R'_\mathrm{HK}$ between $-4.93$ and $-4.83$ \citep{Wright2004, Isaacson2010, Pace2013, BoroSaikia2018a, GomesdaSilva2021}. This means that the star is slightly more active than the Sun, the latter exhibiting $\log R'_\mathrm{HK}=-4.905$ and $-4.984$ at cycle maxima and minima, respectively \citep{Egeland2017}. 

\begin{figure}[t]
    \includegraphics[width=\columnwidth]{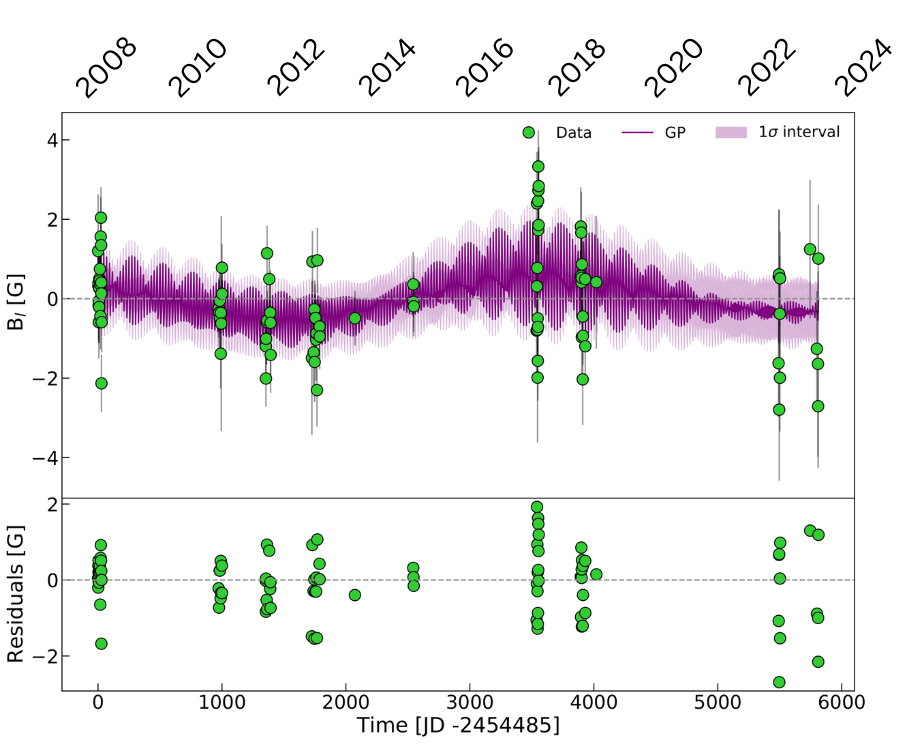}
    \includegraphics[width=\columnwidth]{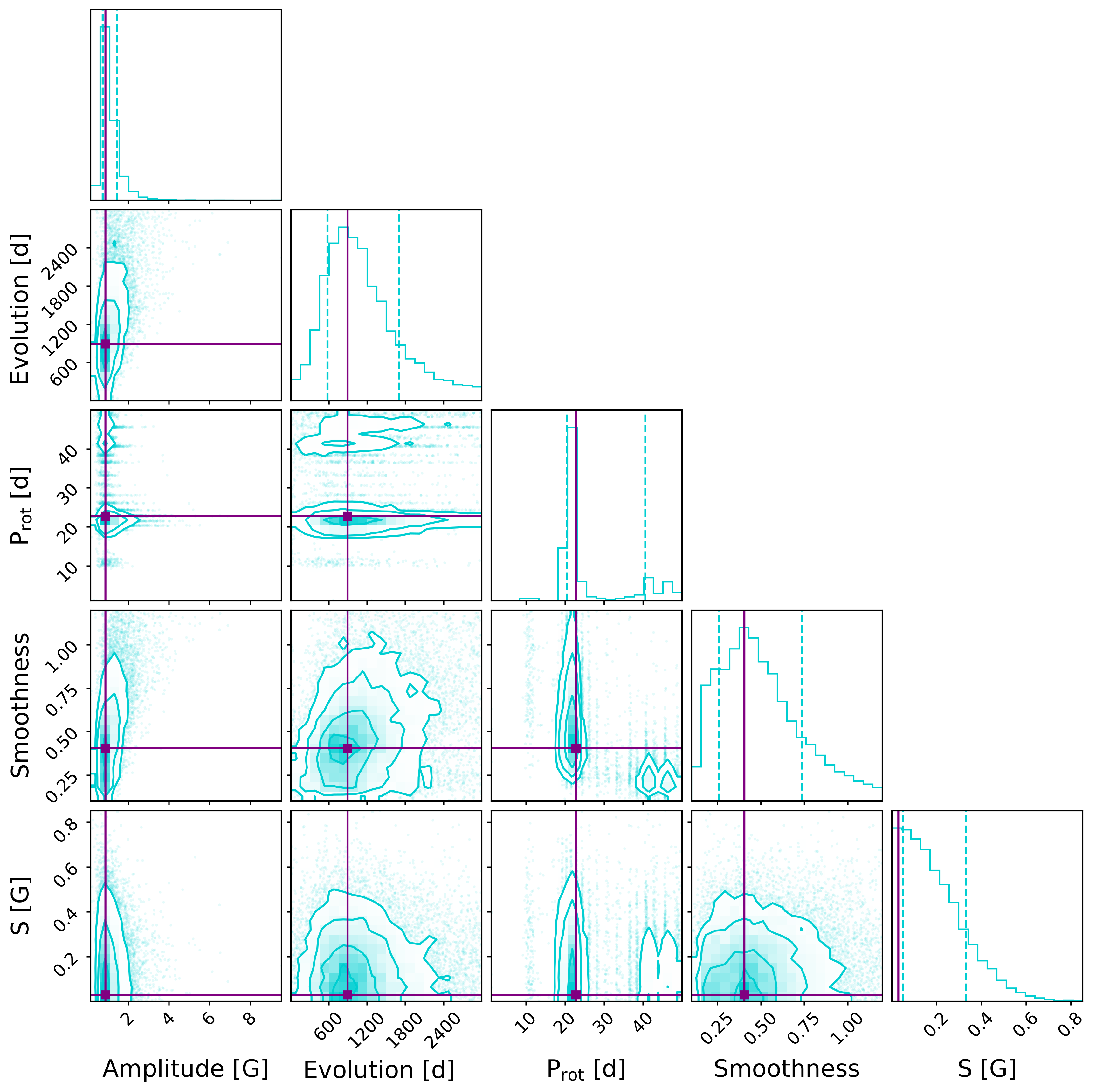}
    \caption{Longitudinal magnetic field measurements for HD\,9986 and GP regression analysis. Top: Gaussian process model of the full time series of B$_l$. The shaded area corresponds to the 1\,$\sigma$ uncertainty interval. The lower panel contains the residuals between the model and the observations. Bottom: Posterior distributions of the hyperparameters characterising the GP. The panels on the diagonal display the 1D marginalised distributions of the hyperparameters, while the other panels contain the 2D posterior distributions. The vertical solid lines indicate the modes of the distributions, while dashed lines indicate the 16th and 84th percentiles.}
    \label{fig:Bl_hd9986}
\end{figure}

Figure~\ref{fig:Bl_hd9986} illustrates the time series of longitudinal field measurements for HD\,9986, from 2008 to 2023. Overall, B$_l$ assumes positive and negative values, spanning between $-2.2$\,G and 3.3\,G, with a median of $-0.2$\,G. We note an oscillation of the median B$_l$ for each epoch, going from 0.3\,G in 2008 to $-0.8$\,G in 2012, up to $1.7$\,G in 2017 and down to $-0.38$\,G in 2023. Likewise, the interval of B$_l$ values goes from $\pm2$\,G, to $\pm1$\,G, and finally between $-2$ and $3$\,G.

The Lomb-Scargle analysis of the B$_l$ data for HD\,9986 was not conclusive, as no significant ($\mathrm{FAP}<0.1\%$) peak was observed (see Fig.~\ref{fig:LSP_stars}). The results of the GP regression are shown in Fig.~\ref{fig:Bl_hd9986}. The model identifies a stellar rotation period of $22.8^{+17.8}_{-2.4}$\,d, which is in good agreement with the reported value of 22.4\,d \citep[see][]{Marsden2014}. The larger upper error bar stems from the presence of harmonic periodicities around 40-50\,d that were sampled by the GP. This can be seen from the posterior distributions in Fig.~\ref{fig:Bl_hd9986}. Given that the posterior distribution is reasonably symmetric around the peak at 22.8\,d, a more realistic upper error bar is 2.4\,d, as reported in Table~\ref{tab:gp}. We also retrieved an amplitude of the variations of 0.8\,G and an excess of uncorrelated noise $S$ of 0.03$^{+0.17}_{-0.11}$\,G, which is consistent with zero, signifying an appropriate estimate of the error bars. Although the retrieved evolution time scale is $852^{+497}_{-375}$\,d (or 2.3\,yr), implying fast evolution of the longitudinal field, the GP captures a long-term sinusoidal trend of $\sim13$~yr (upper panel of Fig.~\ref{fig:Bl_hd9986}), which can be representative of a magnetic cycle.

\begin{figure}[t]
    \includegraphics[width=\columnwidth]{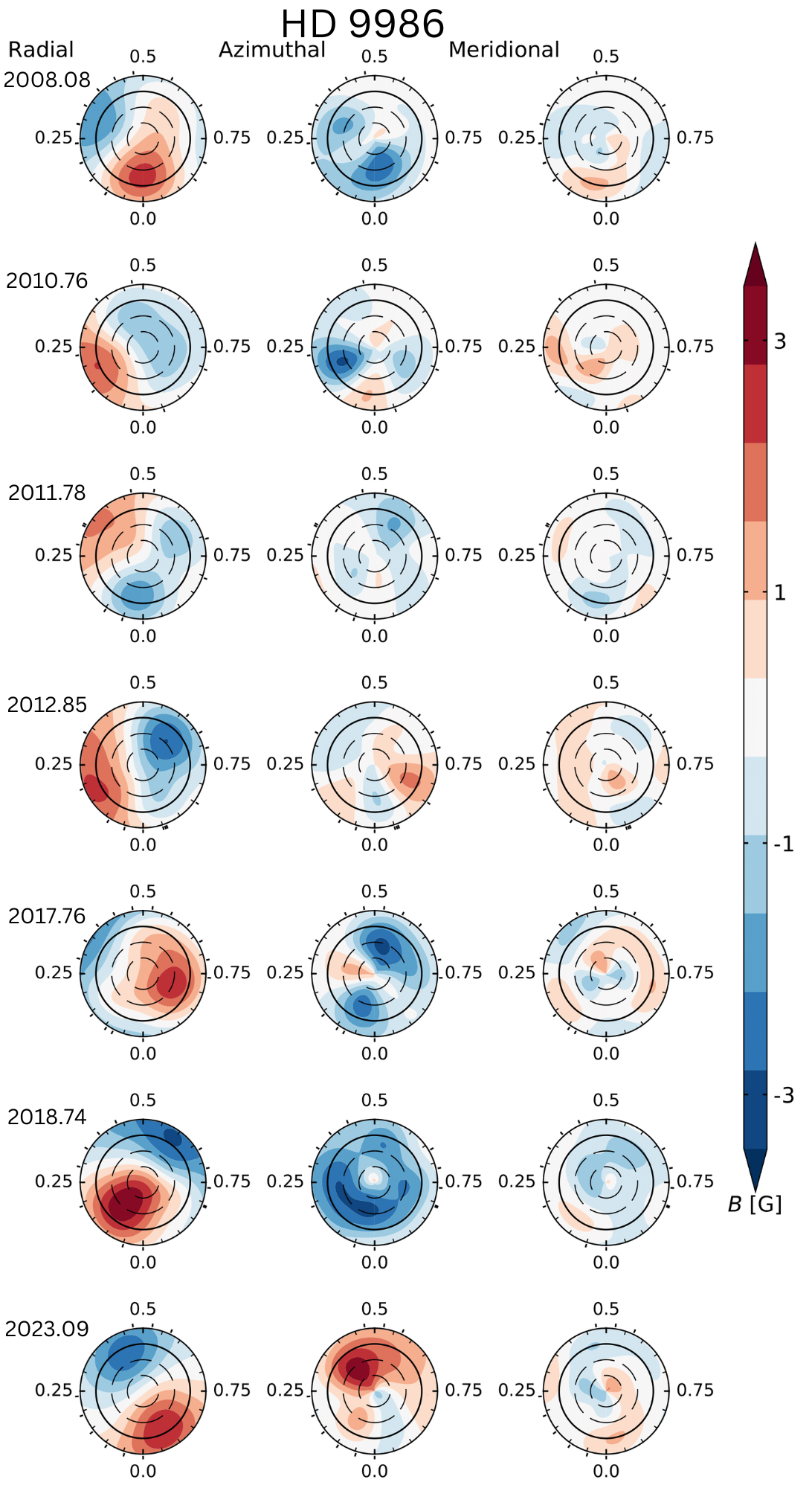}
    \caption{Reconstructed large-scale magnetic field map of HD\,9986, in flattened polar view. From the left, the radial, azimuthal, and meridional components of the magnetic field vector are illustrated. Concentric circles represent different stellar latitudes: -30\,$^{\circ}$, +30\,$^{\circ}$, and +60\,$^{\circ}$ (dashed lines), as well as the equator (solid line). The radial ticks are located at the rotational phases when the observations were collected. The rotational phases are computed with Eq.~\ref{eq:ephemeris} using the first observation of each individual epoch (see Table~\ref{tab:log}). The colour bar indicates the polarity and strength (in G) of the magnetic field. Indications of polarity reversals of the radial field have occurred in 2010.76 and 2017.76 epochs, and of the azimuthal field in 2023.09.}
    \label{fig:zdi_hd9986}
\end{figure}

The ZDI-reconstructed magnetic field maps are presented in Fig.~\ref{fig:zdi_hd9986}, and the line fits are provided in Fig.~\ref{fig:stokesV_hd9986}. For the reconstructions, we assumed an inclination of 60$^{\circ}$ and a projected equatorial velocity $v_\mathrm{eq}\sin(i) = 2.6$\,km\,s$^{-1}$ (see Table~\ref{tab:stars_properties}). The differential rotation search pointed at $d\Omega=0.0$~rad\,d$^{-1}$ in most epochs, that is, consistent with solid body rotation. We then performed a rotation period optimisation (see Sec~\ref{sec:zdi}) for the examined epochs, finding an average of P$_\mathrm{rot}=21.03\pm0.44$\,d. This value is compatible with the literature range: between 19\,d \citep{Isaacson2010}, 22.4\,d \citep{Marsden2014}, and 23.4$\pm$3.4\,d \citep{Lorenzo-Oliveira2019}. 

The properties of the magnetic field maps are listed in Table~\ref{tab:zdi_output}. We fitted the observed Stokes~$V$ LSD profiles down to $\chi^2_r$ of 1.00-1.20, suggesting that in some cases our models do not fully reproduce the observations, likely due to undetected intrinsic variability. The average field strength features a decrease from 1.5 to 1.2\,G in the first years, then rises to 2.6\,G in 2018.74 and drops to 1.9\,G in the latest epoch, showing similarities with the long-term trend captured by the GP in the B$_l$ data.

The topology of HD\,9986's large scale magnetic field is predominantly poloidal, dipolar and non-axisymmetric for all the epochs. The fraction of total magnetic energy stored in the poloidal component started at 75\% in 2008.08, then increased to 99\% in 2012.85, then decreased down to 58\% in 2018.74, and finally it increased to 79\% in 2023.09. In 2012.85, the toroidal fraction is at the lowest value over the time series, and it is largely non-axisymmetric compared to the other epochs. In 2023.09, the the axisymmetric fraction of the poloidal energy is at the minimum value of the time series. The dipolar component accounted for more than 58\% of the poloidal energy, and the fraction of total energy in the axisymmetric component decreased from 38 to 16\%, then increased to 55-60\%, and finally decreased to 19\% in the last epoch. 

There are striking features characterising the evolution of the large-scale field (see Fig.~\ref{fig:zdi_hd9986}). The radial component exhibited a hemisphere dominated by a positive polarity in 2008.08, which then switched to a negative polarity between 2010.76 and 2012.85, to finally revert back to a positive polarity in 2017.76 and 2018.74. This correlates with a decrease of the toroidal energy fraction from 25\% to 1\%, and then a rise to 40\%. The time scale of the double polarity flip of the radial field is on the order of 10-11\,yr, which is half of the Hale cycle period of the Sun. This is consistent with the sinusoidal trend suggested by the GP model of the B$_l$ data (see Fig.~\ref{fig:Bl_hd9986}). The azimuthal component of the field transitioned from a negative-dominated polarity, to a more complex configuration, to a negative sign, and finally to a positive-dominated polarity.

\subsection{HD\,56124 (HIP\,35265)}

HD\,56124 is a G0~dwarf with an age of 3.9\,Gyr and a rotation period of 20.7$\pm$0.2\,d \citep{Marsden2014}. Measurements of the chromospheric activity index $\log R'_\mathrm{HK}$ were reported between $-4.84$ and $-4.65$ \citep{Wright2004, Isaacson2010, Pace2013}, making the star more active than HD\,9986, as expected from the shorter rotation period. 

The time series of B$_l$ measurements is shown in Fig.~\ref{fig:Bl_hd56124}, from 2008 to 2021. The values are initially all positive, with a median value of 2.3\,G, and then transition to a mostly negative sign from 2010 onwards, with a median around $-0.7$\,G. In the latest epoch, the median measurement is 1.7\,G, and the RMS scatter is also visibly increased to a value of 3.7\,G. The generalised Lomb-Scargle periodogram analysis revealed a prominent peak ($\mathrm{FAP}<10^{-2}\%$) at 2870\,d or equivalently 7.9\,yr (see Fig.~\ref{fig:LSP_stars}, together with a forest of peaks between $10^2$-$10^3$\,d. The latter are mirrored in the window function, meaning that they stem from the irregular observational cadence and temporal gaps in the time series. For this reason some of the power may have been injected in the predominant peak.

The GP applied to the B$_l$ time series found an oscillatory trend directed towards negative values of the field at start, and toward positive values at the end of the time series. The lack of data between 2012 and 2017 prevented us from discerning how realistic the oscillation in such time gap is, which is encapsulated by the larger uncertainty band of the GP fit in Fig.~\ref{fig:Bl_hd56124}. Assuming positive values of the magnetic field during this gap would imply an oscillatory trend of 8-10\,yr. The model is characterised by a rotation period of $21.32^{+1.96}_{-5.02}$\,d, which is larger compared to previous estimates \citep{Marsden2014}, but compatible within 1$\sigma$. The largely asymmetric error bar is due to harmonic structure in the posterior distribution, owing to the large scatter in the last epoch, since the model would be able to fit multiple, shorter periodicities. A more realistic lower error bar is $-2.0$\,d. The evolution time scale of B$_l$ is $511^{+390}_{-275}$\,d (or 1.4\,yr), which is roughly about six times shorter than the periodicity measured with the Lomb-Scargle periodogram. 

\begin{figure}[t]
    \includegraphics[width=\columnwidth]{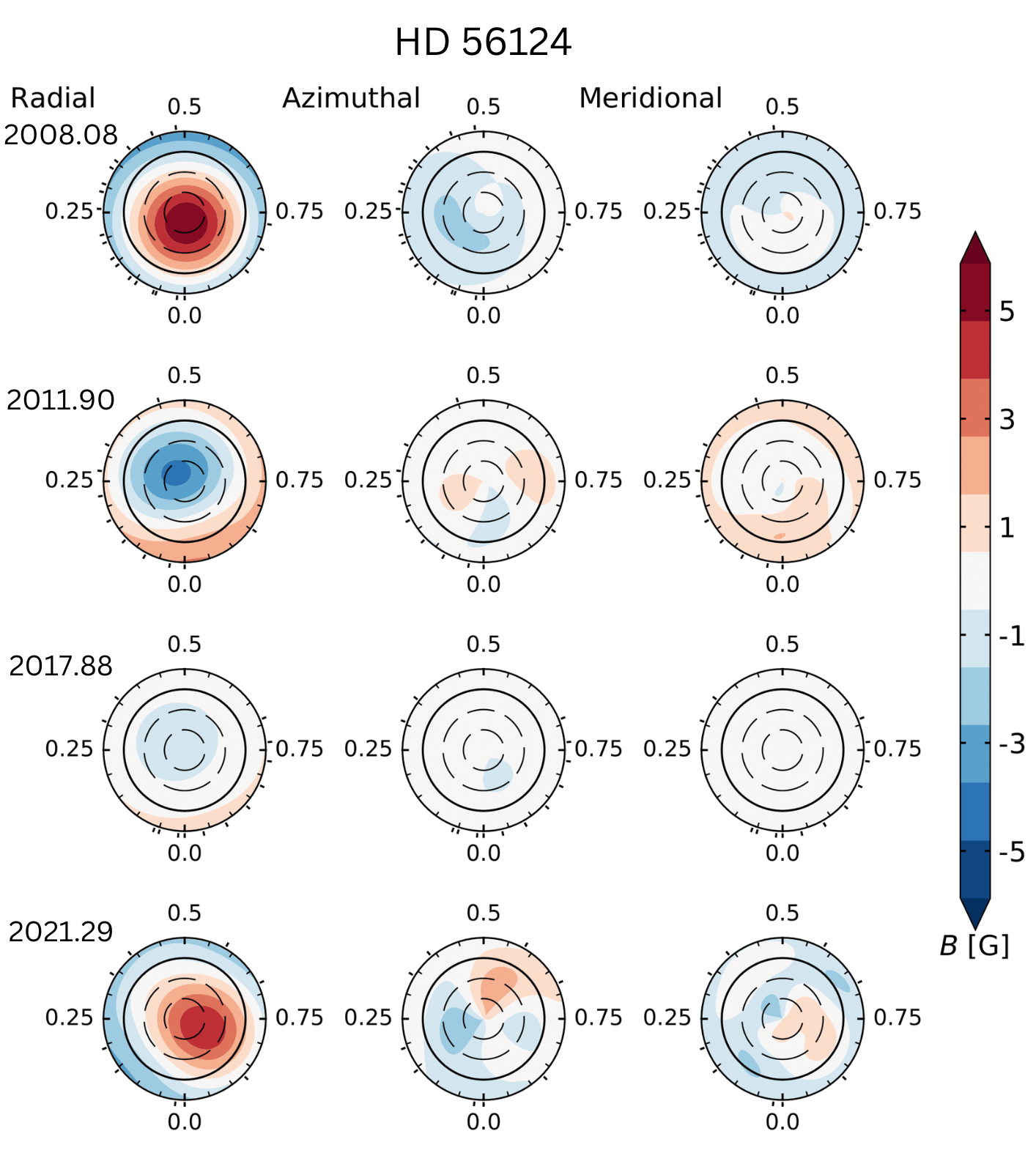}
    \caption{Reconstructed large-scale magnetic field map of HD\,56124, in flattened polar view. The format is the same as Fig.~\ref{fig:zdi_hd9986}.}
    \label{fig:zdi_hd56124}
\end{figure}

The ZDI-reconstructed magnetic field maps are presented in Fig.~\ref{fig:zdi_hd56124} and the properties listed in Table~\ref{tab:zdi_output} for four epochs: 2008.08, 2011.90, 2017.88, and 2021.29. The corresponding ZDI line fits are shown in fig.~\ref{fig:stokesV_hd56124}. We assumed an inclination of 40$^{\circ}$ and $v_\mathrm{eq}\sin(i) = 1.5$\,km\,s$^{-1}$. The differential rotation search was inconclusive in each case since the $\chi^2_r$ landscape built over the $d\Omega$-P$_\mathrm{rot}$ grid (see Sect.~\ref{sec:zdi}) featured multiple, stretched valleys preventing a straightforward identification of a minimum. The optimisation of the rotation period alone yielded a value of $20.749\pm1.028$\,d for 2008.08 epoch, which is highly compatible with the literature value \citep{Marsden2014}. For 2011.90 and 2017.88, the minimum of the $\chi^2_r$ distribution is at lower values (around 5-10\,d), but there is a sharp secondary minimum at $20.898\pm0.476$ and $20.158\pm1.292$\,d, respectively. The 2021.29 data set is not suitable for a rotation period search of this order of magnitude because the observations span around 20\,d. We only find a spurious minimum of the distribution around 9\,d. We therefore decided to fix the rotation period to $20.70\pm0.32$\,d and assume solid body rotation for all epochs. The target $\chi^2_r$ is between 0.97 and 1.15 for the maps, as listed in Table~\ref{tab:zdi_output}.

The ZDI reconstructions of HD\,56124 feature a predominantly poloidal ($>95\%$), dipolar ($>88\%$) and axisymmetric ($>70\%$) field. The maps reveal two evident polarity reversal, since the pole underwent a switch between positive sign in 2008.08 to negative in 2011.90, and then positive again in 2021.29 (see Fig.~\ref{fig:zdi_hd56124}). In 2017.88, we observe a similar topology and polarity as 2011.90, but a weaker average strength from 2.3 to 0.7\,G, and in 2021.29 the axisymmetry is the lowest value reconstructed ($\sim70\%$). With this information, we can see how HD\,56124 experiences a magnetic cycle characterised by a time scale of $\sim3-4$\,yr between polarity reversals. If exactly 3\,yr, we would have expected the same magnetic field strength in 2011.90 and 2017.88, whereas in the latter epoch we most likely observe the onset of a reversal after the peak at negative polarity. The evolution time scale of 1.4\,yr obtained from the GP fit on B$_l$ data would be too fast to explain the polarity reversal, since in this case the same magnetic field configuration would have been observed in 2008.08 and 2011.90.

\subsection{HD\,73350 (HIP\,42333)}

HD\,73350 is a G5~dwarf with an age of 1.4\,Gyr and a rotation period of 14.0\,d \citep{Marsden2014}. Measurements of the chromospheric activity index $\log R'_\mathrm{HK}$ were reported between $-4.61$ and $-4.45$ \citep{Wright2004, Isaacson2010, Pace2013, BoroSaikia2018a}, which are 0.3-0.5 dex larger than the solar values \citep{Egeland2017}. 

The time series of B$_l$ measurements is shown in Fig.~\ref{fig:Bl_hd73350}, from 2007 to 2018. The field has both positive and negative values within the same epoch, ranging between 6 and $-4$\,G. This suggests that the topology is possibly non-axisymmetric or complex. The field has a strength of $-2.0$ and $-2.6$\,G in 2017 and 2018, but these are individual B$_l$ measurements, which prevents us from drawing any conclusion on a possible trend toward negative values. 

The generalised Lomb-Scargle periodogram analysis revealed a marginally significant peak ($\mathrm{FAP}<10^{-1}\%$) at 13.74\,d, compatible with the rotation period reported in the literature. However, we did not detect any significant prominent long-term periodicity (see Fig.~\ref{fig:LSP_stars}). The GP regression produced a model with a rotation period of $14.20^{+13.06}_{-1.79}$\,d, which is on the same order of magnitude as literature values \citep{Petit2008,Marsden2014}, and an evolution timescale of $1497^{+1002}_{-931}$\,d (or 4.1\,yr). The large error bars for both hyperparameters reflect the difficulty of constraining the time scales encapsulated in the data set, due to the multi-peak nature of the posterior distributions (see Fig.~\ref{fig:Bl_hd73350}). In turn, this may be due to the fact that the bulk of our observations span a shorter interval than the evolution time scale, thus we are not able to constrain it robustly. In a similar manner as for HD\,9986 and HD\,56124, a more realistic upper error bar for P$_\mathrm{rot}$ is 2.0\,d.

\begin{figure}[t]
    \includegraphics[width=\columnwidth]{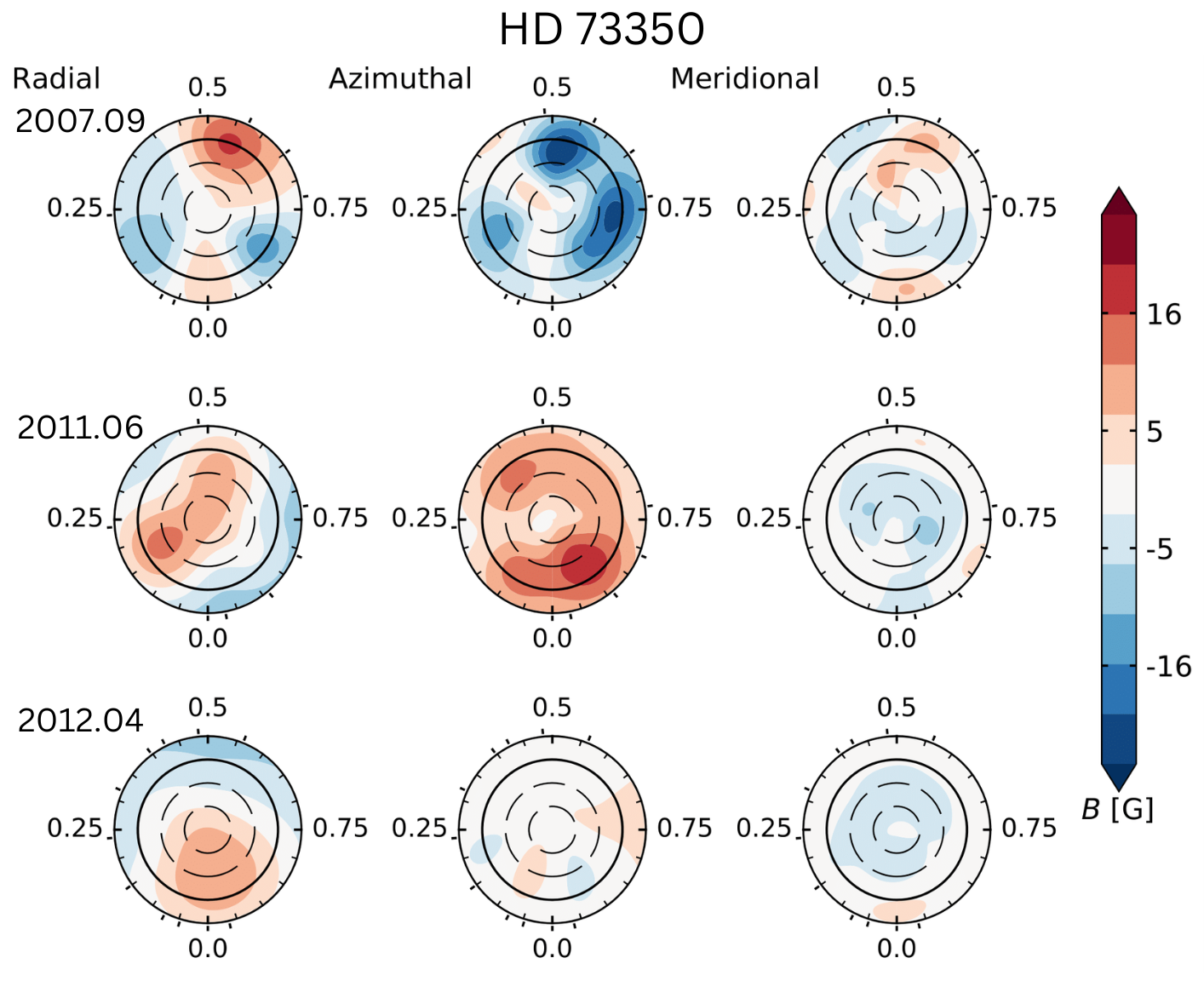}
    \caption{Reconstructed large-scale magnetic field map of HD\,73350, in flattened polar view. The format is the same as Fig.~\ref{fig:zdi_hd9986}.}
    \label{fig:zdi_hd73350}
\end{figure}

We obtained three magnetic field maps corresponding to the 2007.09, 2011.06, and 2012.04 epochs, as illustrated in Fig.~\ref{fig:zdi_hd73350}. The properties are listed in Table~\ref{tab:zdi_output} and the model Stokes~$V$ profiles are shown in Fig.~\ref{fig:stokesV_hd73350}. We only have seven observations for the 2011.06 epoch, but their longitudinal coverage allows for a reliable ZDI reconstruction. As stellar input parameters, we used an inclination of 70$^{\circ}$ and $v_\mathrm{eq}\sin(i) = 4.0$\,km\,s$^{-1}$, and we assumed solid body rotation, since the number of observations per each epoch did not allow a robust estimate of differential rotation. We optimised the stellar rotation period and obtained an average P$_\mathrm{rot}=12.27\pm0.13$\,d, the same as \citet{Petit2008}. By applying ZDI on the 2007.09 time series of Stokes~$V$ LSD profiles, \citet{Petit2008} revealed a complex field with a dominant toroidal component (more than 60\%), and the poloidal component had a substantial amount of energy in the dipolar, quadrupolar and octupolar modes (40\%, 20\%, and 20\%, respectively). Our reconstruction of 2007.09 map is consistent with \citet{Petit2008}.

The field topology is shown in Fig.~\ref{fig:zdi_hd73350}. The poloidal component increases from 54\% to 99\% and the dipolar component from 37\% to 83\%, with a contemporaneous decrease of the quadrupolar (from 27 to 10\%) and octupolar (from 23\% to 5\%) components. The axisymmetric fraction follows the dominant component of the field. In the first epoch, the axisymmetry is 44\% due to the combination of an axisymmetric toroidal component and non-axisymmetric poloidal component. In the second epoch, the field is axisymmetric because both components are also axisymmetric, and the last epoch exhibits the same level of axisymmetry as the significantly dominant poloidal component. Within five years, the average field strength seems to show a decreasing, monotonic trend from 30 to 13\,G. 

Therefore, the magnetic topology of HD\,73350 manifests an initially complex radial field that transitions towards a simple configuration in five years. The azimuthal field is predominantly negative in the first epoch and flips to positive after four years, and it almost switches off one year later. If the polarity switch of the azimuthal field were on a yearly time scale, we would expect the field in 2011.06 to have the same polarity as in 2007.09, so we can exclude it. Instead, if we assume a time scale of the azimuthal field reversal of four years, the two polarity switches become more consistent. These values are consistent with the photometric cycle period of 3.5\,yr reported by \citet{Lehtinen2016}.

\subsection{HD\,76151 (HIP\,43726)}

HD\,76151 is a G2~dwarf with an age of 2.1\,Gyr and a rotation period of 18.6$\pm$0.4\,d \citep{Marsden2014}. Measurements of the chromospheric activity index $\log R'_\mathrm{HK}$ were reported between $-4.82$ and $-4.50$ \citep{Wright2004, Isaacson2010, Pace2013, BoroSaikia2018a, GomesdaSilva2021}. The spectropolarimetric analysis of \citet{Petit2008} on 2007 data showed a predominantly poloidal, dipolar and mostly axisymmetric field.

The time series of B$_l$ measurements is shown in Fig.~\ref{fig:Bl_hd76151}, from 2007 to 2024. In the first part of the time series (until 2012) the values are mostly negative with a slight increasing trend towards positive polarity, since the median value goes from $-3$\,G in 2007 to $-$0.9\,G in 2012. After a gap of almost four years, the field is negative and stronger, with a median of $-4.6$\,G. From 2016 to 2024, we observe rapid variations of the bulk of the data, indicating fast variations of the field. From 2016, there is a rise towards positive values (median of 1.7\,G), then a switch to a median of $-0.6$\,G in 2019 and $-3.1$\,G in 2021, another rise to $-$2.1\,G in 2022 and 6.6\,G in 2023, and finally a decrease to 0.4\,G in 2024. The fast variations of B$_l$ in the second part of the time series illustrate that the observational cadence of the first part of the time series was likely missing the oscillations of the field between positive and negative polarities.

The Lomb-Scargle periodogram applied to the B$_l$ time series is shown in Fig.~\ref{fig:LSP_stars}. It features several significant peaks (FAP$<10^{-2}\%$), but most are mirrored in the window function, signifying signals with periods on the order of months or a year due to aliases of the observing cadence. The most prominent peak is at 1727\,d (or equivalently 4.7\,yr), and has a counterpart in the window function shifted towards longer periods (2000\,d). The quasi-periodic GP model retrieved a well-constrained stellar rotation period of $16.70^{+0.18}_{-0.16}$\,d (see Fig.~\ref{fig:Bl_hd76151}), which is lower than the values of 20.5$\pm$0.3\,d \citep{Petit2008} and 18.6$\pm$0.4\,d \citep{Marsden2014} reported in the literature. We also obtained an evolution time scale of $232^{+40}_{-41}$\,d, or equivalently 0.6\,yr.

\begin{figure}[t!]
\centering
    \includegraphics[width=0.87\columnwidth]{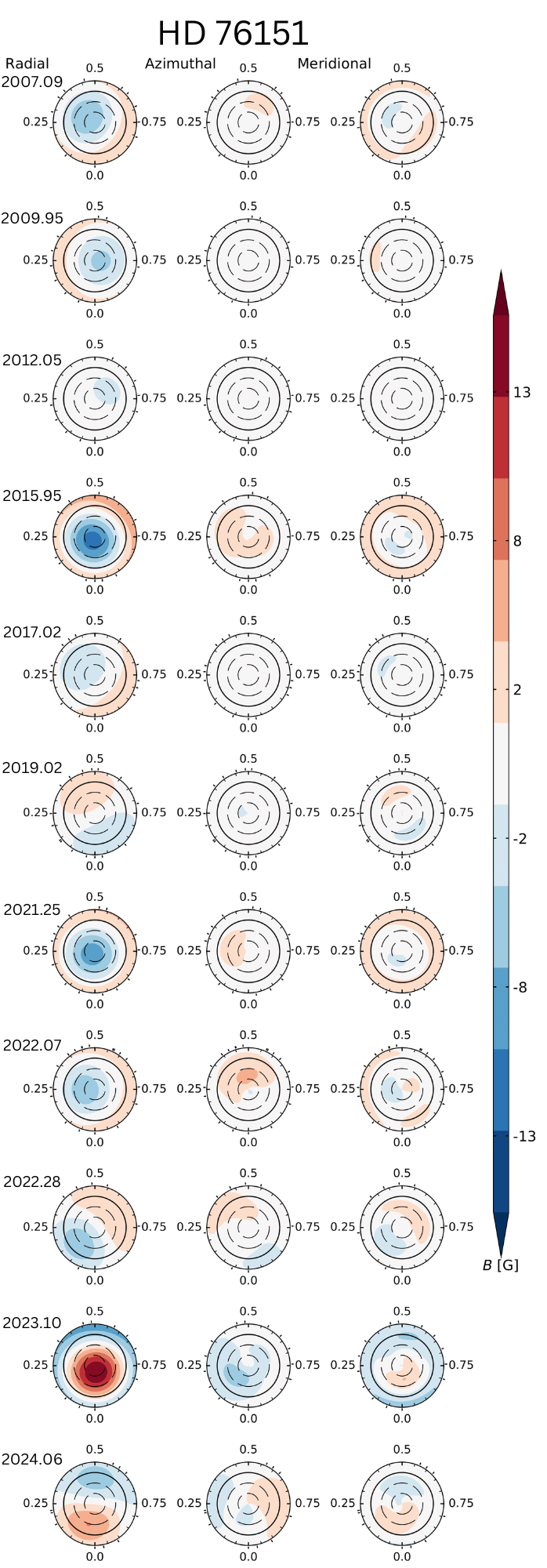}
    \caption{Reconstructed large-scale magnetic field map of HD\,76151, in flattened polar view. The format is the same as Fig.~\ref{fig:zdi_hd9986}.}
    \label{fig:zdi_hd76151}
\end{figure}

The reconstructed maps with ZDI are shown in Fig.~\ref{fig:zdi_hd76151}, and the Stokes~$V$ line fits are illustrated in Fig.~\ref{fig:stokesV_hd76151}. We assumed an inclination of 30$^{\circ}$, $v_\mathrm{eq}\sin(i) = 1.2$\,km\,s$^{-1}$, and solid body rotation, since the differential rotation search was inconclusive. The rotation period optimisation yielded an average of P$_\mathrm{rot}=17.47\pm0.81$\,d, where the larger error bar compared to the other stars stems from a larger dispersion of the epoch-optimised rotation periods. The value falls in the range of the literature measurements of $14.4\pm0.19$\,d \citep{Olspert2018} and $20.5\pm0.3$\,d \citep{Petit2008}. Possibly, we could attribute this range of rotation period values to solar-like differential rotation, with dominant active regions occurring at different latitudes over time, although our data sets cannot capture such signal. Assuming P$_\mathrm{equator}=14.4\pm0.19$ and P$_\mathrm{pole}=20.5\pm0.3$, the corresponding differential rotation rate would be $0.13\pm0.01$\,rad\,d$^{-1}$, which is almost twice as solar.

As reported in Table~\ref{tab:zdi_output}, the Stokes~$V$ LSD profiles were fitted to a $\chi^2_r$ of 1.20-1.90, except for the 2015.95 epoch, for which only $\chi^2_r=2.65$ can be reached before overfitting. The time span of 2015.95 epoch is 20 days, which is not significantly different from the time span of other epochs like 2017.02 or 2019.02 in which a $\chi^2_r$ of 1.5 and 1.6 could be reached. This indicates that the evolution, that is the emergence and decay, of magnetic regions was likely faster during the 2015.95 epoch.

The large scale magnetic field exhibits a dominant (more than $84\%$) poloidal component over the entire time series, with most of the magnetic energy stored in the dipolar mode (more than $80\%$). The average field strength oscillates mostly between 1 and 6\,G, with a peak at 8.5\,G in the 2023.10 epoch. The reconstruction of the 2007.09 epoch is compatible with the map of \citet{Petit2008}. The most striking feature is the fluctuation in axisymmetry, and in particular the poloidal-axisymmetric component since it is the dominant one. In 2007.09, the axisymmetry is large ($75\%$) and it decreases to 44.05\% in 2012.05 and rises again to 90\% in 2015.95. Then, it lowers to 50\% in 2017.02 and to 5\% within 2019.02, before rising again to 95\% in 2021.25. In the latest epochs, we see a rapid decrease from 73\% in 2022.07 to 5\% in 2022.28, then another increase to 88\% in 2023.10 and a decrease to 4\% in 2024.06. The epochs of low axisymmetry generally correlate with an increased amount of magnetic energy in the quadrupolar and octupolar modes of the poloidal component. 

During the 17\,yr of the time series, we observe only one polarity reversal in 2023.10, and a fast variation between axisymmetric and non-axisymmetric configurations, overall deviating from a Hale-like magnetic cycle. The highly non-axisymmetric configurations in 2019.02, 2022.28 and 2024.06 are not sufficient to determine whether additional polarity reversals occurred around such epochs or if only a temporary variation in axisymmetry occurred. As we will discuss in Sect.~\ref{sec:discussion}, we cannot robustly constrain a time scale for the variations of the large-scale topology, since they can be explained by a short-period, magnetic cycle for which we did not capture a polarity reversal or the superposition of two cycles, a shorter one that modulates the axisymmetry and a longer one responsible of polarity reversals.

\subsection{HD\,166435 (HIP\,88945)}

HD\,166435 is a young, fast-rotating, G1~dwarf with an estimated age of 0.2\,Gyr and a rotation period of 4.2\,d \citep{Marsden2014}. The chromospheric activity index $\log R'_\mathrm{HK}$ was measured between $-4.36$ and $-4.20$ \citep{Isaacson2010, Pace2013, Marsden2014, BoroSaikia2018a}, which is approximately 0.7\,dex larger than the Sun. HD\,166435 is the most active star in our sample, and it is a benchmark for the limitations that stellar activity poses on radial velocity searches of exoplanets \citep{Queloz2001}.

The time series of B$_l$ measurements is shown in Fig.~\ref{fig:Bl_hd166435}, from 2007 to 2020. The values oscillate in sign, between $-10$ and 15\,G, but the bulk of measurements is mostly positive. More precisely, the median B$_l$ over individual years varies between 3.5\,G, to 0.5\,G and up to 7\,G in the latest epochs. The evident scatter of B$_l$ data for an individual epoch is stemming from a most likely complex or non-axisymmetric field.

The Lomb-Scargle periodogram, shown in Fig.~\ref{fig:LSP_stars}, did not reveal any significant periodicity in the time series. There is a forest of peaks between 4 and 10\,d which is not reflected in the window function, but the associated FAP is higher than 1\%. We therefore decided to apply the same tool on three different $TESS$ light curves (see Sect.~\ref{sec:tess}), to extract the main periodicity from the light curves. The results are shown in Appendix~\ref{app:time_analysis}. We found a highly significant (FAP$\ll0.01\%$) peak for each light curve, with a mean of $3.47\pm0.10$\,d, where the error bar represents the standard deviation of the three measurements. %This value is 0.7\,d shorter than estimated by \citet{Marsden2014}.

An initial attempt to fit the B$_l$ time series with a GP produced a posterior distribution of the stellar rotation period with a maximum at $\sim30$\,d, but it also showed an additional peak below 10\,d. Considering that 30\,d most likely corresponds to the observational cadence, and that literature estimates of P$_\mathrm{rot}$ are one order of magnitude lower, we restricted the uniform prior on the stellar rotation period between 1 and 10\,d. A shorter rotation period is also more consistent with the activity level of the star \citep[see e.g.][]{Noyes1984} and it is supported by the value obtained from the $TESS$ light curves. We found P$_\mathrm{rot}=3.54^{+0.51}_{-0.29}$\,d, which is consistent with the value obtained from $TESS$ data and literature values \citep{Wright2004}. Given the robust and independent result from the $TESS$ light curves, we decided to set a Gaussian prior on the stellar rotation period centred on $3.47\pm0.10$\,d and perform GP regression again. The results are listed in Table~\ref{tab:gp} and shown in Fig.~\ref{fig:Bl_hd166435}. We found a visually similar GP fit as when using a uniform prior on P$_\mathrm{rot}$, with an evolution time scale of $652^{+541}_{-293}$\,d (or 1.8\,yr).

\begin{figure}[t]
    \includegraphics[width=\columnwidth]{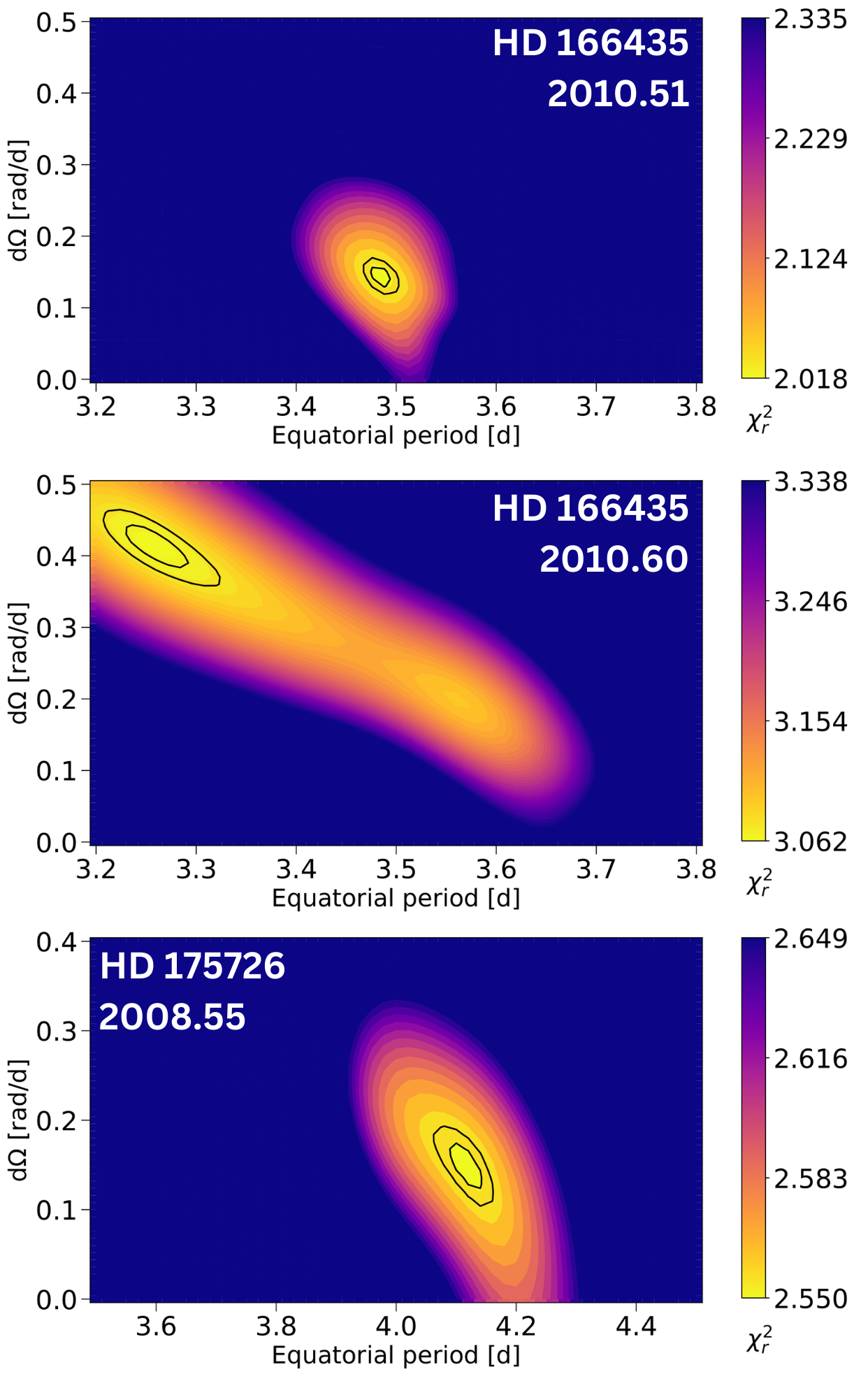}
    \caption{Joint search of differential rotation and equatorial rotation period for HD\,166435 and HD\,175726. Two epochs are shown for HD\,166435 and one for HD\,175726. The panels illustrates the $\chi^2_r$ landscape over a grid of (P$_\mathrm{rot,eq}$,$d\Omega$) pairs, with the $1\sigma$ and $3\sigma$ contours. The best values are obtained by fitting a 2D paraboloid around the minimum, while their error bars are estimated from the projection of the $1\sigma$ contour on the respective axis \citep{Press1992}.}
    \label{fig:domega}
\end{figure}

The Stokes~$V$ models are illustrated in Fig.~\ref{fig:stokesV_hd166435}. We assumed an inclination of 40$^{\circ}$ and $v_\mathrm{eq}\sin(i) = 7.9$\,km\,s$^{-1}$. The search of latitudinal differential rotation resulted in P$_\mathrm{rot}=3.48\pm0.01$\,d and d$\Omega=0.14\pm0.01$~rad\,d$^{-1}$ for 2010.51 and P$_\mathrm{rot}=3.26\pm0.04$\,d and d$\Omega=0.41\pm0.03$~rad\,d$^{-1}$ for 2010.60, as shown in Fig.~\ref{fig:domega}. For the other epochs, the search was inconclusive. With such differential rotation rates, the rotation period at the pole is $3.77\pm0.02$\,d and $4.14\pm0.10$\,d.

Both values of equatorial rotation period are consistent with the average P$_\mathrm{rot}$ of the $TESS$ light curves and the best fit hyperparameter constrained by the GP. Although cases of substantial differential rotation (up to d$\Omega=0.5$~rad\,d$^{-1}$) have been reported before, such as HD\,29615 \citep{Waite2015}, EK~Dra \citep{Waite2017}, V889\,Her \citep{Brown2024}, and $\tau$~Boo \citep{Donati2008b,Fares2009}, the value of d$\Omega=0.41\pm0.03$~rad\,d$^{-1}$ from August 2010 may be spurious. This because the $\chi^2_r$ landscape does not show an individual and well-constrained minimum, rather a more complex shape with an additional (but less pronounced) minimum around d$\Omega=0.15-0.20$~rad\,d$^{-1}$ (see Fig.~\ref{fig:domega}). This secondary minimum would be compatible with the differential rotation rate found in 2010.51, which is a factor of two greater than the solar value. Overall, the measurement of a differential rotation rate greater than the solar value for HD\,166435 is consistent with the increasing trend of differential rotation with stellar photospheric temperature \citep{Barnes2005,CollierCameron2007,Balona2016}.

Since we cannot constrain a reliable value of d$\Omega$ from the other epochs, the ZDI reconstructions were performed fixing P$_\mathrm{rot}=3.48\pm0.01$\,d and d$\Omega=0.14\pm0.01$~rad\,d$^{-1}$, for all epochs. Assuming solid body rotation for epochs other than 2010.51 and 2010.60 would have been contradictory, and would have led to a poorer quality of the Stokes~$V$ models (as quantified by $\chi^2_r$ increases between 1.0 and 5.0 for different epochs). However, using the same value of d$\Omega$ for all the epochs may limit us in accounting for the intrinsic variability of the surface shear and its evolution. Indeed, previous studies on cool stars have shown that the amount of latitudinal differential rotation can change over a time scale of a few years \citep{Donati2003b,BoroSaikia2016}, which was interpreted as the feedback of the magnetic field on the surface shear flow. Given the lack of additional constraints on d$\Omega$ for the other epochs, our choice represents a trade-off.

The Stokes~$V$ LSD profiles were fitted to a $\chi^2_r$ of 1.50-2.50 for most epochs, and to 4.0 for 2016.49. Although a $\chi^2_r=4.0$ represents an improvement compared to the case of assuming solid body rotation (for which only $\chi^2_r=5.5$ could be reached), its high value for the 2016 epoch suggests that significant evolution of the surface magnetic features occurred within the time span of such epoch. This evolution, presumably related to the limited lifetime of magnetic spots, cannot be modelled under the simple assumption of a surface progressively distorted by differential rotation. The equator-pole lap time, representing the amount of time it takes for the magnetic map to be sheared until it is unrecognisable, is indeed shorter ($\sim45$\,d) than the time span of the 2016.49 epoch ($\sim50$\,d)

\begin{figure}[t]
    \includegraphics[width=\columnwidth]{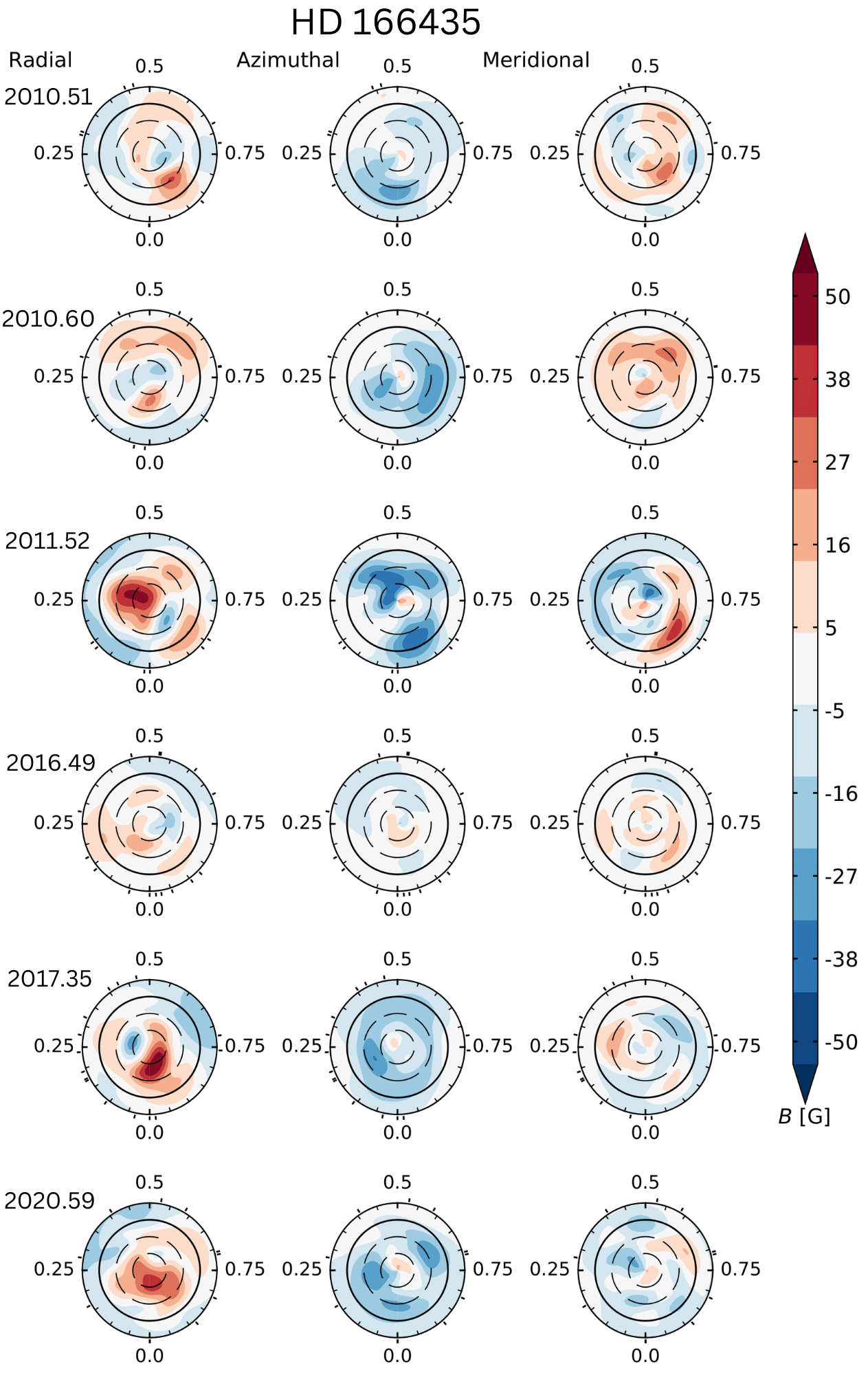}
    \caption{Reconstructed large-scale magnetic field map of HD\,166435, in flattened polar view. The format is the same as Fig.~\ref{fig:zdi_hd9986}.}
    \label{fig:zdi_hd166435}
\end{figure}

The maps of the large-scale magnetic field are shown in Fig.~\ref{fig:zdi_hd166435}. HD\,166435 exhibits a large-scale magnetic field with a complex topology, where the poloidal component accounts for 60\% of the magnetic energy for most of the epochs, with a peak to 84\% in 2016.49. The dipolar, quadrupolar and octupolar modes of the poloidal component start with values between 20-25\% in the first epoch, then the dipolar and octupolar remain reasonably stable around 30\% and 20\% until 2020.59, while the quadrupolar component oscillates between 32\% to 16\% and back to 20\%. In 2020.59, the dipolar component increases to 67\%, and the quadrupolar and octupolar decrease to 10\%. The poloidal component is mostly non-axisymmetric ($10-40\%$) with an increase to 56\% in the latest epoch, while the toroidal component is more axisymmetric ($50-90\%$), making the global axisymmetry oscillate between 20 and 66\%. 

Although it is not straightforward to pinpoint cyclic features in the magnetic field topology of HD\,166435 due to its multipolar nature, we notice that globally the field experiences a decrease in complexity reaching a more poloidal, axisymmetric configuration in the final epoch. The azimuthal field maintains a negative polarity with an oscillating strength throughout. In addition, we note the intermittent presence of a magnetic spot between 30-60 degrees in latitude, with positive polarity and stronger average field. Therefore, the magnetic topology seem to be characterised by various and distributed magnetic spots in certain epochs (2010.60 and 2016.49), and a more concentrated field of positive polarity at others (2010.51, 2011.52, 2017.35, and 2020.59). If corroborated, the time scale of the appearance of such feature is approximately one year.

\subsection{HD\,175726 (HIP\,92984)}

HD\,175726 is a young, fast-rotating, G0~dwarf with an estimated age of 0.6\,Gyr and a rotation period of 5.1\,d \citep{Marsden2014}. The chromospheric activity index $\log R'_\mathrm{HK}$ was measured between between $-4.44$ and $-4.36$ \citep{Isaacson2010, Pace2013, Marsden2014, BoroSaikia2018a}, which makes it the second most active star in our sample. 

Figure~\ref{fig:Bl_hd175726} illustrates the B$_l$ time series, from 2008 to 2024. The field values span between $-23.0$ and $13.1$\,G, and the bulk of the measurements per each epoch does not show significant signs of evolution. In a similar manner to HD\,166435, the fact that the field becomes positive and negative within a stellar rotation indicates a rather non-axisymmetric or complex field. 

The Lomb-Scargle periodogram, shown in Fig.~\ref{fig:LSP_stars}, features a series of peaks around 2-5\,d and no evident long-term periodicity. The most significant peak is at 2.03\,d, with a FAP lower than 0.01\%. This period is lower than the literature values of 3\, \citep{Isaacson2010}, 4.0\,d \citep{Mosser2009}, and 5.1\,d \citep{Marsden2014}, possibly reflecting an alias of the high-frequency observing cadence in 2008. Indeed, during 2008 multiple observations were taken during multiple nights, rather than one observation per night like the other stars. If we restrict the Lomb-Scargle analysis to the 2008 and 2016 epochs separately, we observe the most prominent peaks to be around 2\,d and 4\,d, respectively. Knowing that their surface magnetic field evolves fast, we further restricted the search between 2008.55 and 2008.63 separately, we observe peaks at 2\,d and 4\,d for both subsets. In 2008.55, the two peaks are significant (FAP$<0.01\%$), while in 2008.63 neither peak is significant. The period at 4\,d is closer to the reported literature value. Splitting over the 2008.55 and 2008.63 subsets is performed considering the dense monitoring of the 2008 epoch, and the fact that, owing to an increased spatial resolution correlated to the large value of $v_\mathrm{eq}\sin(i)$, we may be sensitive to faster evolution time scales of inhomogeneities on the stellar surface.

The GP fitting is performed while limiting the uniform prior on the stellar rotation period between 1 and 10\,d, to prevent unnecessary harmonic peaks to emerge. The model is shown in Fig.~\ref{fig:Bl_hd175726} and it is characterised by a stellar rotation period of P$_\mathrm{rot}=4.04^{+4.1}_{-0.11}$\,d. The large upper error bar stems from the multiple peaks of the posterior distribution, in a similar manner as HD\,9986, and a more realistic estimate is 0.11\,d. The retrieved P$_\mathrm{rot}$ is within the range of reported values, and compatible with the estimate of \citet{Mosser2009}. The evolution time scale is not well-constrained, partly because the field may possess a complex and fast-evolving topology between epochs, and additionally because of the large observational gaps in the time series, preventing the GP to probe finely the changes of B$_l$ in the long term.

\begin{figure}[t]
    \includegraphics[width=\columnwidth]{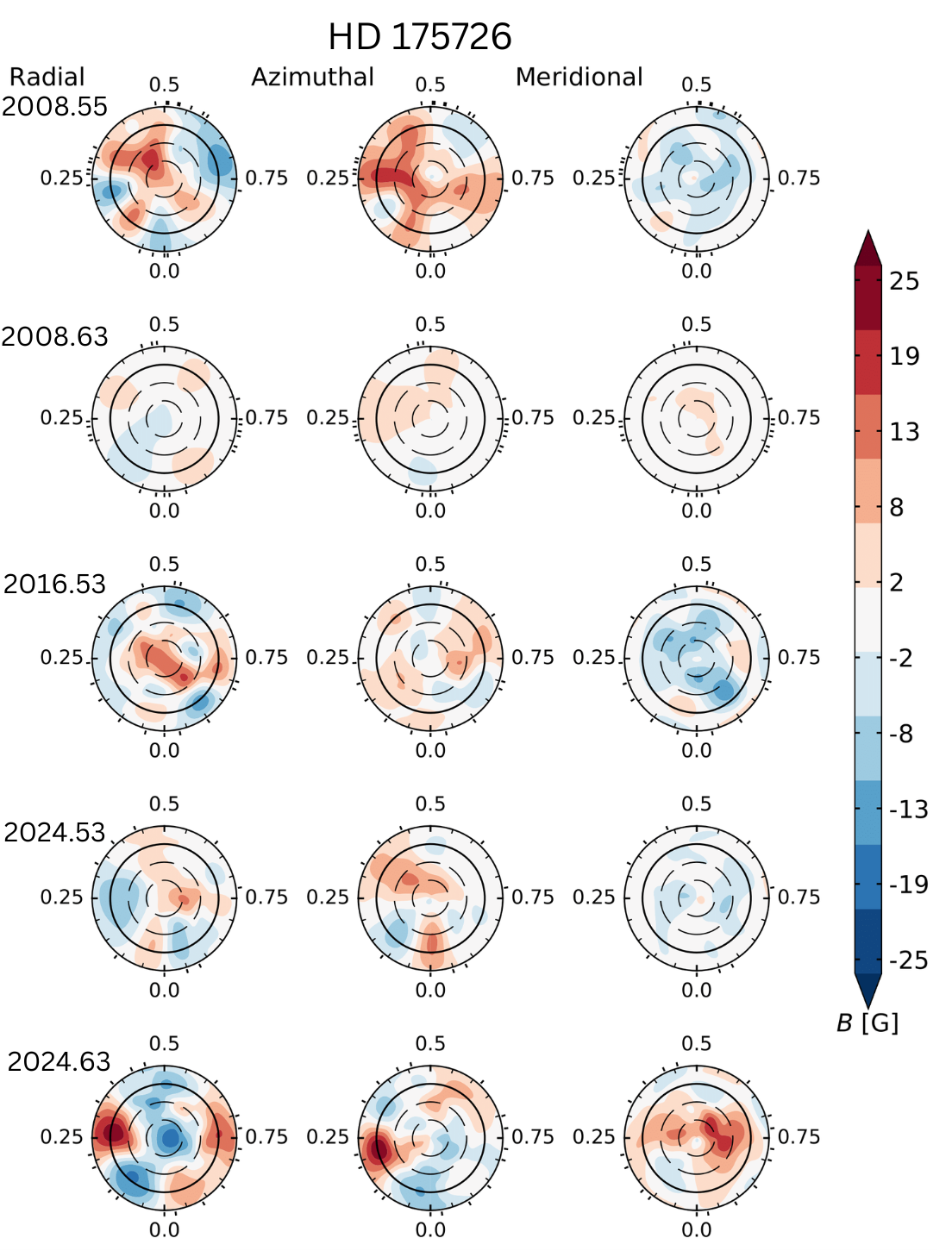}
    \caption{Reconstructed large-scale magnetic field map of HD\,175726, in flattened polar view. The format is the same as Fig.~\ref{fig:zdi_hd9986}.}
    \label{fig:zdi_hd175726}
\end{figure}

The maps of the large-scale magnetic field are shown in Fig.~\ref{fig:zdi_hd175726}, and the Stokes~$V$ models are illustrated in Fig.~\ref{fig:stokesV_hd175726}. We assumed an inclination of 70$^{\circ}$ and $v_\mathrm{eq}\sin(i) = 12.3$\,km\,s$^{-1}$. The latitudinal differential rotation search was conclusive for the 2008.55 epoch, which is not surprising considering that it contains the largest number of observations with an evident and evolving Stokes~$V$ signature (see Fig.~\ref{fig:stokesV_hd175726}). The results of the optimisation process are shown in Fig.~\ref{fig:domega}, and we found a minimum $\chi^2_r$ located at P$_\mathrm{rot}=4.12\pm0.03$ and d$\Omega=0.15\pm0.03$~rad\,d$^{-1}$. 

The differential rotation rate is 2.2 times larger than on the Sun, of the same order of magnitude as the solar-like star HD\,35296 \citep{Waite2015}, but not as extreme as HD\,29615 \citep{Waite2015}, EK~Dra \citep{Waite2017} or $\tau$~Boo \citep{Donati2008b,Fares2009}, reaching up to 0.5\,rad\,d$^{-1}$. Finally, the value of d$\Omega$ we found for HD\,175726 implies a rotation period at the pole of $4.55\pm0.91$\,d. Since we cannot constrain a reliable value of d$\Omega$ from the other epochs, we decided to fix the value of rotation period and differential rotation values to those inferred from 2008.55 epoch, with the same caveats as for HD\,166435. 

Assuming P$_\mathrm{rot}=4.12$\,d and d$\Omega=0.15$~rad\,d$^{-1}$ for all epochs, we fitted the Stokes~$V$ LSD profiles down to a $\chi^2_r$ of 0.80-1.70. The topology of HD\,175726 is complex (see Fig.~\ref{fig:zdi_hd175726}), which also stems from the increased available spatial resolution given the larger value of $v_\mathrm{eq}\sin(i)$ compared to the other stars \citep[see for instance][]{Donati2003c}. The poloidal field accounts for more than 70\% of the magnetic energy in general, with more than 20\%, 20\% and 8\% in the dipolar, the quadrupolar, and the octupolar components, respectively. The toroidal component is also significant, in most cases storing between 11\% and 30\% of the energy, except for the last epoch when it is 5\%. The field is mostly non-axisymmetric, with about 23-47\% of the energy stored in the axisymmetric field. In all epochs the poloidal field is mainly non-axisymmetric (less than 50\%) while the toroidal component oscillates between axisymmetric (in 2008 and 2024.53) and non-axisymmetric (in 2016.53 and 2024.63) configurations.

Owing to the large observational gap in the time series, it is not straightforward to draw conclusions in terms of magnetic cycles for this star. If we consider that its properties are analogous to $\xi$~Boo~A \citep{Morgenthaler2012} and HN~Peg \citep{BoroSaikia2016}, we expect short-term (months to a few years) evolution of the magnetic geometry to occur. This can occur in the form of fluctuating poloidal-to-toroidal energy fraction as well as the change in complexity, that is the distribution of the energy content in the modes of the poloidal component.

\section{Discussion}\label{sec:discussion}

\subsection{Trends from GP evolution time scale}

As described in Sect.~\ref{sec:gp}, we applied a quasi-periodic GP to the time series of B$_l$ data of each star to constrain its temporal variation 5-17~yr. One of the hyperparameters of the model is the evolution timescale ($\theta_2$), which describes how the rotational modulation of B$_l$ varies over time. It is generally associated to the lifetime of active regions on the stellar surface \citep{Nicholson2022,Aigrain2022}, therefore it may not necessarily reflect a putative cycle time scale, as in the case of fast rotators presented here.

We found values mostly between 232 and 852\,d. For HD\,73350 and HD\,175726, we found values of 1497\,d and 148\,d, respectively, with large error bars stemming from unconstrained posterior distributions. Although this is consistent with the expected rapid evolution of the magnetic field for the younger, fast-rotating star HD\,175726 with respect to the older HD\,73350, the GP model cannot put robust constraints on these values owing to the cadence of the observations, as the time series have gaps of 4\,yr. Owing to the limited sample of stars, we did not observe striking trends of $\theta_2$ as a function of stellar rotation period, mass, age, and Rossby number, average $S$-index, and $B_V$.

\citet{Giles2017} computed the decay lifetime from the autocorrelation function of $Kepler$ lightcurves of stars ranging between M- and F-type, and whose rotation period was close to either 10\,d or 20\,d. Compared to the decay lifetime of starspots inferred by the authors, our values of $\theta_2$ for our stars are larger by at most a factor of two. This difference may be a consequence of the different method employed to capture the time scale, that is autocorrelation function or Gaussian Process. Additionally, the physics probed by photometric and spectropolarimetric activity proxies may be distinct. Indeed, B$_l$ is derived from spectropolarimetric data, therefore it is sensitive to polarity cancellation effects (especially at low $v_\mathrm{eq}\sin i$) and may not be modulated over long time scales in the same manner as photometric light curves. An example is V889\,Her, for which brightness oscillations were reported to be twice as fast as the magnetic field variations \citep{Brown2024}. A similar distinction can be made with respect to the results of \citet{Olspert2018}, who used a GP applied to Ca~\textsc{ii} H\&K data with a different formalism, that is with the cycle period as the only time scale in the covariance kernel.

A general complication in using the GP for constraining the cycle period time scale is the sensitivity to short-term variations, which could be misinterpreted for a fast cycle. Indeed, there is evidence for such variations also on the Sun, like the Rieger modulations, on a time scale of $\sim$150\,d \citep{Rieger1984}, and the quasi-biennial oscillations \citep{Mendoza2006,VelascoHerrera2018}. Subsequent `Sun-as-a-star' work in this direction is therefore required to address this point further.

\subsection{Comparisons with the solar cycle}

The magnetic field of the Sun undergoes a polarity reversal in the poloidal and toroidal components during cycle maximum, which is every 11\,yr on average \citep{Richards2009}. The activity is found to increase over 3 to 5\,yr and then to decline over 6 to 8\,yr depending on the cycle strength \citep[e.g.][]{Clette2012}. The behaviour of the solar large-scale magnetic field as it would be reconstructed by ZDI was reported by \citet{Vidotto2018} and \citet{Lehmann2021}: at cycle minimum, the large-scale magnetic field is poloidal, dipolar and axisymmetric, while it is less poloidal, more complex and non-axisymmetric during cycle maximum. This likely stems from the equatorward emergence of sunspots when approaching solar maximum, considering also that the toroidal energy fraction and the sunspot number are correlated. In terms of magnetic energy, the large-scale field intensifies during maximum and decreases during minimum \citep{Vidotto2018,Lehmann2021}. Using this information of the large-scale field topology during solar Hale cycle as benchmark, we compare the magnetic field evolution of the stars in our sample.

For the two slow rotators in our sample, that is HD\,9986 and HD\,56124, we observe some similarities with the solar cycle. As shown in Fig.~\ref{fig:zdi_hd9986}, the large-scale magnetic field of HD\,9986 exhibits two polarity reversals of the radial field within approximately 11\,yr, which is twice as fast as the solar cycle. Our observations of HD\,9986 grasped phases of the magnetic cycle in which the large-scale field was mostly non-axisymmetric, that is 15 to 50\% of the magnetic energy is in the axisymmetric modes. Correspondingly, the obliquity of the positive polarity of the dipole oscillated between 55$^{\circ}$, 125$^{\circ}$ and back to 40$^{\circ}$ (as measured relative to the stellar rotation axis). In 2008.76, the visible hemisphere of the star is dominated by a positive polarity, which flips to negative in 2010.76, suggesting that a maximum of the cycle occurred within this interval. From 2010.76 to 2012.85, the topology sees a rise in poloidal (and dipolar) component, while the axisymmetry remained low, hence it resembles the initial stages of a magnetic cycle descending phase. We observe a second polarity flip in 2017.76, meaning that a second maximum of the cycle likely occurred between 2012.85 and 2017.76. The maps of 2017.76 and 2018.74 show a poloidal, stronger and more axisymmetric topology, so they could be placed in the second descending phase of the cycle. Finally the map of 2023.09 features a low axisymmetry like 2012.85, hence it may reflect the start of another descending phase.

For HD\,56124, our observations capture mostly-axisymmetric configurations of the large-scale magnetic field (see Fig.~\ref{fig:zdi_hd56124}), in 2008.08, 2011.90, 2017.88, and 2021.29. We inferred a polarity switch of the radial component time scale of around 3\,yr, considering that the topology in 2008.08 is similar to 2011.90 but with opposite sign. The maps are not exactly identical apart from the polarity reversal, therefore the timescale may be larger than 3\,yr. This is supported by the fact that if the time scale is 3\,yr, the map of 2017.88 would have a stronger magnetic field as seen in 2011.90. Therefore the sought magnetic switch time scale is a factor of 3-4 shorter than the solar polarity reversal timescale. If we assume the Hale cycle paradigm for HD\,56124, then the maps of 2008.08 and 2011.90 may represent two cycle minima, in which the field is poloidal, dipolar and axisymmetric. Then, the 2017.88 map may be capturing the final stages of a descending phase or the initial stages of an ascending phase, given the weak, poloidal and axisymmetric topology. Finally, the 2021.29 map could be placed in the middle of an ascending phase given the strong, poloidal and less axisymmetric field.

The case of HD\,73350 is also similar to the Sun, although the star rotates in 14\,d (twice as fast as the Sun). From the three reconstructed ZDI maps (see Fig.~\ref{fig:zdi_hd73350}), we note that the radial magnetic field starts in 2007.09 from a complex geometry and then becomes weaker and more poloidal, dipolar, and axisymmetric in the subsequent five years. This resembles a descending phase of the solar cycle, for which high-order harmonics dominate during activity maximum and the dipolar mode during minimum. This findings suggest a putative magnetic polarity reversal time scale between 10-15\,yr, given the similarity to the Sun, but we cannot exclude a faster evolution, given the restricted number of available data and the polarity reversal of the toroidal field.

For HD\,76151, which rotates in roughly 18\,d, the ZDI reconstructions captured one clear polarity reversal of the large-scale topology, which is mostly dipolar and with fast-varying levels of axisymmetry. One possible scenario is that the polarity reversal occurs on time scales of 2-2.5\,yr and our observing cadence did not capture this phenomenon clearly. Indeed, we can consider an axisymmetric dipolar topology with a dominant negative polarity in 2006, 2011, 2016, and 2021, and the same with a positive polarity in 2008, 2013, 2018, and 2023. This is supported by the magnetic topology with opposite polarity reconstructed in 2015.95 and 2023.10 (consistently with the B$_l$ measurements, see Fig.~\ref{fig:Bl_hd76151}). This scenario would be similar to a fast, solar-like magnetic cycle. 

An alternative scenario for HD\,76151 could be that there is a superposition of a short- and long-term cycle that affects the large-scale magnetic field. The short cycle would be responsible for the fast variations in axisymmetry, and the long term for the polarity reversal. We indeed saw a drop of the field axisymmetry to 45\% in 2012.05 and 2017.02 and a more substantial decrease down to 5\% in 2019.02, 2022.28, and 2024.06, possibly due to a combination of the short-term and long-term cycles. Having detected only one polarity reversal does not allow us to constrain the time scale for the longer cycle. Although the configurations in 2015.95 and 2023.10 are opposite, suggesting a magnetic cycle of 16\,yr, reconstructing a map with either of these configurations would be more definitive. Previous work on the chromospheric and photometric variability of HD\,76151 revealed a long-term cycle of 16-18\,yr \citep{Olspert2018,BoroSaikia2018a} and a fast cycle of 2.5-3\,yr \citep{Baliunas1995,Brandenburg2017} or 5\,yr \citep{Olah2016,Egeland2017b}, therefore this non-solar cycle scenario cannot be ruled out. Such variations may resemble other cases like $\varepsilon$ Eri \citep{Jeffers2022} and V889\,Her \citep{Brown2024}, and can be also interpreted as the equivalent of the biennial variations observed on the Sun \citep{Fletcher2010,Bazilevskaya2014}.

Our two remaining stars, HD\,166435 and HD\,175726, are the fastest rotators in our sample, with a rotation period 7.7 and 6.5 times shorter than the Sun, respectively. They exhibit somewhat discordant behaviour relative to the solar cycle, with complex field topologies and mainly an oscillation in strength.

\subsection{Comparisons with other Sun-like stars}

The application of ZDI to a time series of spectropolarimetric observations has revealed solar-like magnetic cycles for other stars in the past: notable examples of solar analogs are HD\,190771 and $\kappa$\,Cet. In this section, we compare and discuss the magnetic cycles reported for these stars to the patterns observed for our sample. 

HD\,190771 has a temperature of 5834\,K, a mass of 0.96\,M$_\odot$, and a rotation period of 8.8\,d \citep{Morgenthaler2011}, therefore it lies close to HD\,73350. There is a striking resemblance of the evolution of HD\,73350's field relative to HD\,190711. More precisely, the ZDI reconstructions of HD\,190771 by \citet{Petit2009} and \citet{Morgenthaler2011} showed a polarity reversal of the azimuthal field, the transformation of a toroidal-dominated geometry into a poloidal-dominated one, and finally a polarity reversal of the radial field. Our observations of HD\,73350 capture the first two stages of this evolution, since the azimuthal field switches polarity from 2007.09 to 2011.06 and becomes mainly poloidal (see Fig.~\ref{fig:zdi_hd73350}). The map of 2012.04 shows a poloidal field with lower axisymmetry than 2011.06, thus additional monitoring is required to potentially observe the polarity flip of the radial field.

The ZDI reconstructions of HD\,56124 (see Fig.~\ref{fig:zdi_hd56124}) show two evident polarity reversals of the radial field as well as an increased complexity of the toroidal field from a mostly-negative configuration at the start of our time series. These characteristics are similar to the field maps of $\kappa$~Cet \citep{BoroSaikia2022}, which is a G5 dwarf with a mass of 0.95\,M$_\odot$ and a rotation period of 9.2\,d. The observations of $\kappa$~Cet spanned between 2012 and 2018, and exhibited two polarity flips of the radial field separated by epochs with a highly complex field. The azimuthal field showed one polarity reversal with a transition phase of high complexity as well. The inferred time scale for the Hale-like cycle of $\kappa$~Cet is approximately 10\,yr. Our observations of HD\,56124 captured only phases in which the radial field was dipolar and axisymmetric, and possibly missed phases of evident high complexity. Together with the similarity of the maps in 2008.08 and 2021.29 and the polarity reversal between 2008.08 and 2011.90, this comparison supports a magnetic cycle for HD\,56124 with a time scale shorter than $\kappa$~Cet. 

Considering the radial field component of the large-scale magnetic field, the ZDI reconstructions of our fast rotators (HD~166435 and HD~175726) suggest fast evolution in the polar regions. Examples are the epochs 2010.51-2010.60 for HD~166435, and 2008.55-2008.63 and 2024.53-2024.63 for HD~175726. Magnetic polarity reversals of polar regions were observed in young Sun-like stars such as V~1385~Ori and possibly HD~35296 \citep{Waite2015,Rosen2016,Willamo2022}.

\subsection{Correlations with chromospheric cycles}

The wavelength coverage of ESPaDOnS and Narval gives access to useful chromospheric diagnostics, that is the Ca~\textsc{ii} H\&K lines at 3968.470 and 3933.661\,{\AA}. By normalising the unpolarised flux contained within these lines with respect to the nearby continuum, it is possible to define the $S$-index \citep{Vaughan1978}. This is a canonical proxy to gauge the activity level of a star and it has been used extensively to search for stellar activity cycles since the Mt. Wilson Project \citep{Wilson1968,Duncan1991,Baliunas1995}. 

The $S$-index is computed from unpolarised spectra by definition, therefore it is expected to be sensitive to magnetic fields of both small and large spatial scales. Studies on the temporal variability of the $S$-index of the Sun over the Schwabe cycle have shown a direct correlation with photometric time series \citep{Radick2018}. Furthermore, the variations of the solar $S$-index are also connected to the evolution of the large-scale field geometry over the Hale cycle, since the $S$-index correlates with the number of spots and active regions. In turn, the spot number is correlated to the large-scale magnetic field. Over the cycle, the magnetic field topology is complex when the $S$-index is at maximum, and it is a simple dipole when the $S$-index is at minimum. This correlation was also observed for 61~Cyg~A over the course of its 7.3\,yr Hale-like cycle \citep{BoroSaikia2016,BoroSaikia2018}, as well as for the 120-d cycle of $\tau$ Boo \citep{Jeffers2022} and the 1.1\,yr cycle of HD\,75332 \citep{Brown2021}. \citet{Jeffers2022} recently showed that, for 61~Cyg~A and $\varepsilon$~Eri, the axisymmetric component of the toroidal field (which is a proxy for flux emergence) follows the respective $S$-index cycle, in a similar manner as for the Sun \citep{Cameron2018}.

\begin{figure}[!ht]
    \centering
    \includegraphics[width=0.92\columnwidth]{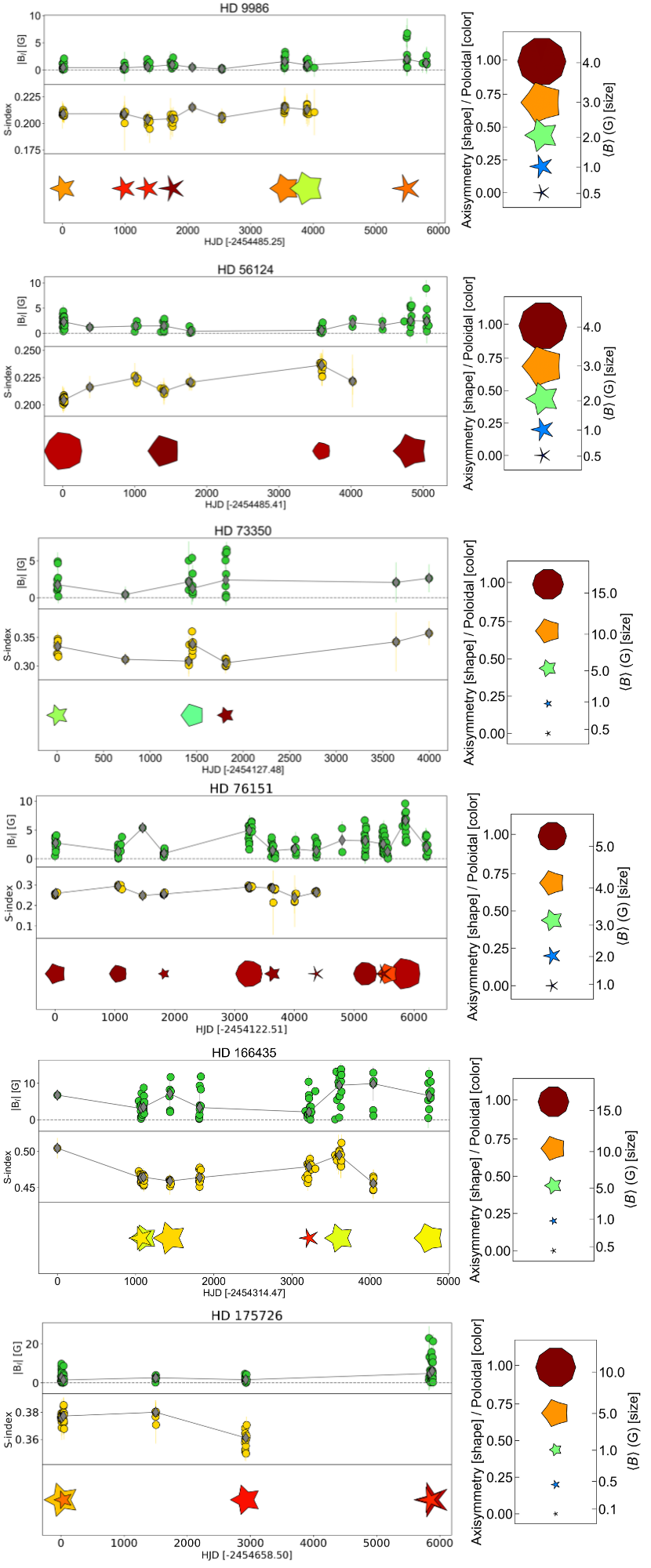}
    \caption{Long-term evolution of activity indices and large-scale magnetic field topology for all our stars. Each panel corresponds to a star and contains the time series of $|$B$_l|$ (top), $S$-index (middle) and large-scale topology reconstructed with ZDI (bottom). In the $|$B$_l|$ and $S$-index panels, the epoch-averaged values are shown as grey diamonds, and they are connected by a solid grey line. In the topology panels, the symbol size, colour and shape encodes the ZDI average field strength, poloidal/toroidal energy fraction and axisymmetry, as illustrated in the side bar.}
    \label{fig:sindex}
\end{figure}

To provide additional insight on the relation between chromospheric diagnostics and the large-scale field geometry, and their temporal evolution, we computed the $S$-index for all the ESPaDOnS and Narval observations following the prescription of \citet{Marsden2014}. We then compared the variations of the $S$-index with respect to the longitudinal magnetic field and the magnetic field topology. The time series of $S$-index, absolute value of B$_l$, and main properties of magnetic field topologies are shown in Fig.~\ref{fig:sindex}. The extraction of the blue spectral orders (below approximately 400\,nm) from Neo-Narval observations encountered problems \citep{LopezAriste2022}, thus we did not compute the $S$-index for these observations.

For the first half of the time series of HD\,9986, the epoch-averaged values of $S$-index and $|$B$_l|$ exhibit an anti-correlated modulation, whereas for the second half the two quantities show a correlation. We note that the amplitude of variations of the $S$-index is around 0.02, which is similar to the Sun \citep{Egeland2017}. When compared to the evolution of the large-scale field, we notice that the $S$-index decreased from 2008.08 to 2012.85 when the poloidal (toroidal) fraction increased (decreased), and in the latest epochs the $S$-index increased when the poloidal (toroidal) fraction decreased (increased). This may be a first hint at a correlation between toroidal flux emergence and the $S$-index, as expected for the Sun \citep[as conveyed by the butterfly diagram, see e.g.][]{Maunder1904,Vidotto2018,Lehmann2021}. It is also interesting to point out that the 5-6\,yr long-term evolution of the $S$-index of HD\,9986 is of the same order of magnitude as the one of 18\,Sco \citep{doNascimento2023}, which is a solar analog with similar stellar properties to HD\,9986.

Another notable star in our sample is HD\,76151 (see Fig.~\ref{fig:sindex}). The long-term modulation of the average $S$-index appears anti-correlated with $|$B$_l|$ in the first half of the time series and reasonably correlated in the rest, but the interpretation is not straightforward. We checked where our observations fall in the context of previous work on the long-term behaviour of chromospheric activity indices \citep{Baliunas1995,Olspert2018,BoroSaikia2018a}. These studies analysed $S$-index time series for observations collected between 1965 and 1995 within the Mt. Wilson project, and found multiple cycle timescales, of 2.5-5\,yr and 16-18\,yr, for HD\,76151. Considering the long-term modulation, for which there are cycle minima in 1970 and 1988 \citep{BoroSaikia2018a}, we would expect the next ones in 2006 and 2024. Likewise, the cycle maxima were recorded in 1978 and 1977, hence they would occur again in 2015 and 2033. Although our data set is not as dense as \citet{Baliunas1995} and \citet{BoroSaikia2018a}, we observe that our $S$-index values in 2007.09 are at minimum, and the values in 2015.95 are at maximum. This is consistent with what is expected from the time scales reported in \citet{BoroSaikia2018a}. We also note that our time series shows a maximum of $S$-index values in 2009.95 followed by a rapid decrease, similar to the fast variation after the maximum in 2015.95. This is compatible with the fast cycle of 2.5\,yr.

When compared to the evolution of the large-scale magnetic field of HD\,76151, the maximum of $S$-index values in our time series corresponds to when the field was dipolar, axisymmetric and the most intense (6\,G on average). This is at odds with respect to the solar cycle, since the solar $S$-index maximum correlates with a complex magnetic field topology. The situation may resemble the case of $\varepsilon$ Eri, which does not show a polarity reversal at $S$-index maxima every 3\,yr \citep{Jeffers2022}, but rather it is synchronised to the long-term chromospheric cycle of 13\,yr \citep{Metcalfe2013}. 

For HD\,56124 and HD\,73350 respectively, we observe an overall anti-correlation and correlation between $S$-index and $|$B$_l|$, but no striking connection with the large-scale field topology. For HD\,166435 and HD\,175726, we do not observe specific patterns between the $S$-index and $|$B$_l|$ or the magnetic topology evolution. For HD\,73350 and HD\,175726, we may still observe a hint of the correlation between toroidal energy fraction and $S$-index (like for HD\,9986), albeit with fewer ZDI reconstructions. HD\,56124 almost supports this feature, but the amount of toroidal field reconstructed during activity maximum in our time series (2017.88 epoch) is not substantial. Considering that the inclination of HD\,56124 is $40^{\circ}$ and it has a low $v_\mathrm{eq}\sin i$ at 1.5~km~s$^{-1}$, perhaps we may be losing some sensitivity to the axisymmetric toroidal field in such conditions. A similar consideration could also apply for HD~76151 (inclination of $30^{\circ}$ and $v_\mathrm{eq}\sin i$ of 1.2~km~s$^{-1}$).

The lack of specific patterns between magnetic maps and activity indices is partly due to a time series with scarce sampling (especially for HD\,175726). The complex temporal variations of the activity indices for our fastest rotators (HD\,166435 and HD\,175726) are somewhat consistent with that expected from chromospheric activity analyses of cool stars, in particular the absence of clear cycles for P$_\mathrm{rot}<10$\,d \citep{BoroSaikia2018a}, as well as with photometric monitoring of young, fast-rotating stars \citep{Olah2016}. A clear example for this is AB~Dor \citep{Donati2003c}, whose magnetic maps do not display obvious polarity reversals but rather show an erratic evolution.

\subsection{Connection to dynamo simulations}

Studying differential rotation and magnetic cycles on stars other than the Sun provides additional observational constraints for numerical simulations of dynamo models \citep{BrunBrowning2017} and self-consistent models of convection, differential rotation and dynamo driving \citep{Kapyla2023}. In this section, we contextualise our findings with respect to some trends that have emerged and that have been reproduced by these simulations.

Differential rotation is expected to vary with stellar rotation period and spectral type, with larger values for F-types and lower for M-types \citep{Barnes2005,CollierCameron2007,Balona2016}. This sensitivity to stellar mass and effective temperature was also captured by simulations \citep{Brown2008,Brun2017}. Consistently with the expected trend, we measured values of $d\Omega=0.14$ and 0.15\,rad\,d$^{-1}$ (i.e. twice than solar) for HD\,166435 and HD\,175726, which are the hottest stars in our sample ($\sim100-200$\,K hotter than the Sun). 

Furthermore, \citet{Gastine2014} reported solar-like differential rotation ($d\Omega>0$) for rapidly rotating stars with a small convective Rossby number, and anti-solar cases ($d\Omega<0$) for slowly rotating stars with a large convective Rossby number. At the transition region between these two regimes, both cases of differential rotation can occur. While we cannot directly compare the differential rotation rates we derived with the simulations, owing to a different Rossby number formalism \citep[see][for a discussion]{Brun2017}, we note that $d\Omega>0$ for our fast rotators, namely HD\,166435 and HD\,175726. For our slow rotators ($Ro\gtrsim1.0$), our search of differential rotation does not yield conclusive constraints, but does not exclude the presence of differential rotation for these stars. This could stem from two main reasons \citep[as already pointed out by][]{Petit2002}: i) the large-scale magnetic field topology is not favourable because it does not possess multiple magnetic features probing different latitudes, and ii) the span of the observations may limit our monitoring of active regions, that decay before we are able to grasp their influence on Stokes~$V$ over multiple stellar rotations.

Turning to the magnetic topology, our observations of HD\,166435 and HD\,175726 did not capture an evident magnetic cycle (see Fig.~\ref{fig:zdi_hd166435} and \ref{fig:zdi_hd175726}), rather a fast evolution of the magnetic features in the polar regions. The emergence of magnetic flux at the pole is expected for fast rotators \citep{Schuessler1992}, and it was observed on, for instance, AB~Dor \citep{Mackay2004}. In particular, stars rotating 4-8 times faster than the Sun show polar magnetic regions from flux transport simulations \citep{Isik2007,Isik2018}, therefore our ZDI reconstructions are consistent (HD\,166435 and HD\,175726 rotate 7.8 and 6.6 times faster). The lack of Hale-like cycle signatures could be due to an underlying cycle time scale longer than the time span of our observations: about 10\,yr and 16\,yr for the two stars, respectively. This is compatible with magnetohydrodynamical simulations based on a Babcock-Leighton model \citep{Jouve2010,Karak2014,Hazra2019}, for which we expect fast rotating stars to possess a slower meridional circulation, and ultimately a longer magnetic cycle. Additional simulations by \citet{Brun2022} of solar-like convective dynamos found long magnetic cycles for small fluid Rossby numbers, while other studies obtained irregular patterns for fast rotators \citep{Vashishth2023}. Recent numerical simulations by \citet{Noraz2024} showed that fast rotators tend to exhibit rapid evolution and local polarity reversal. In our sample, we could potentially observe this behaviour for HD\,166435, since the polar region flipped polarity between 2010.51 and 2010.60.

For the slower rotators in our sample, that is HD\,9986, HD\,56124, and HD\,73350, they exhibit an evolution of the large-scale magnetic field topology that resembles a Sun-like magnetic cycle, although with faster time scales. \citet{Viviani2018} presented global magnetohydrodynamic convection simulations of solar-like stars with rotation rates between 1 and 31 times the solar rotation rate. They reported the presence of magnetic cycles with polarity reversals in the slow rotation regime (that is with rotation rate larger than 1.8 the solar value), and the absence of cycles in the fast-rotating regime, in overall agreement with our observational results.

In accordance with the flux emergence simulations and corresponding ZDI reconstructions of large-scale field for solar-like stars by \citet{Lehmann2019} and \citet{Lehmann2021}, we observe that the reconstructed axisymmetry (the shape of the data points in Fig.~\ref{fig:sindex}) has a reasonable correlation with the $S$-index, that is the field topology is more axisymmetric when the $S$-index increases. The sharp decrease of toroidal fraction for HD\,9986 in 2012.85 epoch can be indicative of activity minimum \citep{Lehmann2021}. We also observed a concentration of large-scale azimuthal field at low latitudes, up to approximately $30^{\circ}$, as illustrated in Fig.~\ref{fig:zdi_hd9986}, \ref{fig:zdi_hd56124}, and \ref{fig:zdi_hd73350}. The interpretation of such observation as solar-like cycles is supported by the magnetohydrodynamic simulations of \citet{Strugarek2017}. They modelled solar-type stars with rotation periods between 14 and 29\,d (i.e. the same range as HD\,9986, HD\,56124, and HD\,73350), and captured regular polarity reversals with time scales of $\sim$10\,yr, together with an equatorial propagation of the large-scale magnetic field. Our observations are also in agreement with simulations of solar convection zones reporting decade-long polarity switches \citep{Ghizaru2010,Kapyla2012,Augustson2015,Noraz2024}.

The case of HD\,76151 is more complicated because our observations could be explained by short-term attempts at polarity reversals modulated over a long-term cycle, in a similar manner to $\varepsilon$ Eri \citep{Jeffers2022} and V889\,Her \citep{Brown2024}. As pointed out by the authors, short-term variations with polarity reversals were reproduced by magnetohydrodynamic simulations of dynamo in the upper part of the convection zone \citep{Kapyla2016,Strugarek2018,Brun2022}. These short cycles can be explained by a near-surface $\alpha\Omega$ dynamo, in contrast to the long-term cycles that would require a deep-seated dynamo \citep{Brun2022}. Additional spectropolarimetric observations of HD\,76151 are required to further investigate its cyclic variations.

\section{Conclusions}\label{sec:conclusions}

We have carried out a long-term spectropolarimetric monitoring of six solar-like stars, in order to search for evolution of the large-scale magnetic field in the form of a magnetic cycle. The Sun is the primary benchmark. Our stars were observed as part of the BCool program \citep{Marsden2014} and possess analogous properties to the Sun. The masses are at most 6\% larger than the solar value and the temperatures are at most 200\,K larger. The rotation periods range between 3.5 and 21\,d, which makes our stars a suitable sample to probe different levels of stellar magnetic activity and investigate the nature of dynamo cycles based primarily on the rotation period.

With a baseline covering 17 years (between 2007 and 2024) of high-resolution, circularly-polarised spectra, we computed a time series of longitudinal magnetic field values for each star in our sample. We analysed the temporal content of these time series using a generalised Lomb-Scargle periodogram as well as a quasi-periodic Gaussian process. Correspondingly, we reconstructed the large-scale magnetic field topology via Zeeman-Doppler imaging to analyse in detail the evolution of the field properties. For four stars, that is HD\,9986, HD\,56124, HD\,166435, and HD\,175726 we reconstructed field maps for the first time. Finally, we computed the chromospheric $S$-index from Ca~\textsc{ii} H\&K lines to compare its long-term behaviour to the evolution of the magnetic field. 

Our conclusions are the following:
\begin{enumerate}
    \item There is a variety in the long-term evolution of the field topology which depends on the stellar rotation period. For HD\,9986, HD\,56124, and HD\,73350 (rotation rate up to 2.2 times faster than the Sun, and $Ro$ between 0.93 and 1.80), the stars exhibit cyclic variations with polarity reversals. These are observed in both the poloidal and toroidal components, but not simultaneously. The star HD\,76151 (rotation rate equal to 1.5 times solar and $Ro=1.50$) may represent an exception, with short-term cyclic oscillations in axisymmetry modulated by a long-term cycle for which only one polarity reversal was captured. HD\,166435 and HD\,175726 are the fastest rotators in our sample (6.6 and 7.8 times solar and $Ro=0.3-0.5$), and they did not manifest magnetic cycles, but a persistently complex magnetic topology over the time span of our observations with possibly fast polarity reversals in the polar regions.
    \item For stars showing cyclic evolution, the time scale between polarity reversals is shorter than for the Sun. In particular, for HD\,9986 we have P$_\mathrm{rot}=21.03$~d and P$_\mathrm{cyc}=5-6$~yr, and for HD\,56124 we have P$_\mathrm{rot}=20.70$~d and P$_\mathrm{cyc}=2-3$~yr. For HD\,73350 (P$_\mathrm{rot}=12.27$~d), the field topology seems to have a cycle as well, but we most probably captured only the descending phase, thus a robust time scale cannot be constrained.
    \item In a similar manner to the Sun, the variations of $S$-index for HD\,9986 seem to follow the fluctuations in toroidal energy fraction, possibly hinting at a connection between $S$-index and toroidal flux emergence. The long-term evolution of the epoch-averaged $S$-index and unsigned longitudinal field exhibit an overall anti-correlation for HD\,56124 and a correlation for HD\,73350. For the other stars, the evolution of these quantities is less straightforward to interpret because either the time series are partly correlated and anti-correlated, or the observational gaps in the time series prevent us from following the evolution efficiently.
    \item The quasi-periodic GP modelling of the longitudinal field time series allowed us to obtain evolution time scales for most of our stars between 230 and 850\,d, with two unconstrained cases corresponding to stars with scarce sampling and long observational gaps. For HD\,56124, the time scale retrieved by the GP is half the time scale topology evolution and for HD\,76151, it is on the same order of magnitude as the short-term cycle. As opposed to the ZDI analysis, it is not straightforward to identify magnetic cycles from the GP analysis on B$_l$ alone because, despite being sensitive to polarity reversals, information related to the main magnetic field component, complexity, and axisymmetry is not recovered in detail.
\end{enumerate}

Considering that magnetic cycles induce a seismic signal that alters the parameters of p-mode oscillations in the stellar interior, namely frequency, amplitude, and energy \citep{Garcia2010,Basu2016,Kiefer2018}, our findings provide an interesting set of stars for multi-technique follow-up \citep{Karoff2009,Chaplin2014}. In these regards, the future space-based mission PLAnetary Transits and Oscillations of stars \citep[PLATO;][]{Rauer2014} will play a crucial role, provided that the observational baseline will be long enough to grasp long-term modulations of photometric light curves for hundreds of solar-like stars \citep{Breton2024}.

More generally, our findings provide additional motivation for tailored campaigns targeting solar-like stars since combined studies can bring more insights on the connection between the large-scale magnetic field at the stellar surface and other observables such as cycle-induced internal signatures. Long-term spectropolarimetric monitoring of solar-like stars is also paramount to provide reliable information on stellar activity to aid extreme precision radial velocity searches of exoplanets \citep[see e.g.][]{Rescigno2024}, which is particularly relevant, for instance, in light of the development of HARPS3 \citep{Thompson2016,Hall2018}.

Finally, stellar magnetic fields govern the environment in which exoplanets are embedded, and ultimately affect the conditions of climate \citep{Edmonds2024}, and habitability \citep[e.g.][]{Vidotto2013,Vidotto2014,Airapetian2017}. Information regarding the large-scale magnetic field is crucial for the accurate modelling of stellar magnetospheres, space weather and star-planet interactions \citep[e.g.][]{Vidotto2014,Villareal2018,Kavanagh2021,RodgersLee2023,Bellotti2023a,Bellotti2024b}, which can be modulated by magnetic cycles \citep{Hazra2020}. In this context, our findings provide additional observational constraints to the evolution of these environments.

\begin{acknowledgements}

We thank the referee for the comments that improved the manuscript. This publication is part of the project `Exo-space weather and contemporaneous signatures of star-planet interactions' (with project number OCENW.M.22.215 of the research programme `Open Competition Domain Science- M'), which is financed by the Dutch Research Council (NWO). AAV acknowledges funding from the European Research Council (ERC) under the European Union's Horizon 2020 research and innovation programme (grant agreement No 817540, ASTROFLOW). 
CPF acknowledges funding from the European Union's Horizon Europe research and innovation programme under grant agreement No. 101079231 (EXOHOST), and from the United Kingdom Research and Innovation (UKRI) Horizon Europe Guarantee Scheme (grant number 10051045). 
VS acknowledges support from the European Space Agency (ESA) as an ESA Research Fellow.
Based on observations obtained at the Canada-France-Hawaii Telescope (CFHT) which is operated by the National Research Council of Canada, the Institut National des Sciences de l'Univers of the Centre National de la Recherche Scientique of France, and the University of Hawaii. We thank the TBL team for providing service observing with Neo-Narval. This work has made use of the VALD database, operated at Uppsala University, the Institute of Astronomy RAS in Moscow, and the University of Vienna; Astropy, 12 a community-developed core Python package for Astronomy \citep{Astropy2013,Astropy2018}; NumPy \citep{VanderWalt2011}; Matplotlib: Visualization with Python \citep{Hunter2007}; SciPy \citep{Virtanen2020} and PyAstronomy \citep{Czesla2019}.

\end{acknowledgements}


\bibliographystyle{aa}
\bibliography{biblio}

\begin{thebibliography}{208}
\expandafter\ifx\csname natexlab\endcsname\relax\def\natexlab#1{#1}\fi

\bibitem[{{Aigrain} \& {Foreman-Mackey}(2023)}]{Aigrain2022}
{Aigrain}, S. \& {Foreman-Mackey}, D. 2023, \araa, 61, 329

\bibitem[{{Airapetian} {et~al.}(2017){Airapetian}, {Glocer}, {Khazanov}, {Loyd}, {France}, {Sojka}, {Danchi}, \& {Liemohn}}]{Airapetian2017}
{Airapetian}, V.~S., {Glocer}, A., {Khazanov}, G.~V., {et~al.} 2017, \apjl, 836, L3

\bibitem[{{Angus} {et~al.}(2018){Angus}, {Morton}, {Aigrain}, {Foreman-Mackey}, \& {Rajpaul}}]{Angus2018}
{Angus}, R., {Morton}, T., {Aigrain}, S., {Foreman-Mackey}, D., \& {Rajpaul}, V. 2018, \mnras, 474, 2094

\bibitem[{{Astropy Collaboration} {et~al.}(2018){Astropy Collaboration}, {Price-Whelan}, {Sip{\H{o}}cz}, {G{\"u}nther}, {Lim}, {Crawford}, {Conseil}, {Shupe}, {Craig}, {Dencheva}, {Ginsburg}, {VanderPlas}, {Bradley}, {P{\'e}rez-Su{\'a}rez}, {de Val-Borro}, {Aldcroft}, {Cruz}, {Robitaille}, {Tollerud}, {Ardelean}, {Babej}, {Bach}, {Bachetti}, {Bakanov}, {Bamford}, {Barentsen}, {Barmby}, {Baumbach}, {Berry}, {Biscani}, {Boquien}, {Bostroem}, {Bouma}, {Brammer}, {Bray}, {Breytenbach}, {Buddelmeijer}, {Burke}, {Calderone}, {Cano Rodr{\'\i}guez}, {Cara}, {Cardoso}, {Cheedella}, {Copin}, {Corrales}, {Crichton}, {D'Avella}, {Deil}, {Depagne}, {Dietrich}, {Donath}, {Droettboom}, {Earl}, {Erben}, {Fabbro}, {Ferreira}, {Finethy}, {Fox}, {Garrison}, {Gibbons}, {Goldstein}, {Gommers}, {Greco}, {Greenfield}, {Groener}, {Grollier}, {Hagen}, {Hirst}, {Homeier}, {Horton}, {Hosseinzadeh}, {Hu}, {Hunkeler}, {Ivezi{\'c}}, {Jain}, {Jenness}, {Kanarek}, {Kendrew}, {Kern}, {Kerzendorf}, {Khvalko}, {King}, {Kirkby}, {Kulkarni},
  {Kumar}, {Lee}, {Lenz}, {Littlefair}, {Ma}, {Macleod}, {Mastropietro}, {McCully}, {Montagnac}, {Morris}, {Mueller}, {Mumford}, {Muna}, {Murphy}, {Nelson}, {Nguyen}, {Ninan}, {N{\"o}the}, {Ogaz}, {Oh}, {Parejko}, {Parley}, {Pascual}, {Patil}, {Patil}, {Plunkett}, {Prochaska}, {Rastogi}, {Reddy Janga}, {Sabater}, {Sakurikar}, {Seifert}, {Sherbert}, {Sherwood-Taylor}, {Shih}, {Sick}, {Silbiger}, {Singanamalla}, {Singer}, {Sladen}, {Sooley}, {Sornarajah}, {Streicher}, {Teuben}, {Thomas}, {Tremblay}, {Turner}, {Terr{\'o}n}, {van Kerkwijk}, {de la Vega}, {Watkins}, {Weaver}, {Whitmore}, {Woillez}, {Zabalza}, \& {Astropy Contributors}}]{Astropy2018}
{Astropy Collaboration}, {Price-Whelan}, A.~M., {Sip{\H{o}}cz}, B.~M., {et~al.} 2018, AJ, 156, 123

\bibitem[{{Astropy Collaboration} {et~al.}(2013){Astropy Collaboration}, {Robitaille}, {Tollerud}, {Greenfield}, {Droettboom}, {Bray}, {Aldcroft}, {Davis}, {Ginsburg}, {Price-Whelan}, {Kerzendorf}, {Conley}, {Crighton}, {Barbary}, {Muna}, {Ferguson}, {Grollier}, {Parikh}, {Nair}, {Unther}, {Deil}, {Woillez}, {Conseil}, {Kramer}, {Turner}, {Singer}, {Fox}, {Weaver}, {Zabalza}, {Edwards}, {Azalee Bostroem}, {Burke}, {Casey}, {Crawford}, {Dencheva}, {Ely}, {Jenness}, {Labrie}, {Lim}, {Pierfederici}, {Pontzen}, {Ptak}, {Refsdal}, {Servillat}, \& {Streicher}}]{Astropy2013}
{Astropy Collaboration}, {Robitaille}, T.~P., {Tollerud}, E.~J., {et~al.} 2013, A\&A, 558, A33

\bibitem[{{Augustson} {et~al.}(2015){Augustson}, {Brun}, {Miesch}, \& {Toomre}}]{Augustson2015}
{Augustson}, K., {Brun}, A.~S., {Miesch}, M., \& {Toomre}, J. 2015, \apj, 809, 149

\bibitem[{{Babcock}(1961)}]{Babcock1961}
{Babcock}, H.~W. 1961, ApJ, 133, 572

\bibitem[{{Bagnulo} {et~al.}(2009){Bagnulo}, {Landolfi}, {Landstreet}, {Landi Degl'Innocenti}, {Fossati}, \& {Sterzik}}]{Bagnulo2009}
{Bagnulo}, S., {Landolfi}, M., {Landstreet}, J.~D., {et~al.} 2009, PASP, 121, 993

\bibitem[{{Baliunas} {et~al.}(1995){Baliunas}, {Donahue}, {Soon}, {Horne}, {Frazer}, {Woodard-Eklund}, {Bradford}, {Rao}, {Wilson}, {Zhang}, {Bennett}, {Briggs}, {Carroll}, {Duncan}, {Figueroa}, {Lanning}, {Misch}, {Mueller}, {Noyes}, {Poppe}, {Porter}, {Robinson}, {Russell}, {Shelton}, {Soyumer}, {Vaughan}, \& {Whitney}}]{Baliunas1995}
{Baliunas}, S.~L., {Donahue}, R.~A., {Soon}, W.~H., {et~al.} 1995, ApJ, 438, 269

\bibitem[{{Balona} \& {Abedigamba}(2016)}]{Balona2016}
{Balona}, L.~A. \& {Abedigamba}, O.~P. 2016, \mnras, 461, 497

\bibitem[{{Barnes} {et~al.}(2005){Barnes}, {Collier Cameron}, {Donati}, {James}, {Marsden}, \& {Petit}}]{Barnes2005}
{Barnes}, J.~R., {Collier Cameron}, A., {Donati}, J.~F., {et~al.} 2005, \mnras, 357, L1

\bibitem[{{Basu}(2016)}]{Basu2016}
{Basu}, S. 2016, Living Reviews in Solar Physics, 13, 2

\bibitem[{{Baum} {et~al.}(2022){Baum}, {Wright}, {Luhn}, \& {Isaacson}}]{Baum2022}
{Baum}, A.~C., {Wright}, J.~T., {Luhn}, J.~K., \& {Isaacson}, H. 2022, \aj, 163, 183

\bibitem[{{Bazilevskaya} {et~al.}(2014){Bazilevskaya}, {Broomhall}, {Elsworth}, \& {Nakariakov}}]{Bazilevskaya2014}
{Bazilevskaya}, G., {Broomhall}, A.~M., {Elsworth}, Y., \& {Nakariakov}, V.~M. 2014, \ssr, 186, 359

\bibitem[{{Bellotti} {et~al.}(2024{\natexlab{a}}){Bellotti}, {Evensberget}, {Vidotto}, {Lavail}, {L{\"u}ftinger}, {Hussain}, {Morin}, {Petit}, {Boro Saikia}, {Danielski}, \& {Micela}}]{Bellotti2024b}
{Bellotti}, S., {Evensberget}, D., {Vidotto}, A.~A., {et~al.} 2024{\natexlab{a}}, \aap, 688, A63

\bibitem[{{Bellotti} {et~al.}(2023{\natexlab{a}}){Bellotti}, {Fares}, {Vidotto}, {Morin}, {Petit}, {Hussain}, {Bourrier}, {Donati}, {Moutou}, \& {H{\'e}brard}}]{Bellotti2023a}
{Bellotti}, S., {Fares}, R., {Vidotto}, A.~A., {et~al.} 2023{\natexlab{a}}, \aap, 676, A139

\bibitem[{{Bellotti} {et~al.}(2023{\natexlab{b}}){Bellotti}, {Morin}, {Lehmann}, {Folsom}, {Hussain}, {Petit}, {Donati}, {Lavail}, {Carmona}, {Martioli}, {Romano Zaire}, {Alecian}, {Moutou}, {Fouqu{\'e}}, {Alencar}, {Artigau}, {Boisse}, {Bouchy}, {Cadieux}, {Cloutier}, {Cook}, {Delfosse}, {Doyon}, {H{\'e}brard}, {Kochukhov}, \& {Wade}}]{Bellotti2023b}
{Bellotti}, S., {Morin}, J., {Lehmann}, L.~T., {et~al.} 2023{\natexlab{b}}, \aap, 676, A56

\bibitem[{{Bellotti} {et~al.}(2024{\natexlab{b}}){Bellotti}, {Morin}, {Lehmann}, {Petit}, {Hussain}, {Donati}, {Folsom}, {Carmona}, {Martioli}, {Klein}, {Fouqu{\'e}}, {Moutou}, {Alencar}, {Artigau}, {Boisse}, {Bouchy}, {Bouvier}, {Cook}, {Delfosse}, {Doyon}, \& {H{\'e}brard}}]{Bellotti2024}
{Bellotti}, S., {Morin}, J., {Lehmann}, L.~T., {et~al.} 2024{\natexlab{b}}, \aap, 686, A66

\bibitem[{{Bloot} {et~al.}(2024){Bloot}, {Callingham}, {Vedantham}, {Kavanagh}, {Pope}, {Climent}, {Guirado}, {Pe{\~n}a-Mo{\~n}ino}, \& {P{\'e}rez-Torres}}]{Bloot2024}
{Bloot}, S., {Callingham}, J.~R., {Vedantham}, H.~K., {et~al.} 2024, \aap, 682, A170

\bibitem[{{Boro Saikia} {et~al.}(2016){Boro Saikia}, {Jeffers}, {Morin}, {Petit}, {Folsom}, {Marsden}, {Donati}, {Cameron}, {Hall}, {Perdelwitz}, {Reiners}, \& {Vidotto}}]{BoroSaikia2016}
{Boro Saikia}, S., {Jeffers}, S.~V., {Morin}, J., {et~al.} 2016, A\&A, 594, A29

\bibitem[{{Boro Saikia} {et~al.}(2015){Boro Saikia}, {Jeffers}, {Petit}, {Marsden}, {Morin}, \& {Folsom}}]{BoroSaikia2015}
{Boro Saikia}, S., {Jeffers}, S.~V., {Petit}, P., {et~al.} 2015, \aap, 573, A17

\bibitem[{{Boro Saikia} {et~al.}(2018{\natexlab{a}}){Boro Saikia}, {Lueftinger}, {Jeffers}, {Folsom}, {See}, {Petit}, {Marsden}, {Vidotto}, {Morin}, {Reiners}, {Guedel}, \& {BCool Collaboration}}]{BoroSaikia2018}
{Boro Saikia}, S., {Lueftinger}, T., {Jeffers}, S.~V., {et~al.} 2018{\natexlab{a}}, \aap, 620, L11

\bibitem[{{Boro Saikia} {et~al.}(2022){Boro Saikia}, {L{\"u}ftinger}, {Folsom}, {Antonova}, {Alecian}, {Donati}, {Guedel}, {Hall}, {Jeffers}, {Kochukhov}, {Marsden}, {Metodieva}, {Mittag}, {Morin}, {Perdelwitz}, {Petit}, {Schmid}, \& {Vidotto}}]{BoroSaikia2022}
{Boro Saikia}, S., {L{\"u}ftinger}, T., {Folsom}, C.~P., {et~al.} 2022, A\&A, 658, A16

\bibitem[{{Boro Saikia} {et~al.}(2018{\natexlab{b}}){Boro Saikia}, {Marvin}, {Jeffers}, {Reiners}, {Cameron}, {Marsden}, {Petit}, {Warnecke}, \& {Yadav}}]{BoroSaikia2018a}
{Boro Saikia}, S., {Marvin}, C.~J., {Jeffers}, S.~V., {et~al.} 2018{\natexlab{b}}, \aap, 616, A108

\bibitem[{{Brandenburg} {et~al.}(2017){Brandenburg}, {Mathur}, \& {Metcalfe}}]{Brandenburg2017}
{Brandenburg}, A., {Mathur}, S., \& {Metcalfe}, T.~S. 2017, \apj, 845, 79

\bibitem[{{Breton} {et~al.}(2024){Breton}, {Lanza}, {Messina}, {Pagano}, {Bugnet}, {Corsaro}, {Garc{\'\i}a}, {Mathur}, {Santos}, {Aigrain}, {Amard}, {Brun}, {Degott}, {Noraz}, {Palakkatharappil}, {Panetier}, {Strugarek}, {Belkacem}, {Goupil}, {Ouazzani}, {Philidet}, {Reni{\'e}}, \& {Roth}}]{Breton2024}
{Breton}, S.~N., {Lanza}, A.~F., {Messina}, S., {et~al.} 2024, \aap, 689, A229

\bibitem[{{Brown} {et~al.}(2008){Brown}, {Browning}, {Brun}, {Miesch}, \& {Toomre}}]{Brown2008}
{Brown}, B.~P., {Browning}, M.~K., {Brun}, A.~S., {Miesch}, M.~S., \& {Toomre}, J. 2008, \apj, 689, 1354

\bibitem[{{Brown} {et~al.}(2024){Brown}, {Marsden}, {Jeffers}, {Heitzmann}, {Barnes}, \& {Folsom}}]{Brown2024}
{Brown}, E.~L., {Marsden}, S.~C., {Jeffers}, S.~V., {et~al.} 2024, \mnras, 528, 4092

\bibitem[{{Brown} {et~al.}(2021){Brown}, {Marsden}, {Mengel}, {Jeffers}, {Millburn}, {Mittag}, {Petit}, {Vidotto}, {Morin}, {See}, {Jardine}, {Gonz{\'a}lez-P{\'e}rez}, {Gonz{\'a}lez-P{\'e}rez}, \& {BCool Collaboration}}]{Brown2021}
{Brown}, E.~L., {Marsden}, S.~C., {Mengel}, M.~W., {et~al.} 2021, \mnras, 501, 3981

\bibitem[{{Brun} \& {Browning}(2017)}]{BrunBrowning2017}
{Brun}, A.~S. \& {Browning}, M.~K. 2017, Living Reviews in Solar Physics, 14, 4

\bibitem[{{Brun} {et~al.}(2022){Brun}, {Strugarek}, {Noraz}, {Perri}, {Varela}, {Augustson}, {Charbonneau}, \& {Toomre}}]{Brun2022}
{Brun}, A.~S., {Strugarek}, A., {Noraz}, Q., {et~al.} 2022, \apj, 926, 21

\bibitem[{{Brun} {et~al.}(2017){Brun}, {Strugarek}, {Varela}, {Matt}, {Augustson}, {Emeriau}, {DoCao}, {Brown}, \& {Toomre}}]{Brun2017}
{Brun}, A.~S., {Strugarek}, A., {Varela}, J., {et~al.} 2017, \apj, 836, 192

\bibitem[{{Cameron} {et~al.}(2018){Cameron}, {Duvall}, {Sch{\"u}ssler}, \& {Schunker}}]{Cameron2018}
{Cameron}, R.~H., {Duvall}, T.~L., {Sch{\"u}ssler}, M., \& {Schunker}, H. 2018, \aap, 609, A56

\bibitem[{{Carmona} {et~al.}(2023){Carmona}, {Delfosse}, {Bellotti}, {Cort{\'e}s-Zuleta}, {Ould-Elhkim}, {Heidari}, {Mignon}, {Donati}, {Moutou}, {Cook}, {Artigau}, {Fouqu{\'e}}, {Martioli}, {Cadieux}, {Morin}, {Forveille}, {Boisse}, {H{\'e}brard}, {D{\'\i}az}, {Lafreni{\`e}re}, {Kiefer}, {Petit}, {Doyon}, {Acu{\~n}a}, {Arnold}, {Bonfils}, {Bouchy}, {Bourrier}, {Dalal}, {Deleuil}, {Demangeon}, {Dumusque}, {Hara}, {Hoyer}, {Mousis}, {Santerne}, {S{\'e}grasan}, {Stalport}, \& {Udry}}]{Carmona2023}
{Carmona}, A., {Delfosse}, X., {Bellotti}, S., {et~al.} 2023, \aap, 674, A110

\bibitem[{{Chaplin} \& {Basu}(2014)}]{Chaplin2014}
{Chaplin}, W.~J. \& {Basu}, S. 2014, \ssr, 186, 437

\bibitem[{{Charbonneau}(2010)}]{Charbonneau2010}
{Charbonneau}, P. 2010, Living Reviews in Solar Physics, 7, 3

\bibitem[{{Charbonneau}(2020)}]{Charbonneau2020}
{Charbonneau}, P. 2020, Living Reviews in Solar Physics, 17, 4

\bibitem[{{Charbonneau} \& {Sokoloff}(2023)}]{Charbonneau2023}
{Charbonneau}, P. \& {Sokoloff}, D. 2023, \ssr, 219, 35

\bibitem[{{Chatterjee} {et~al.}(2011){Chatterjee}, {Mitra}, {Rheinhardt}, \& {Brandenburg}}]{Chatterjee2011}
{Chatterjee}, P., {Mitra}, D., {Rheinhardt}, M., \& {Brandenburg}, A. 2011, \aap, 534, A46

\bibitem[{{Claret} \& {Bloemen}(2011)}]{Claret2011}
{Claret}, A. \& {Bloemen}, S. 2011, A\&A, 529, A75

\bibitem[{{Clements} {et~al.}(2017){Clements}, {Henry}, {Hosey}, {Jao}, {Silverstein}, {Winters}, {Dieterich}, \& {Riedel}}]{Clements2017}
{Clements}, T.~D., {Henry}, T.~J., {Hosey}, A.~D., {et~al.} 2017, \aj, 154, 124

\bibitem[{{Clette} \& {Lef{\`e}vre}(2012)}]{Clette2012}
{Clette}, F. \& {Lef{\`e}vre}, L. 2012, Journal of Space Weather and Space Climate, 2, A06

\bibitem[{{Coffaro} {et~al.}(2020){Coffaro}, {Stelzer}, {Orlando}, {Hall}, {Metcalfe}, {Wolter}, {Mittag}, {Sanz-Forcada}, {Schneider}, \& {Ducci}}]{Coffaro2020}
{Coffaro}, M., {Stelzer}, B., {Orlando}, S., {et~al.} 2020, \aap, 636, A49

\bibitem[{{Collier Cameron}(2007)}]{CollierCameron2007}
{Collier Cameron}, A. 2007, Astronomische Nachrichten, 328, 1030

\bibitem[{{Czesla} {et~al.}(2019){Czesla}, {Schr{\"o}ter}, {Schneider}, {Huber}, {Pfeifer}, {Andreasen}, \& {Zechmeister}}]{Czesla2019}
{Czesla}, S., {Schr{\"o}ter}, S., {Schneider}, C.~P., {et~al.} 2019, {PyA: Python astronomy-related packages}

\bibitem[{{Datson} {et~al.}(2015){Datson}, {Flynn}, \& {Portinari}}]{Datson2015}
{Datson}, J., {Flynn}, C., \& {Portinari}, L. 2015, \aap, 574, A124

\bibitem[{{Del Pozzo} \& {Veitch}(2022)}]{DelPozzo2022}
{Del Pozzo}, W. \& {Veitch}, J. 2022, {CPNest: Parallel nested sampling}, Astrophysics Source Code Library, record ascl:2205.021

\bibitem[{{DeRosa} {et~al.}(2012){DeRosa}, {Brun}, \& {Hoeksema}}]{DeRosa2012}
{DeRosa}, M.~L., {Brun}, A.~S., \& {Hoeksema}, J.~T. 2012, \apj, 757, 96

\bibitem[{{DeWarf} {et~al.}(2010){DeWarf}, {Datin}, \& {Guinan}}]{DeWarf2010}
{DeWarf}, L.~E., {Datin}, K.~M., \& {Guinan}, E.~F. 2010, \apj, 722, 343

\bibitem[{{Dikpati} {et~al.}(2009){Dikpati}, {Gilman}, {Cally}, \& {Miesch}}]{Dikpati2009}
{Dikpati}, M., {Gilman}, P.~A., {Cally}, P.~S., \& {Miesch}, M.~S. 2009, \apj, 692, 1421

\bibitem[{{do Nascimento} {et~al.}(2016){do Nascimento}, {Vidotto}, {Petit}, {Folsom}, {Castro}, {Marsden}, {Morin}, {Porto de Mello}, {Meibom}, {Jeffers}, {Guinan}, \& {Ribas}}]{doNascimento2016}
{do Nascimento}, J.~D., J., {Vidotto}, A.~A., {Petit}, P., {et~al.} 2016, \apjl, 820, L15

\bibitem[{{do Nascimento} {et~al.}(2023){do Nascimento}, {Barnes}, {Saar}, {de Mello}, {Hall}, {Anthony}, {de Almeida}, {Velloso}, {da Costa}, {Petit}, {Strugarek}, {Wargelin}, {Castro}, {Strassmeier}, \& {Brun}}]{doNascimento2023}
{do Nascimento}, J.~D., {Barnes}, S.~A., {Saar}, S.~H., {et~al.} 2023, \apj, 958, 57

\bibitem[{{Donati}(2003)}]{Donati2003}
{Donati}, J.~F. 2003, in Astronomical Society of the Pacific Conference Series, Vol. 307, Solar Polarization, ed. J.~{Trujillo-Bueno} \& J.~{Sanchez Almeida}, 41

\bibitem[{{Donati} \& {Brown}(1997)}]{DonatiBrown1997}
{Donati}, J.~F. \& {Brown}, S.~F. 1997, A\&A, 326, 1135

\bibitem[{{Donati} {et~al.}(2003{\natexlab{a}}){Donati}, {Collier Cameron}, \& {Petit}}]{Donati2003b}
{Donati}, J.~F., {Collier Cameron}, A., \& {Petit}, P. 2003{\natexlab{a}}, \mnras, 345, 1187

\bibitem[{{Donati} {et~al.}(2003{\natexlab{b}}){Donati}, {Collier Cameron}, {Semel}, {Hussain}, {Petit}, {Carter}, {Marsden}, {Mengel}, {L{\'o}pez Ariste}, {Jeffers}, \& {Rees}}]{Donati2003c}
{Donati}, J.~F., {Collier Cameron}, A., {Semel}, M., {et~al.} 2003{\natexlab{b}}, \mnras, 345, 1145

\bibitem[{{Donati} {et~al.}(2023){Donati}, {Cristofari}, {Finociety}, {Klein}, {Moutou}, {Gaidos}, {Cadieux}, {Artigau}, {Correia}, {Bou{\'e}}, {Cook}, {Carmona}, {Lehmann}, {Bouvier}, {Martioli}, {Morin}, {Fouqu{\'e}}, {Delfosse}, {Doyon}, {H{\'e}brard}, {Alencar}, {Laskar}, {Arnold}, {Petit}, {K{\'o}sp{\'a}l}, {Vidotto}, {Folsom}, \& {collaboration}}]{Donati2023}
{Donati}, J.~F., {Cristofari}, P.~I., {Finociety}, B., {et~al.} 2023, \mnras, 525, 455

\bibitem[{{Donati} {et~al.}(2000){Donati}, {Mengel}, {Carter}, {Marsden}, {Collier Cameron}, \& {Wichmann}}]{Donati2000}
{Donati}, J.~F., {Mengel}, M., {Carter}, B.~D., {et~al.} 2000, MNRAS, 316, 699

\bibitem[{{Donati} {et~al.}(2008){Donati}, {Moutou}, {Far{\`e}s}, {Bohlender}, {Catala}, {Deleuil}, {Shkolnik}, {Collier Cameron}, {Jardine}, \& {Walker}}]{Donati2008b}
{Donati}, J.~F., {Moutou}, C., {Far{\`e}s}, R., {et~al.} 2008, \mnras, 385, 1179

\bibitem[{{Donati} {et~al.}(1997){Donati}, {Semel}, {Carter}, {Rees}, \& {Collier Cameron}}]{Donati1997}
{Donati}, J.~F., {Semel}, M., {Carter}, B.~D., {Rees}, D.~E., \& {Collier Cameron}, A. 1997, MNRAS, 291, 658

\bibitem[{{Duncan} {et~al.}(1991){Duncan}, {Vaughan}, {Wilson}, {Preston}, {Frazer}, {Lanning}, {Misch}, {Mueller}, {Soyumer}, {Woodard}, {Baliunas}, {Noyes}, {Hartmann}, {Porter}, {Zwaan}, {Middelkoop}, {Rutten}, \& {Mihalas}}]{Duncan1991}
{Duncan}, D.~K., {Vaughan}, A.~H., {Wilson}, O.~C., {et~al.} 1991, \apjs, 76, 383

\bibitem[{{Edmonds}(2024)}]{Edmonds2024}
{Edmonds}, I.~R. 2024, arXiv e-prints, arXiv:2404.13542

\bibitem[{{Egeland}(2017)}]{Egeland2017b}
{Egeland}, R. 2017, PhD thesis, Montana State University, Bozeman

\bibitem[{{Egeland} {et~al.}(2017){Egeland}, {Soon}, {Baliunas}, {Hall}, {Pevtsov}, \& {Bertello}}]{Egeland2017}
{Egeland}, R., {Soon}, W., {Baliunas}, S., {et~al.} 2017, \apj, 835, 25

\bibitem[{{Fares} {et~al.}(2009){Fares}, {Donati}, {Moutou}, {Bohlender}, {Catala}, {Deleuil}, {Shkolnik}, {Collier Cameron}, {Jardine}, \& {Walker}}]{Fares2009}
{Fares}, R., {Donati}, J.~F., {Moutou}, C., {et~al.} 2009, \mnras, 398, 1383

\bibitem[{{Fares} {et~al.}(2013){Fares}, {Moutou}, {Donati}, {Catala}, {Shkolnik}, {Jardine}, {Cameron}, \& {Deleuil}}]{Fares2013}
{Fares}, R., {Moutou}, C., {Donati}, J.~F., {et~al.} 2013, \mnras, 435, 1451

\bibitem[{{Feinstein} {et~al.}(2024){Feinstein}, {Seligman}, {France}, {Gagn{\'e}}, \& {Kowalski}}]{Feinstein2024}
{Feinstein}, A.~D., {Seligman}, D.~Z., {France}, K., {Gagn{\'e}}, J., \& {Kowalski}, A. 2024, \aj, 168, 60

\bibitem[{{Ferreira Lopes} {et~al.}(2015){Ferreira Lopes}, {Le{\~a}o}, {de Freitas}, {Canto Martins}, {Catelan}, \& {De Medeiros}}]{Ferreira-Lopes2015}
{Ferreira Lopes}, C.~E., {Le{\~a}o}, I.~C., {de Freitas}, D.~B., {et~al.} 2015, \aap, 583, A134

\bibitem[{{Finley} \& {Brun}(2023)}]{Finley2023}
{Finley}, A.~J. \& {Brun}, A.~S. 2023, \aap, 679, A29

\bibitem[{{Fletcher} {et~al.}(2010){Fletcher}, {Broomhall}, {Salabert}, {Basu}, {Chaplin}, {Elsworth}, {Garcia}, \& {New}}]{Fletcher2010}
{Fletcher}, S.~T., {Broomhall}, A.-M., {Salabert}, D., {et~al.} 2010, \apjl, 718, L19

\bibitem[{{Folsom} {et~al.}(2018){Folsom}, {Bouvier}, {Petit}, {L{\`e}bre}, {Amard}, {Palacios}, {Morin}, {Donati}, \& {Vidotto}}]{Folsom2018}
{Folsom}, C.~P., {Bouvier}, J., {Petit}, P., {et~al.} 2018, MNRAS, 474, 4956

\bibitem[{{Folsom} {et~al.}(2016){Folsom}, {Petit}, {Bouvier}, {L{\`e}bre}, {Amard}, {Palacios}, {Morin}, {Donati}, {Jeffers}, {Marsden}, \& {Vidotto}}]{Folsom2016}
{Folsom}, C.~P., {Petit}, P., {Bouvier}, J., {et~al.} 2016, MNRAS, 457, 580

\bibitem[{{Fouqu{\'e}} {et~al.}(2023){Fouqu{\'e}}, {Martioli}, {Donati}, {Lehmann}, {Zaire}, {Bellotti}, {Gaidos}, {Morin}, {Moutou}, {Petit}, {Alencar}, {Arnold}, {Artigau}, {Cang}, {Carmona}, {Cook}, {Cort{\'e}s-Zuleta}, {Cristofari}, {Delfosse}, {Doyon}, {H{\'e}brard}, {Malo}, {Reyl{\'e}}, \& {Usher}}]{Fouque2023}
{Fouqu{\'e}}, P., {Martioli}, E., {Donati}, J.~F., {et~al.} 2023, \aap, 672, A52

\bibitem[{{Gaia Collaboration}(2020)}]{GaiaCollaboration2020}
{Gaia Collaboration}. 2020, VizieR Online Data Catalog, I/350

\bibitem[{{Garc{\'\i}a} {et~al.}(2010){Garc{\'\i}a}, {Mathur}, {Salabert}, {Ballot}, {R{\'e}gulo}, {Metcalfe}, \& {Baglin}}]{Garcia2010}
{Garc{\'\i}a}, R.~A., {Mathur}, S., {Salabert}, D., {et~al.} 2010, Science, 329, 1032

\bibitem[{{Gastine} {et~al.}(2014){Gastine}, {Yadav}, {Morin}, {Reiners}, \& {Wicht}}]{Gastine2014}
{Gastine}, T., {Yadav}, R.~K., {Morin}, J., {Reiners}, A., \& {Wicht}, J. 2014, \mnras, 438, L76

\bibitem[{{Ghizaru} {et~al.}(2010){Ghizaru}, {Charbonneau}, \& {Smolarkiewicz}}]{Ghizaru2010}
{Ghizaru}, M., {Charbonneau}, P., \& {Smolarkiewicz}, P.~K. 2010, \apjl, 715, L133

\bibitem[{{Giles} {et~al.}(2017){Giles}, {Collier Cameron}, \& {Haywood}}]{Giles2017}
{Giles}, H. A.~C., {Collier Cameron}, A., \& {Haywood}, R.~D. 2017, MNRAS, 472, 1618

\bibitem[{{Gomes da Silva} {et~al.}(2021){Gomes da Silva}, {Santos}, {Adibekyan}, {Sousa}, {Campante}, {Figueira}, {Bossini}, {Delgado-Mena}, {Monteiro}, {de Laverny}, {Recio-Blanco}, \& {Lovis}}]{GomesdaSilva2021}
{Gomes da Silva}, J., {Santos}, N.~C., {Adibekyan}, V., {et~al.} 2021, \aap, 646, A77

\bibitem[{{G{\"u}del}(2004)}]{Gudel2004}
{G{\"u}del}, M. 2004, \aapr, 12, 71

\bibitem[{{Guenther}(1989)}]{Guenter1989}
{Guenther}, D.~B. 1989, \apj, 339, 1156

\bibitem[{{Hahlin} {et~al.}(2023){Hahlin}, {Kochukhov}, {Rains}, {Lavail}, {Hatzes}, {Piskunov}, {Reiners}, {Seemann}, {Boldt-Christmas}, {Guenther}, {Heiter}, {Nortmann}, {Yan}, {Shulyak}, {Smoker}, {Rodler}, {Bristow}, {Dorn}, {Jung}, {Marquart}, \& {Stempels}}]{Hahlin2023}
{Hahlin}, A., {Kochukhov}, O., {Rains}, A.~D., {et~al.} 2023, \aap, 675, A91

\bibitem[{{Hale} {et~al.}(1919){Hale}, {Ellerman}, {Nicholson}, \& {Joy}}]{Hale1919}
{Hale}, G.~E., {Ellerman}, F., {Nicholson}, S.~B., \& {Joy}, A.~H. 1919, ApJ, 49, 153

\bibitem[{{Hall}(2008)}]{Hall2008}
{Hall}, J.~C. 2008, Living Reviews in Solar Physics, 5, 2

\bibitem[{{Hall} {et~al.}(2018){Hall}, {Thompson}, {Handley}, \& {Queloz}}]{Hall2018}
{Hall}, R.~D., {Thompson}, S.~J., {Handley}, W., \& {Queloz}, D. 2018, \mnras, 479, 2968

\bibitem[{{Hathaway}(2010)}]{Hathaway2010}
{Hathaway}, D.~H. 2010, Living Reviews in Solar Physics, 7, 1

\bibitem[{{Hathaway}(2015)}]{Hathaway2015}
{Hathaway}, D.~H. 2015, Living Reviews in Solar Physics, 12, 4

\bibitem[{{Haywood} {et~al.}(2014){Haywood}, {Collier Cameron}, {Queloz}, {Barros}, {Deleuil}, {Fares}, {Gillon}, {Lanza}, {Lovis}, {Moutou}, {Pepe}, {Pollacco}, {Santerne}, {S{\'e}gransan}, \& {Unruh}}]{Haywood2014}
{Haywood}, R.~D., {Collier Cameron}, A., {Queloz}, D., {et~al.} 2014, MNRAS, 443, 2517

\bibitem[{{Hazra} {et~al.}(2019){Hazra}, {Jiang}, {Karak}, \& {Kitchatinov}}]{Hazra2019}
{Hazra}, G., {Jiang}, J., {Karak}, B.~B., \& {Kitchatinov}, L. 2019, \apj, 884, 35

\bibitem[{{Hazra} {et~al.}(2020){Hazra}, {Vidotto}, \& {D'Angelo}}]{Hazra2020}
{Hazra}, G., {Vidotto}, A.~A., \& {D'Angelo}, C.~V. 2020, MNRAS, 496, 4017

\bibitem[{{H{\'e}brard} {et~al.}(2016){H{\'e}brard}, {Donati}, {Delfosse}, {Morin}, {Moutou}, \& {Boisse}}]{Hebrard2016}
{H{\'e}brard}, {\'E}.~M., {Donati}, J.~F., {Delfosse}, X., {et~al.} 2016, MNRAS, 461, 1465

\bibitem[{{Hempelmann} {et~al.}(2006){Hempelmann}, {Robrade}, {Schmitt}, {Favata}, {Baliunas}, \& {Hall}}]{Hempelmann2006}
{Hempelmann}, A., {Robrade}, J., {Schmitt}, J.~H.~M.~M., {et~al.} 2006, \aap, 460, 261

\bibitem[{{Hunter}(2007)}]{Hunter2007}
{Hunter}, J.~D. 2007, Computing in Science and Engineering, 9, 90

\bibitem[{{I{\c{s}}ik} {et~al.}(2007){I{\c{s}}ik}, {Sch{\"u}ssler}, \& {Solanki}}]{Isik2007}
{I{\c{s}}ik}, E., {Sch{\"u}ssler}, M., \& {Solanki}, S.~K. 2007, \aap, 464, 1049

\bibitem[{{I{\c{s}}{\i}k} {et~al.}(2018){I{\c{s}}{\i}k}, {Solanki}, {Krivova}, \& {Shapiro}}]{Isik2018}
{I{\c{s}}{\i}k}, E., {Solanki}, S.~K., {Krivova}, N.~A., \& {Shapiro}, A.~I. 2018, \aap, 620, A177

\bibitem[{{Isaacson} \& {Fischer}(2010)}]{Isaacson2010}
{Isaacson}, H. \& {Fischer}, D. 2010, \apj, 725, 875

\bibitem[{{Isaacson} {et~al.}(2024){Isaacson}, {Howard}, {Fulton}, {Petigura}, {Weiss}, {Kane}, {Carter}, {Beard}, {Giacalone}, {Van Zandt}, {Murphy}, {Dai}, {Chontos}, {Polanski}, {Rice}, {Lubin}, {Brinkman}, {Rubenzahl}, {Blunt}, {Yee}, {MacDougall}, {Dalba}, {Tyler}, {Behmard}, {Angelo}, {Pidhorodetska}, {Mayo}, {Holcomb}, {Turtelboom}, {Hill}, {Bouma}, {Zhang}, {Crossfield}, \& {Saunders}}]{Isaacson2024}
{Isaacson}, H., {Howard}, A.~W., {Fulton}, B., {et~al.} 2024, \apjs, 274, 35

\bibitem[{{Jeffers} {et~al.}(2022){Jeffers}, {Cameron}, {Marsden}, {Boro Saikia}, {Folsom}, {Jardine}, {Morin}, {Petit}, {See}, {Vidotto}, {Wolter}, \& {Mittag}}]{Jeffers2022}
{Jeffers}, S.~V., {Cameron}, R.~H., {Marsden}, S.~C., {et~al.} 2022, A\&A, 661, A152

\bibitem[{{Jeffers} \& {Donati}(2008)}]{JeffersDonati2008}
{Jeffers}, S.~V. \& {Donati}, J.~F. 2008, \mnras, 390, 635

\bibitem[{{Jeffers} {et~al.}(2011){Jeffers}, {Donati}, {Alecian}, \& {Marsden}}]{Jeffers2011}
{Jeffers}, S.~V., {Donati}, J.~F., {Alecian}, E., \& {Marsden}, S.~C. 2011, \mnras, 411, 1301

\bibitem[{{Jeffers} {et~al.}(2023){Jeffers}, {Kiefer}, \& {Metcalfe}}]{Jeffers2023}
{Jeffers}, S.~V., {Kiefer}, R., \& {Metcalfe}, T.~S. 2023, \ssr, 219, 54

\bibitem[{{Jeffers} {et~al.}(2018){Jeffers}, {Mengel}, {Moutou}, {Marsden}, {Barnes}, {Jardine}, {Petit}, {Schmitt}, {See}, {Vidotto}, \& {BCool Collaboration}}]{Jeffers2018}
{Jeffers}, S.~V., {Mengel}, M., {Moutou}, C., {et~al.} 2018, MNRAS, 479, 5266

\bibitem[{{Jouve} {et~al.}(2010){Jouve}, {Brown}, \& {Brun}}]{Jouve2010}
{Jouve}, L., {Brown}, B.~P., \& {Brun}, A.~S. 2010, A\&A, 509, A32

\bibitem[{{K{\"a}pyl{\"a}} {et~al.}(2016){K{\"a}pyl{\"a}}, {K{\"a}pyl{\"a}}, {Olspert}, {Brandenburg}, {Warnecke}, {Karak}, \& {Pelt}}]{Kapyla2016}
{K{\"a}pyl{\"a}}, M.~J., {K{\"a}pyl{\"a}}, P.~J., {Olspert}, N., {et~al.} 2016, A\&A, 589, A56

\bibitem[{{K{\"a}pyl{\"a}} {et~al.}(2023){K{\"a}pyl{\"a}}, {Browning}, {Brun}, {Guerrero}, \& {Warnecke}}]{Kapyla2023}
{K{\"a}pyl{\"a}}, P.~J., {Browning}, M.~K., {Brun}, A.~S., {Guerrero}, G., \& {Warnecke}, J. 2023, \ssr, 219, 58

\bibitem[{{K{\"a}pyl{\"a}} {et~al.}(2012){K{\"a}pyl{\"a}}, {Mantere}, \& {Brandenburg}}]{Kapyla2012}
{K{\"a}pyl{\"a}}, P.~J., {Mantere}, M.~J., \& {Brandenburg}, A. 2012, \apjl, 755, L22

\bibitem[{{Karak} {et~al.}(2014){Karak}, {Kitchatinov}, \& {Choudhuri}}]{Karak2014}
{Karak}, B.~B., {Kitchatinov}, L.~L., \& {Choudhuri}, A.~R. 2014, \apj, 791, 59

\bibitem[{{Karoff} {et~al.}(2009){Karoff}, {Metcalfe}, {Chaplin}, {Elsworth}, {Kjeldsen}, {Arentoft}, \& {Buzasi}}]{Karoff2009}
{Karoff}, C., {Metcalfe}, T.~S., {Chaplin}, W.~J., {et~al.} 2009, \mnras, 399, 914

\bibitem[{{Kavanagh} {et~al.}(2021){Kavanagh}, {Vidotto}, {Klein}, {Jardine}, {Donati}, \& {{\'O} Fionnag{\'a}in}}]{Kavanagh2021}
{Kavanagh}, R.~D., {Vidotto}, A.~A., {Klein}, B., {et~al.} 2021, MNRAS, 504, 1511

\bibitem[{{Kiefer} {et~al.}(2018){Kiefer}, {Komm}, {Hill}, {Broomhall}, \& {Roth}}]{Kiefer2018}
{Kiefer}, R., {Komm}, R., {Hill}, F., {Broomhall}, A.-M., \& {Roth}, M. 2018, \solphys, 293, 151

\bibitem[{{Klein} {et~al.}(2021){Klein}, {Donati}, {Moutou}, {Delfosse}, {Bonfils}, {Martioli}, {Fouqu{\'e}}, {Cloutier}, {Artigau}, {Doyon}, {H{\'e}brard}, {Morin}, {Rameau}, {Plavchan}, \& {Gaidos}}]{Klein2021}
{Klein}, B., {Donati}, J.-F., {Moutou}, C., {et~al.} 2021, MNRAS, 502, 188

\bibitem[{{Kochukhov} {et~al.}(2020){Kochukhov}, {Hackman}, {Lehtinen}, \& {Wehrhahn}}]{Kochukhov2020}
{Kochukhov}, O., {Hackman}, T., {Lehtinen}, J.~J., \& {Wehrhahn}, A. 2020, \aap, 635, A142

\bibitem[{{Kochukhov} {et~al.}(2010){Kochukhov}, {Makaganiuk}, \& {Piskunov}}]{Kochukhov2010a}
{Kochukhov}, O., {Makaganiuk}, V., \& {Piskunov}, N. 2010, A\&A, 524, A5

\bibitem[{{Landi Degl'Innocenti}(1992)}]{Landi1992}
{Landi Degl'Innocenti}, E. 1992, {Magnetic field measurements.}, ed. F.~{Sanchez}, M.~{Collados}, \& M.~{Vazquez}, 71

\bibitem[{{Lehmann} \& {Donati}(2022)}]{Lehmann2022}
{Lehmann}, L.~T. \& {Donati}, J.~F. 2022, MNRAS, 514, 2333

\bibitem[{{Lehmann} {et~al.}(2024){Lehmann}, {Donati}, {Fouqu{\'e}}, {Moutou}, {Bellotti}, {Delfosse}, {Petit}, {Carmona}, {Morin}, {Vidotto}, \& {the SLS consortium}}]{Lehmann2024}
{Lehmann}, L.~T., {Donati}, J.~F., {Fouqu{\'e}}, P., {et~al.} 2024, \mnras, 527, 4330

\bibitem[{{Lehmann} {et~al.}(2019){Lehmann}, {Hussain}, {Jardine}, {Mackay}, \& {Vidotto}}]{Lehmann2019}
{Lehmann}, L.~T., {Hussain}, G.~A.~J., {Jardine}, M.~M., {Mackay}, D.~H., \& {Vidotto}, A.~A. 2019, \mnras, 483, 5246

\bibitem[{{Lehmann} {et~al.}(2021){Lehmann}, {Hussain}, {Vidotto}, {Jardine}, \& {Mackay}}]{Lehmann2021}
{Lehmann}, L.~T., {Hussain}, G.~A.~J., {Vidotto}, A.~A., {Jardine}, M.~M., \& {Mackay}, D.~H. 2021, MNRAS, 500, 1243

\bibitem[{{Lehtinen} {et~al.}(2016){Lehtinen}, {Jetsu}, {Hackman}, {Kajatkari}, \& {Henry}}]{Lehtinen2016}
{Lehtinen}, J., {Jetsu}, L., {Hackman}, T., {Kajatkari}, P., \& {Henry}, G.~W. 2016, \aap, 588, A38

\bibitem[{{Leighton}(1959)}]{Leighton1959}
{Leighton}, R.~B. 1959, \apj, 130, 366

\bibitem[{{Leighton}(1969)}]{Leighton1969}
{Leighton}, R.~B. 1969, ApJ, 156, 1

\bibitem[{{L{\'o}pez Ariste} {et~al.}(2022){L{\'o}pez Ariste}, {Georgiev}, {Mathias}, {L{\`e}bre}, {Wavasseur}, {Josselin}, {Konstantinova-Antova}, \& {Roudier}}]{LopezAriste2022}
{L{\'o}pez Ariste}, A., {Georgiev}, S., {Mathias}, P., {et~al.} 2022, \aap, 661, A91

\bibitem[{{Lorenzo-Oliveira} {et~al.}(2019){Lorenzo-Oliveira}, {Mel{\'e}ndez}, {Yana Galarza}, {Ponte}, {dos Santos}, {Spina}, {Bedell}, {Ram{\'\i}rez}, {Bean}, \& {Asplund}}]{Lorenzo-Oliveira2019}
{Lorenzo-Oliveira}, D., {Mel{\'e}ndez}, J., {Yana Galarza}, J., {et~al.} 2019, \mnras, 485, L68

\bibitem[{{Mackay} {et~al.}(2004){Mackay}, {Jardine}, {Collier Cameron}, {Donati}, \& {Hussain}}]{Mackay2004}
{Mackay}, D.~H., {Jardine}, M., {Collier Cameron}, A., {Donati}, J.~F., \& {Hussain}, G.~A.~J. 2004, \mnras, 354, 737

\bibitem[{{Marsden} {et~al.}(2006){Marsden}, {Donati}, {Semel}, {Petit}, \& {Carter}}]{Marsden2006}
{Marsden}, S.~C., {Donati}, J.~F., {Semel}, M., {Petit}, P., \& {Carter}, B.~D. 2006, \mnras, 370, 468

\bibitem[{{Marsden} {et~al.}(2023){Marsden}, {Evensberget}, {Brown}, {Neiner}, {Seach}, {Morin}, {Petit}, {Jeffers}, \& {Folsom}}]{Marsden2023}
{Marsden}, S.~C., {Evensberget}, D., {Brown}, E.~L., {et~al.} 2023, \mnras, 522, 792

\bibitem[{{Marsden} {et~al.}(2014){Marsden}, {Petit}, {Jeffers}, {Morin}, {Fares}, {Reiners}, {do Nascimento}, {Auri{\`e}re}, {Bouvier}, {Carter}, {Catala}, {Dintrans}, {Donati}, {Gastine}, {Jardine}, {Konstantinova-Antova}, {Lanoux}, {Ligni{\`e}res}, {Morgenthaler}, {Ram{\`\i}rez-V{\`e}lez}, {Th{\'e}ado}, {Van Grootel}, \& {BCool Collaboration}}]{Marsden2014}
{Marsden}, S.~C., {Petit}, P., {Jeffers}, S.~V., {et~al.} 2014, MNRAS, 444, 3517

\bibitem[{{Mathias} {et~al.}(2018){Mathias}, {Auri{\`e}re}, {L{\'o}pez Ariste}, {Petit}, {Tessore}, {Josselin}, {L{\`e}bre}, {Morin}, {Wade}, {Herpin}, {Chiavassa}, {Montarg{\`e}s}, {Konstantinova-Antova}, {Kervella}, {Perrin}, {Donati}, \& {Grunhut}}]{Mathias2018}
{Mathias}, P., {Auri{\`e}re}, M., {L{\'o}pez Ariste}, A., {et~al.} 2018, \aap, 615, A116

\bibitem[{{Mathur} {et~al.}(2013){Mathur}, {Garc{\'\i}a}, {Morgenthaler}, {Salabert}, {Petit}, {Ballot}, {R{\'e}gulo}, \& {Catala}}]{Mathur2013}
{Mathur}, S., {Garc{\'\i}a}, R.~A., {Morgenthaler}, A., {et~al.} 2013, \aap, 550, A32

\bibitem[{{Maunder}(1904)}]{Maunder1904}
{Maunder}, E.~W. 1904, MNRAS, 64, 747

\bibitem[{{Mendoza} {et~al.}(2006){Mendoza}, {Velasco}, \& {Vald{\'e}s-Galicia}}]{Mendoza2006}
{Mendoza}, B., {Velasco}, V.~M., \& {Vald{\'e}s-Galicia}, J.~F. 2006, \solphys, 233, 319

\bibitem[{{Mengel} {et~al.}(2016){Mengel}, {Fares}, {Marsden}, {Carter}, {Jeffers}, {Petit}, {Donati}, {Folsom}, \& {BCool Collaboration}}]{Mengel2016}
{Mengel}, M.~W., {Fares}, R., {Marsden}, S.~C., {et~al.} 2016, \mnras, 459, 4325

\bibitem[{{Metcalfe} {et~al.}(2013){Metcalfe}, {Buccino}, {Brown}, {Mathur}, {Soderblom}, {Henry}, {Mauas}, {Petrucci}, {Hall}, \& {Basu}}]{Metcalfe2013}
{Metcalfe}, T.~S., {Buccino}, A.~P., {Brown}, B.~P., {et~al.} 2013, \apjl, 763, L26

\bibitem[{{Morgenthaler} {et~al.}(2011){Morgenthaler}, {Petit}, {Morin}, {Auri{\`e}re}, {Dintrans}, {Konstantinova-Antova}, \& {Marsden}}]{Morgenthaler2011}
{Morgenthaler}, A., {Petit}, P., {Morin}, J., {et~al.} 2011, Astronomische Nachrichten, 332, 866

\bibitem[{{Morgenthaler} {et~al.}(2012){Morgenthaler}, {Petit}, {Saar}, {Solanki}, {Morin}, {Marsden}, {Auri{\`e}re}, {Dintrans}, {Fares}, {Gastine}, {Lanoux}, {Ligni{\`e}res}, {Paletou}, {Ram{\'\i}rez V{\'e}lez}, {Th{\'e}ado}, \& {Van Grootel}}]{Morgenthaler2012}
{Morgenthaler}, A., {Petit}, P., {Saar}, S., {et~al.} 2012, A\&A, 540, A138

\bibitem[{{Mosser} {et~al.}(2009){Mosser}, {Michel}, {Appourchaux}, {Barban}, {Baudin}, {Boumier}, {Bruntt}, {Catala}, {Deheuvels}, {Garc{\'\i}a}, {Gaulme}, {Regulo}, {Roxburgh}, {Samadi}, {Verner}, {Auvergne}, {Baglin}, {Ballot}, {Benomar}, \& {Mathur}}]{Mosser2009}
{Mosser}, B., {Michel}, E., {Appourchaux}, T., {et~al.} 2009, \aap, 506, 33

\bibitem[{{Nicholson} \& {Aigrain}(2022)}]{Nicholson2022}
{Nicholson}, B.~A. \& {Aigrain}, S. 2022, MNRAS, 515, 5251

\bibitem[{{Noraz} {et~al.}(2024){Noraz}, {Brun}, \& {Strugarek}}]{Noraz2024}
{Noraz}, Q., {Brun}, A.~S., \& {Strugarek}, A. 2024, \aap, 684, A156

\bibitem[{{Noyes} {et~al.}(1984){Noyes}, {Hartmann}, {Baliunas}, {Duncan}, \& {Vaughan}}]{Noyes1984}
{Noyes}, R.~W., {Hartmann}, L.~W., {Baliunas}, S.~L., {Duncan}, D.~K., \& {Vaughan}, A.~H. 1984, ApJ, 279, 763

\bibitem[{{Ol{\'a}h} {et~al.}(2016){Ol{\'a}h}, {K{\H{o}}v{\'a}ri}, {Petrovay}, {Soon}, {Baliunas}, {Koll{\'a}th}, \& {Vida}}]{Olah2016}
{Ol{\'a}h}, K., {K{\H{o}}v{\'a}ri}, Z., {Petrovay}, K., {et~al.} 2016, A\&A, 590, A133

\bibitem[{{Ol{\'a}h} {et~al.}(2009){Ol{\'a}h}, {Koll{\'a}th}, {Granzer}, {Strassmeier}, {Lanza}, {J{\"a}rvinen}, {Korhonen}, {Baliunas}, {Soon}, {Messina}, \& {Cutispoto}}]{Olah2009}
{Ol{\'a}h}, K., {Koll{\'a}th}, Z., {Granzer}, T., {et~al.} 2009, \aap, 501, 703

\bibitem[{{Olspert} {et~al.}(2018){Olspert}, {Lehtinen}, {K{\"a}pyl{\"a}}, {Pelt}, \& {Grigorievskiy}}]{Olspert2018}
{Olspert}, N., {Lehtinen}, J.~J., {K{\"a}pyl{\"a}}, M.~J., {Pelt}, J., \& {Grigorievskiy}, A. 2018, \aap, 619, A6

\bibitem[{{{\"O}zdarcan} {et~al.}(2010){{\"O}zdarcan}, {Evren}, {Strassmeier}, {Granzer}, \& {Henry}}]{Ozdarcan2010}
{{\"O}zdarcan}, O., {Evren}, S., {Strassmeier}, K.~G., {Granzer}, T., \& {Henry}, G.~W. 2010, Astronomische Nachrichten, 331, 794

\bibitem[{{Pace}(2013)}]{Pace2013}
{Pace}, G. 2013, \aap, 551, L8

\bibitem[{{Parker}(1955)}]{Parker1955}
{Parker}, E.~N. 1955, ApJ, 122, 293

\bibitem[{{Petit} {et~al.}(2009){Petit}, {Dintrans}, {Morgenthaler}, {Van Grootel}, {Morin}, {Lanoux}, {Auri{\`e}re}, \& {Konstantinova-Antova}}]{Petit2009}
{Petit}, P., {Dintrans}, B., {Morgenthaler}, A., {et~al.} 2009, A\&A, 508, L9

\bibitem[{{Petit} {et~al.}(2008){Petit}, {Dintrans}, {Solanki}, {Donati}, {Auri{\`e}re}, {Ligni{\`e}res}, {Morin}, {Paletou}, {Ramirez Velez}, {Catala}, \& {Fares}}]{Petit2008}
{Petit}, P., {Dintrans}, B., {Solanki}, S.~K., {et~al.} 2008, \mnras, 388, 80

\bibitem[{{Petit} {et~al.}(2002){Petit}, {Donati}, \& {Collier Cameron}}]{Petit2002}
{Petit}, P., {Donati}, J.~F., \& {Collier Cameron}, A. 2002, MNRAS, 334, 374

\bibitem[{{Petit} {et~al.}(2021){Petit}, {Folsom}, {Donati}, {Yu}, {do Nascimento}, {Jeffers}, {Marsden}, {Morin}, \& {Vidotto}}]{Petit2021}
{Petit}, P., {Folsom}, C.~P., {Donati}, J.~F., {et~al.} 2021, A\&A, 648, A55

\bibitem[{{Petit} {et~al.}(2014){Petit}, {Louge}, {Th{\'e}ado}, {Paletou}, {Manset}, {Morin}, {Marsden}, \& {Jeffers}}]{Petit2014}
{Petit}, P., {Louge}, T., {Th{\'e}ado}, S., {et~al.} 2014, PASP, 126, 469

\bibitem[{{Porto de Mello} {et~al.}(2014){Porto de Mello}, {da Silva}, {da Silva}, \& {de Nader}}]{PortodeMello2014}
{Porto de Mello}, G.~F., {da Silva}, R., {da Silva}, L., \& {de Nader}, R.~V. 2014, \aap, 563, A52

\bibitem[{{Press} {et~al.}(1992){Press}, {Teukolsky}, {Vetterling}, \& {Flannery}}]{Press1992}
{Press}, W.~H., {Teukolsky}, S.~A., {Vetterling}, W.~T., \& {Flannery}, B.~P. 1992, {Numerical recipes in FORTRAN. The art of scientific computing}

\bibitem[{{Queloz} {et~al.}(2001){Queloz}, {Henry}, {Sivan}, {Baliunas}, {Beuzit}, {Donahue}, {Mayor}, {Naef}, {Perrier}, \& {Udry}}]{Queloz2001}
{Queloz}, D., {Henry}, G.~W., {Sivan}, J.~P., {et~al.} 2001, A\&A, 379, 279

\bibitem[{{Radick} {et~al.}(2018){Radick}, {Lockwood}, {Henry}, {Hall}, \& {Pevtsov}}]{Radick2018}
{Radick}, R.~R., {Lockwood}, G.~W., {Henry}, G.~W., {Hall}, J.~C., \& {Pevtsov}, A.~A. 2018, \apj, 855, 75

\bibitem[{{Rauer} {et~al.}(2014){Rauer}, {Catala}, {Aerts}, {Appourchaux}, {Benz}, {Brandeker}, {Christensen-Dalsgaard}, {Deleuil}, {Gizon}, {Goupil}, {G{\"u}del}, {Janot-Pacheco}, {Mas-Hesse}, {Pagano}, {Piotto}, {Pollacco}, {Santos}, {Smith}, {Su{\'a}rez}, {Szab{\'o}}, {Udry}, {Adibekyan}, {Alibert}, {Almenara}, {Amaro-Seoane}, {Eiff}, {Asplund}, {Antonello}, {Barnes}, {Baudin}, {Belkacem}, {Bergemann}, {Bihain}, {Birch}, {Bonfils}, {Boisse}, {Bonomo}, {Borsa}, {Brand{\~a}o}, {Brocato}, {Brun}, {Burleigh}, {Burston}, {Cabrera}, {Cassisi}, {Chaplin}, {Charpinet}, {Chiappini}, {Church}, {Csizmadia}, {Cunha}, {Damasso}, {Davies}, {Deeg}, {D{\'\i}az}, {Dreizler}, {Dreyer}, {Eggenberger}, {Ehrenreich}, {Eigm{\"u}ller}, {Erikson}, {Farmer}, {Feltzing}, {de Oliveira Fialho}, {Figueira}, {Forveille}, {Fridlund}, {Garc{\'\i}a}, {Giommi}, {Giuffrida}, {Godolt}, {Gomes da Silva}, {Granzer}, {Grenfell}, {Grotsch-Noels}, {G{\"u}nther}, {Haswell}, {Hatzes}, {H{\'e}brard}, {Hekker}, {Helled}, {Heng}, {Jenkins},
  {Johansen}, {Khodachenko}, {Kislyakova}, {Kley}, {Kolb}, {Krivova}, {Kupka}, {Lammer}, {Lanza}, {Lebreton}, {Magrin}, {Marcos-Arenal}, {Marrese}, {Marques}, {Martins}, {Mathis}, {Mathur}, {Messina}, {Miglio}, {Montalban}, {Montalto}, {Monteiro}, {Moradi}, {Moravveji}, {Mordasini}, {Morel}, {Mortier}, {Nascimbeni}, {Nelson}, {Nielsen}, {Noack}, {Norton}, {Ofir}, {Oshagh}, {Ouazzani}, {P{\'a}pics}, {Parro}, {Petit}, {Plez}, {Poretti}, {Quirrenbach}, {Ragazzoni}, {Raimondo}, {Rainer}, {Reese}, {Redmer}, {Reffert}, {Rojas-Ayala}, {Roxburgh}, {Salmon}, {Santerne}, {Schneider}, {Schou}, {Schuh}, {Schunker}, {Silva-Valio}, {Silvotti}, {Skillen}, {Snellen}, {Sohl}, {Sousa}, {Sozzetti}, {Stello}, {Strassmeier}, {{\v{S}}vanda}, {Szab{\'o}}, {Tkachenko}, {Valencia}, {Van Grootel}, {Vauclair}, {Ventura}, {Wagner}, {Walton}, {Weingrill}, {Werner}, {Wheatley}, \& {Zwintz}}]{Rauer2014}
{Rauer}, H., {Catala}, C., {Aerts}, C., {et~al.} 2014, Experimental Astronomy, 38, 249

\bibitem[{{Rees} \& {Semel}(1979)}]{Rees1979}
{Rees}, D.~E. \& {Semel}, M.~D. 1979, \aap, 74, 1

\bibitem[{{R{\'e}gulo} {et~al.}(2016){R{\'e}gulo}, {Garc{\'\i}a}, \& {Ballot}}]{Regulo2016}
{R{\'e}gulo}, C., {Garc{\'\i}a}, R.~A., \& {Ballot}, J. 2016, \aap, 589, A103

\bibitem[{{Reiners} {et~al.}(2022){Reiners}, {Shulyak}, {K{\"a}pyl{\"a}}, {Ribas}, {Nagel}, {Zechmeister}, {Caballero}, {Shan}, {Fuhrmeister}, {Quirrenbach}, {Amado}, {Montes}, {Jeffers}, {Azzaro}, {B{\'e}jar}, {Chaturvedi}, {Henning}, {K{\"u}rster}, \& {Pall{\'e}}}]{Reiners2022}
{Reiners}, A., {Shulyak}, D., {K{\"a}pyl{\"a}}, P.~J., {et~al.} 2022, \aap, 662, A41

\bibitem[{{Reinhold} {et~al.}(2017){Reinhold}, {Cameron}, \& {Gizon}}]{Reinhold2017}
{Reinhold}, T., {Cameron}, R.~H., \& {Gizon}, L. 2017, \aap, 603, A52

\bibitem[{{Rescigno} {et~al.}(2024){Rescigno}, {Mortier}, {Dumusque}, {Lakeland}, {Haywood}, {Piskunov}, {Nicholson}, {L{\'o}pez-Morales}, {Dalal}, {Cretignier}, {Klein}, {Cameron}, {Ghedina}, {Gonzalez}, {Cosentino}, {Sozzetti}, \& {Saar}}]{Rescigno2024}
{Rescigno}, F., {Mortier}, A., {Dumusque}, X., {et~al.} 2024, \mnras, 532, 2741

\bibitem[{{Richards} {et~al.}(2009){Richards}, {Rogers}, \& {Richards}}]{Richards2009}
{Richards}, M.~T., {Rogers}, M.~L., \& {Richards}, D. S.~P. 2009, \pasp, 121, 797

\bibitem[{{Ricker} {et~al.}(2015){Ricker}, {Winn}, {Vanderspek}, {Latham}, {Bakos}, {Bean}, {Berta-Thompson}, {Brown}, {Buchhave}, {Butler}, {Butler}, {Chaplin}, {Charbonneau}, {Christensen-Dalsgaard}, {Clampin}, {Deming}, {Doty}, {De Lee}, {Dressing}, {Dunham}, {Endl}, {Fressin}, {Ge}, {Henning}, {Holman}, {Howard}, {Ida}, {Jenkins}, {Jernigan}, {Johnson}, {Kaltenegger}, {Kawai}, {Kjeldsen}, {Laughlin}, {Levine}, {Lin}, {Lissauer}, {MacQueen}, {Marcy}, {McCullough}, {Morton}, {Narita}, {Paegert}, {Palle}, {Pepe}, {Pepper}, {Quirrenbach}, {Rinehart}, {Sasselov}, {Sato}, {Seager}, {Sozzetti}, {Stassun}, {Sullivan}, {Szentgyorgyi}, {Torres}, {Udry}, \& {Villasenor}}]{Ricker2014}
{Ricker}, G.~R., {Winn}, J.~N., {Vanderspek}, R., {et~al.} 2015, Journal of Astronomical Telescopes, Instruments, and Systems, 1, 014003

\bibitem[{{Rieger} {et~al.}(1984){Rieger}, {Share}, {Forrest}, {Kanbach}, {Reppin}, \& {Chupp}}]{Rieger1984}
{Rieger}, E., {Share}, G.~H., {Forrest}, D.~J., {et~al.} 1984, \nat, 312, 623

\bibitem[{{Robinson} {et~al.}(1980){Robinson}, {Worden}, \& {Harvey}}]{Robinson1980}
{Robinson}, R.~D., {Worden}, S.~P., \& {Harvey}, J.~W. 1980, \apjl, 236, L155

\bibitem[{{Robrade} {et~al.}(2012){Robrade}, {Schmitt}, \& {Favata}}]{Robrade2012}
{Robrade}, J., {Schmitt}, J.~H.~M.~M., \& {Favata}, F. 2012, \aap, 543, A84

\bibitem[{{Rodgers-Lee} {et~al.}(2023){Rodgers-Lee}, {Rimmer}, {Vidotto}, {Louca}, {Taylor}, {Mesquita}, {Miguel}, {Venot}, {Helling}, {Barth}, \& {Lacy}}]{RodgersLee2023}
{Rodgers-Lee}, D., {Rimmer}, P.~B., {Vidotto}, A.~A., {et~al.} 2023, \mnras, 521, 5880

\bibitem[{{Ros{\'e}n} {et~al.}(2016){Ros{\'e}n}, {Kochukhov}, {Hackman}, \& {Lehtinen}}]{Rosen2016}
{Ros{\'e}n}, L., {Kochukhov}, O., {Hackman}, T., \& {Lehtinen}, J. 2016, \aap, 593, A35

\bibitem[{{Route}(2016)}]{Route2016}
{Route}, M. 2016, ApJl, 830, L27

\bibitem[{{Ryabchikova} {et~al.}(2015){Ryabchikova}, {Piskunov}, {Kurucz}, {Stempels}, {Heiter}, {Pakhomov}, \& {Barklem}}]{Ryabchikova2015}
{Ryabchikova}, T., {Piskunov}, N., {Kurucz}, R.~L., {et~al.} 2015, Phys. Scr., 90, 054005

\bibitem[{{Sanderson} {et~al.}(2003){Sanderson}, {Appourchaux}, {Hoeksema}, \& {Harvey}}]{Sanderson2003}
{Sanderson}, T.~R., {Appourchaux}, T., {Hoeksema}, J.~T., \& {Harvey}, K.~L. 2003, Journal of Geophysical Research (Space Physics), 108, 1035

\bibitem[{{Sanz-Forcada} {et~al.}(2013){Sanz-Forcada}, {Stelzer}, \& {Metcalfe}}]{SanzForcada2013}
{Sanz-Forcada}, J., {Stelzer}, B., \& {Metcalfe}, T.~S. 2013, \aap, 553, L6

\bibitem[{{Schuessler} \& {Solanki}(1992)}]{Schuessler1992}
{Schuessler}, M. \& {Solanki}, S.~K. 1992, A\&A, 264, L13

\bibitem[{{Sch{\"u}ssler} \& {Ferriz-Mas}(2003)}]{Schussler2003}
{Sch{\"u}ssler}, M. \& {Ferriz-Mas}, A. 2003, in Advances in Nonlinear Dynamics, ed. A.~{Ferriz-Mas} \& M.~{N{\'u}{\~n}ez}, 123

\bibitem[{{Schwabe}(1844)}]{Schwabe1844}
{Schwabe}, H. 1844, Astronomische Nachrichten, 21, 233

\bibitem[{{See} {et~al.}(2019){See}, {Matt}, {Folsom}, {Boro Saikia}, {Donati}, {Fares}, {Finley}, {H{\'e}brard}, {Jardine}, {Jeffers}, {Lehmann}, {Marsden}, {Mengel}, {Morin}, {Petit}, {Vidotto}, {Waite}, \& {BCool Collaboration}}]{See2019}
{See}, V., {Matt}, S.~P., {Folsom}, C.~P., {et~al.} 2019, \apj, 876, 118

\bibitem[{{Semel}(1989)}]{Semel1989}
{Semel}, M. 1989, A\&A, 225, 456

\bibitem[{{Skilling}(2004)}]{Skilling2004}
{Skilling}, J. 2004, in American Institute of Physics Conference Series, Vol. 735, Bayesian Inference and Maximum Entropy Methods in Science and Engineering: 24th International Workshop on Bayesian Inference and Maximum Entropy Methods in Science and Engineering, ed. R.~{Fischer}, R.~{Preuss}, \& U.~V. {Toussaint}, 395--405

\bibitem[{{Skilling} \& {Bryan}(1984)}]{Skilling1984}
{Skilling}, J. \& {Bryan}, R.~K. 1984, MNRAS, 211, 111

\bibitem[{{Spiegel} \& {Zahn}(1992)}]{Spiegel1992}
{Spiegel}, E.~A. \& {Zahn}, J.~P. 1992, \aap, 265, 106

\bibitem[{{Strassmeier}(2009)}]{Strassmeier2009}
{Strassmeier}, K.~G. 2009, A\&Ar, 17, 251

\bibitem[{{Strugarek} {et~al.}(2018){Strugarek}, {Beaudoin}, {Charbonneau}, \& {Brun}}]{Strugarek2018}
{Strugarek}, A., {Beaudoin}, P., {Charbonneau}, P., \& {Brun}, A.~S. 2018, ApJ, 863, 35

\bibitem[{{Strugarek} {et~al.}(2017){Strugarek}, {Beaudoin}, {Charbonneau}, {Brun}, \& {do Nascimento}}]{Strugarek2017}
{Strugarek}, A., {Beaudoin}, P., {Charbonneau}, P., {Brun}, A.~S., \& {do Nascimento}, J.~D. 2017, Science, 357, 185

\bibitem[{{Su{\'a}rez Mascare{\~n}o} {et~al.}(2016){Su{\'a}rez Mascare{\~n}o}, {Rebolo}, \& {Gonz{\'a}lez Hern{\'a}ndez}}]{SuarezMascareno2016}
{Su{\'a}rez Mascare{\~n}o}, A., {Rebolo}, R., \& {Gonz{\'a}lez Hern{\'a}ndez}, J.~I. 2016, A\&A, 595, A12

\bibitem[{{Tessore} {et~al.}(2017){Tessore}, {L{\`e}bre}, {Morin}, {Mathias}, {Josselin}, \& {Auri{\`e}re}}]{Tessore2017}
{Tessore}, B., {L{\`e}bre}, A., {Morin}, J., {et~al.} 2017, \aap, 603, A129

\bibitem[{{Thompson} {et~al.}(2016){Thompson}, {Queloz}, {Baraffe}, {Brake}, {Dolgopolov}, {Fisher}, {Fleury}, {Geelhoed}, {Hall}, {Gonz{\'a}lez Hern{\'a}ndez}, {ter Horst}, {Kragt}, {Navarro}, {Naylor}, {Pepe}, {Piskunov}, {Rebolo}, {Sander}, {S{\'e}gransan}, {Seneta}, {Sing}, {Snellen}, {Snik}, {Spronck}, {Stempels}, {Sun}, {Santana Tschudi}, \& {Young}}]{Thompson2016}
{Thompson}, S.~J., {Queloz}, D., {Baraffe}, I., {et~al.} 2016, in Society of Photo-Optical Instrumentation Engineers (SPIE) Conference Series, Vol. 9908, Ground-based and Airborne Instrumentation for Astronomy VI, ed. C.~J. {Evans}, L.~{Simard}, \& H.~{Takami}, 99086F

\bibitem[{{Usoskin}(2008)}]{Usoskin2008}
{Usoskin}, I.~G. 2008, Living Reviews in Solar Physics, 5, 3

\bibitem[{{van der Walt} {et~al.}(2011){van der Walt}, {Colbert}, \& {Varoquaux}}]{VanderWalt2011}
{van der Walt}, S., {Colbert}, S.~C., \& {Varoquaux}, G. 2011, Computing in Science and Engineering, 13, 22

\bibitem[{VanderPlas(2018)}]{VanderPlas2018}
VanderPlas, J.~T. 2018, The Astrophysical Journal Supplement Series, 236, 16

\bibitem[{{Vashishth} {et~al.}(2023){Vashishth}, {Karak}, \& {Kitchatinov}}]{Vashishth2023}
{Vashishth}, V., {Karak}, B.~B., \& {Kitchatinov}, L. 2023, \mnras, 522, 2601

\bibitem[{{Vaughan} {et~al.}(1978){Vaughan}, {Preston}, \& {Wilson}}]{Vaughan1978}
{Vaughan}, A.~H., {Preston}, G.~W., \& {Wilson}, O.~C. 1978, PASP, 90, 267

\bibitem[{{Velasco Herrera} {et~al.}(2018){Velasco Herrera}, {P{\'e}rez-Peraza}, {Soon}, \& {M{\'a}rquez-Adame}}]{VelascoHerrera2018}
{Velasco Herrera}, V.~M., {P{\'e}rez-Peraza}, J., {Soon}, W., \& {M{\'a}rquez-Adame}, J.~C. 2018, \na, 60, 7

\bibitem[{{Vidotto} {et~al.}(2013){Vidotto}, {Jardine}, {Morin}, {Donati}, {Lang}, \& {Russell}}]{Vidotto2013}
{Vidotto}, A.~A., {Jardine}, M., {Morin}, J., {et~al.} 2013, A\&A, 557, A67

\bibitem[{{Vidotto} {et~al.}(2014){Vidotto}, {Jardine}, {Morin}, {Donati}, {Opher}, \& {Gombosi}}]{Vidotto2014}
{Vidotto}, A.~A., {Jardine}, M., {Morin}, J., {et~al.} 2014, MNRAS, 438, 1162

\bibitem[{{Vidotto} {et~al.}(2018){Vidotto}, {Lehmann}, {Jardine}, \& {Pevtsov}}]{Vidotto2018}
{Vidotto}, A.~A., {Lehmann}, L.~T., {Jardine}, M., \& {Pevtsov}, A.~A. 2018, \mnras, 480, 477

\bibitem[{{Villarreal D'Angelo} {et~al.}(2018){Villarreal D'Angelo}, {Esquivel}, {Schneiter}, \& {Sgr{\'o}}}]{Villareal2018}
{Villarreal D'Angelo}, C., {Esquivel}, A., {Schneiter}, M., \& {Sgr{\'o}}, M.~A. 2018, MNRAS, 479, 3115

\bibitem[{{Virtanen} {et~al.}(2020){Virtanen}, {Gommers}, {Burovski}, {Oliphant}, {Weckesser}, {Cournapeau}, {Alexbrc}, {Peterson}, {Reddy}, {Wilson}, {Haberland}, {Mayorov}, {Endolith}, {Nelson}, {Der Van Walt}, {Laxalde}, {Brett}, {Polat}, {Larson}, {Millman}, {Lars}, {Van Mulbregt}, {Eric-Jones}, {Carey}, {Moore}, {Kern}, {Leslie}, {Perktold}, {Striega}, \& {Feng}}]{Virtanen2020}
{Virtanen}, P., {Gommers}, R., {Burovski}, E., {et~al.} 2020, {scipy/scipy: SciPy 1.5.3}

\bibitem[{{Viviani} {et~al.}(2018){Viviani}, {Warnecke}, {K{\"a}pyl{\"a}}, {K{\"a}pyl{\"a}}, {Olspert}, {Cole-Kodikara}, {Lehtinen}, \& {Brandenburg}}]{Viviani2018}
{Viviani}, M., {Warnecke}, J., {K{\"a}pyl{\"a}}, M.~J., {et~al.} 2018, \aap, 616, A160

\bibitem[{{Wainer} {et~al.}(2024){Wainer}, {Davenport}, {Tovar Mendoza}, {Feinstein}, \& {Wagg}}]{Wainer2024}
{Wainer}, T.~M., {Davenport}, J. R.~A., {Tovar Mendoza}, G., {Feinstein}, A.~D., \& {Wagg}, T. 2024, \aj, 168, 232

\bibitem[{{Waite} {et~al.}(2015){Waite}, {Marsden}, {Carter}, {Petit}, {Donati}, {Jeffers}, \& {Boro Saikia}}]{Waite2015}
{Waite}, I.~A., {Marsden}, S.~C., {Carter}, B.~D., {et~al.} 2015, \mnras, 449, 8

\bibitem[{{Waite} {et~al.}(2017){Waite}, {Marsden}, {Carter}, {Petit}, {Jeffers}, {Morin}, {Vidotto}, {Donati}, \& {BCool Collaboration}}]{Waite2017}
{Waite}, I.~A., {Marsden}, S.~C., {Carter}, B.~D., {et~al.} 2017, \mnras, 465, 2076

\bibitem[{{Wenger} {et~al.}(2000){Wenger}, {Ochsenbein}, {Egret}, {Dubois}, {Bonnarel}, {Borde}, {Genova}, {Jasniewicz}, {Lalo{\"e}}, {Lesteven}, \& {Monier}}]{Wenger2000}
{Wenger}, M., {Ochsenbein}, F., {Egret}, D., {et~al.} 2000, \aaps, 143, 9

\bibitem[{{White} \& {Livingston}(1981)}]{White1981}
{White}, O.~R. \& {Livingston}, W.~C. 1981, \apj, 249, 798

\bibitem[{{Willamo} {et~al.}(2022){Willamo}, {Lehtinen}, {Hackman}, {K{\"a}pyl{\"a}}, {Kochukhov}, {Jeffers}, {Korhonen}, \& {Marsden}}]{Willamo2022}
{Willamo}, T., {Lehtinen}, J.~J., {Hackman}, T., {et~al.} 2022, \aap, 659, A71

\bibitem[{{Wilson}(1968)}]{Wilson1968}
{Wilson}, O.~C. 1968, ApJ, 153, 221

\bibitem[{{Wright} {et~al.}(2004){Wright}, {Marcy}, {Butler}, \& {Vogt}}]{Wright2004}
{Wright}, J.~T., {Marcy}, G.~W., {Butler}, R.~P., \& {Vogt}, S.~S. 2004, \apjs, 152, 261

\bibitem[{{Wright} {et~al.}(2018){Wright}, {Newton}, {Williams}, {Drake}, \& {Yadav}}]{Wright2018}
{Wright}, N.~J., {Newton}, E.~R., {Williams}, P. K.~G., {Drake}, J.~J., \& {Yadav}, R.~K. 2018, MNRAS, 479, 2351

\bibitem[{{Yu} {et~al.}(2019){Yu}, {Donati}, {Grankin}, {Collier Cameron}, {Moutou}, {Hussain}, {Baruteau}, {Jouve}, \& {MaTYSSE Collaboration}}]{Yu2019}
{Yu}, L., {Donati}, J.~F., {Grankin}, K., {et~al.} 2019, MNRAS, 489, 5556

\bibitem[{{Zechmeister} \& {K{\"u}rster}(2009)}]{Zechmeister2009}
{Zechmeister}, M. \& {K{\"u}rster}, M. 2009, A\&A, 496, 577

\end{thebibliography}

\begin{appendix}

\section{Additional figures temporal analysis}\label{app:time_analysis}

In this appendix, we provide complete information on the application of the generalised Lomb-Scargle periodogram to the time series of longitudinal magnetic field data for all stars. In the case of HD\,166435, we also provide the temporal analysis of the $TESS$ light curves collected in 2020, 2021, and 2022. Finally, we include the results of the Gaussian process regression for all stars, showing the GP model overplotted on the B$_l$ time series and the corner plots illustrating the posterior distributions of the GP hyperparameters (see Sec.~\ref{sec:gp} for more details).

%----------- LSP -----------------

\begin{figure}[t]
    \includegraphics[width=\columnwidth]{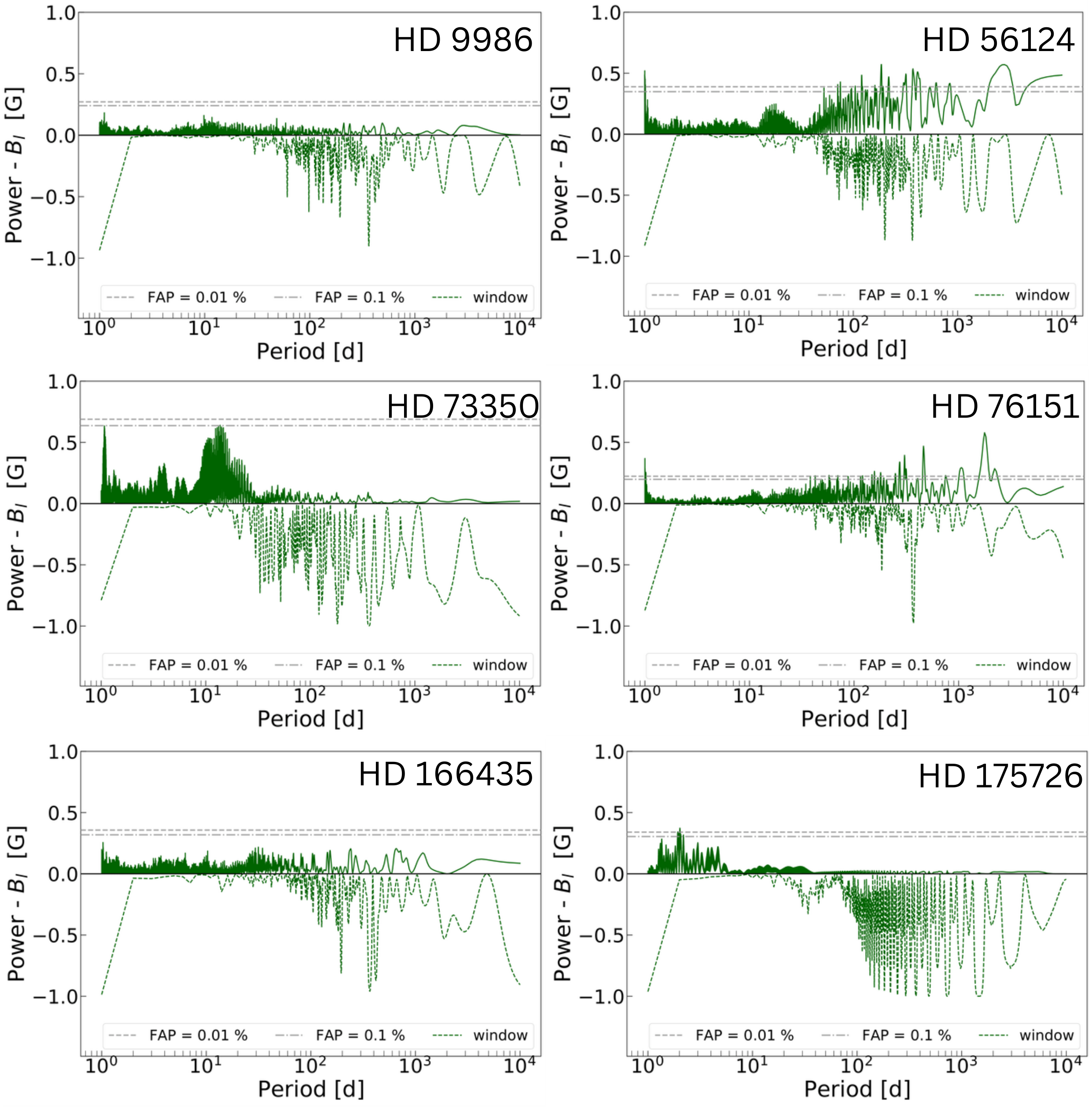}
    \caption{Generalised Lomb-Scargle periodogram analysis for our stars. The power spectrum is shown as a green solid line and the window function as a green dashed line mirrored with respect to the $x$ axis. The horizontal dashed, and dash-dot grey lines represent the 0.01\% and 0.1\% FAP threshold, respectively.}
    \label{fig:LSP_stars}
\end{figure}

%----------- TESS -----------------

\begin{figure}[t]
    \includegraphics[width=\columnwidth]{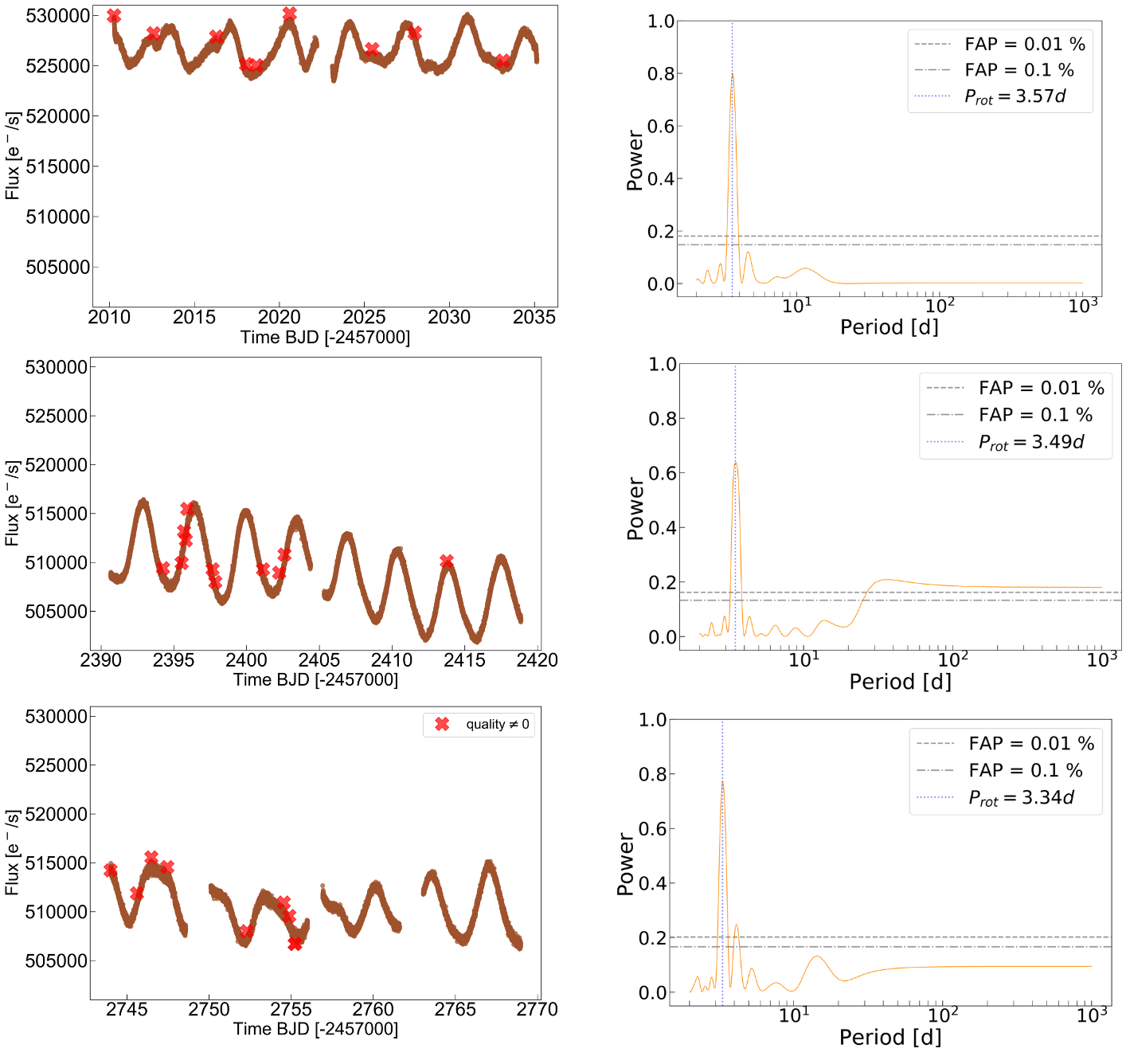}
    \caption{Photometric light curves of HD\,166435 for the 2020, 2021, and 2022 epochs. Left: $TESS$ light curves. The data points with a quality factor flag different than zero are shown as red crosses and removed from the temporal analysis. Right: generalised Lomb-Scargle periodogram applied to the light curves. The format of the periodogram panels is the same as Fig.~\ref{fig:LSP_stars}.}
    \label{fig:LSP_tess}
\end{figure}

%----------- GP -----------------

\begin{figure}[t]
    \includegraphics[width=\columnwidth]{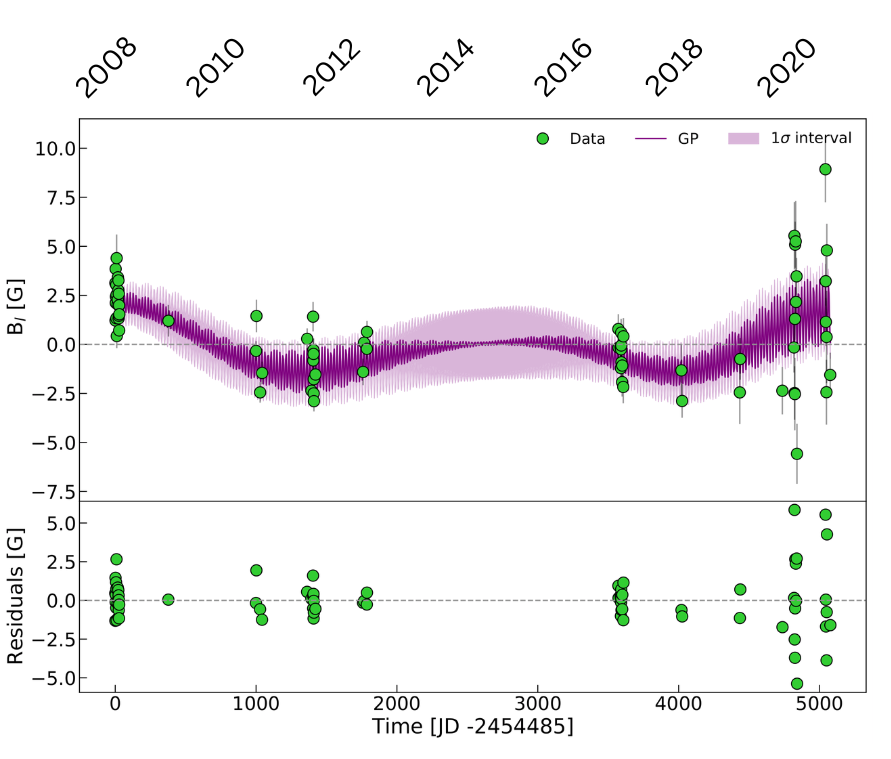}
    \includegraphics[width=\columnwidth]{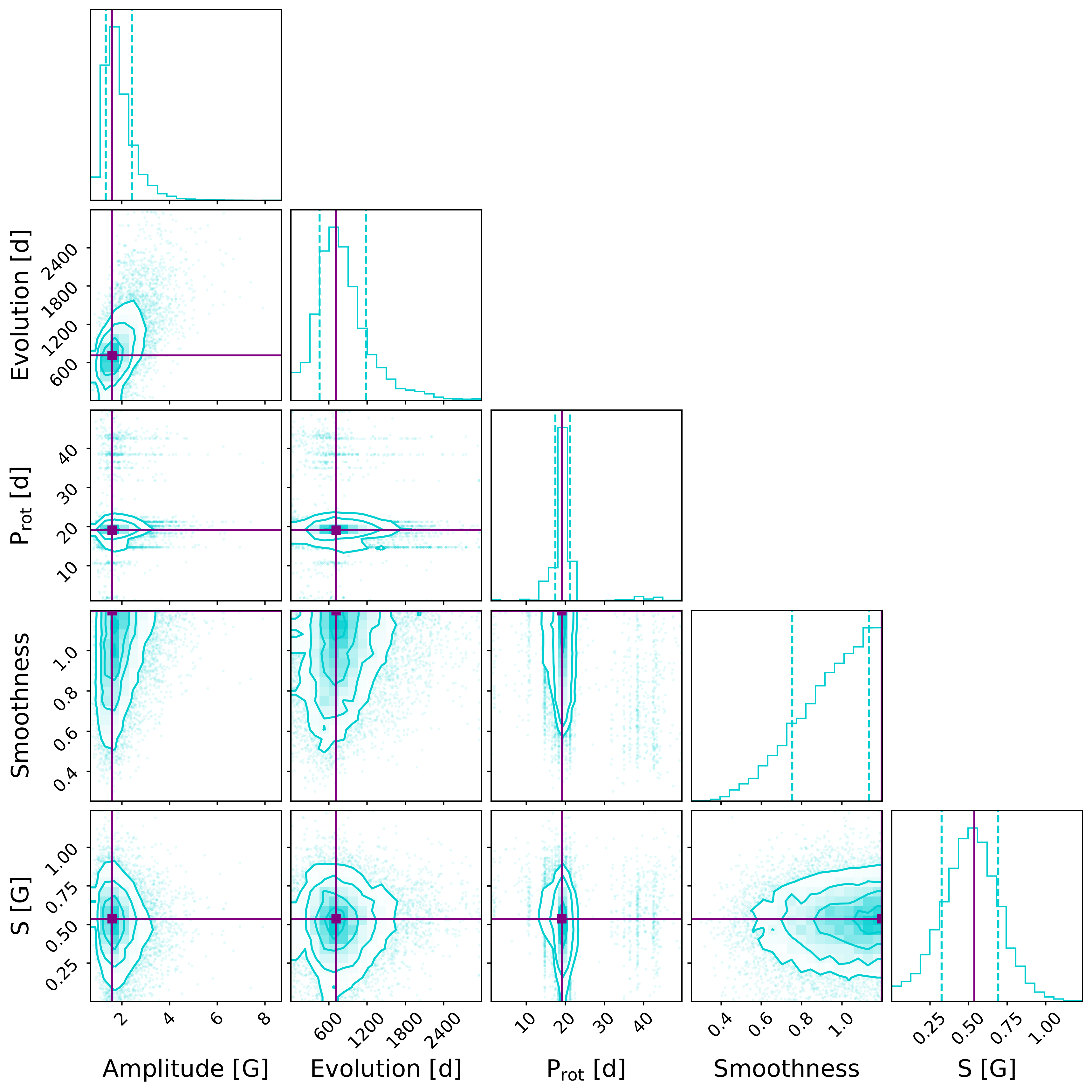}
    \caption{Time series of longitudinal magnetic field measurements for HD\,56124 and posterior distribution of the GP regression. The format is the same as Fig.~\ref{fig:Bl_hd9986}.}
    \label{fig:Bl_hd56124}
\end{figure}

\begin{figure}[t]
    \includegraphics[width=\columnwidth]{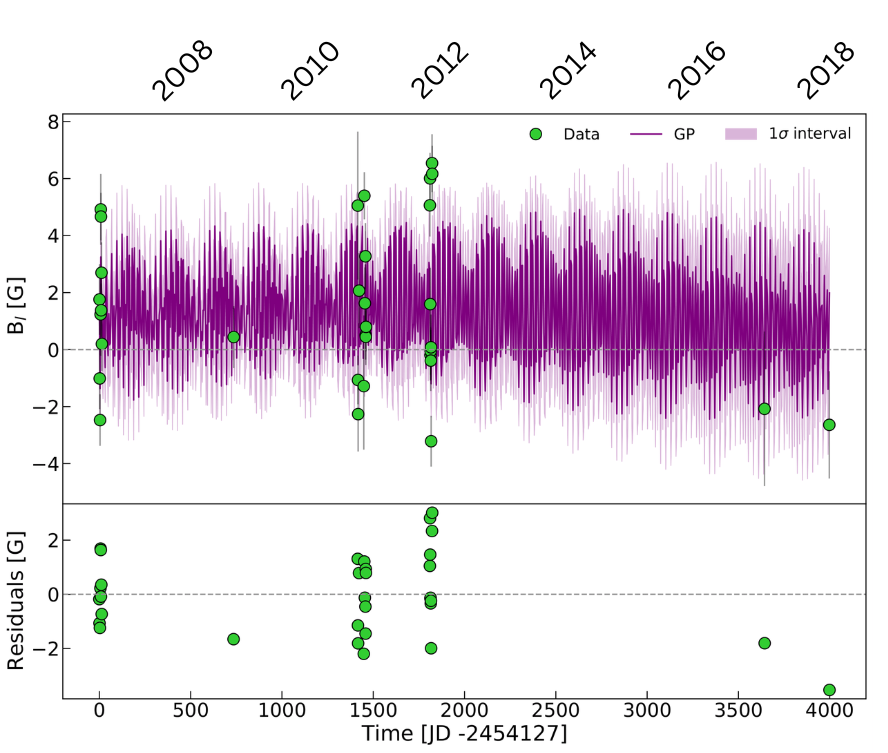}
    \includegraphics[width=\columnwidth]{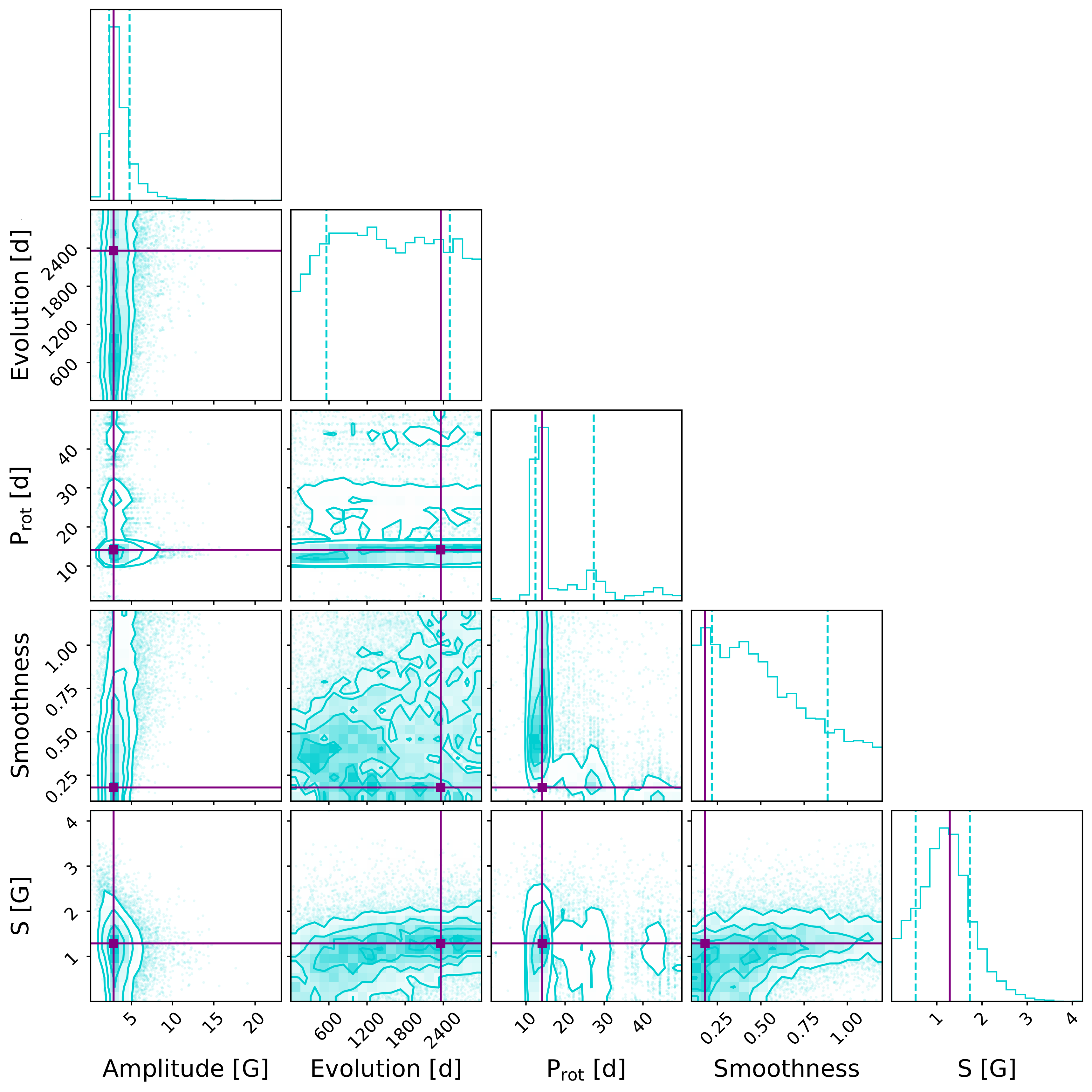}
    \caption{Time series of longitudinal magnetic field measurements for HD\,73350 and posterior distribution of the GP regression. The format is the same as Fig.~\ref{fig:Bl_hd9986}.}
    \label{fig:Bl_hd73350}
\end{figure}

\begin{figure}[t]
    \includegraphics[width=\columnwidth]{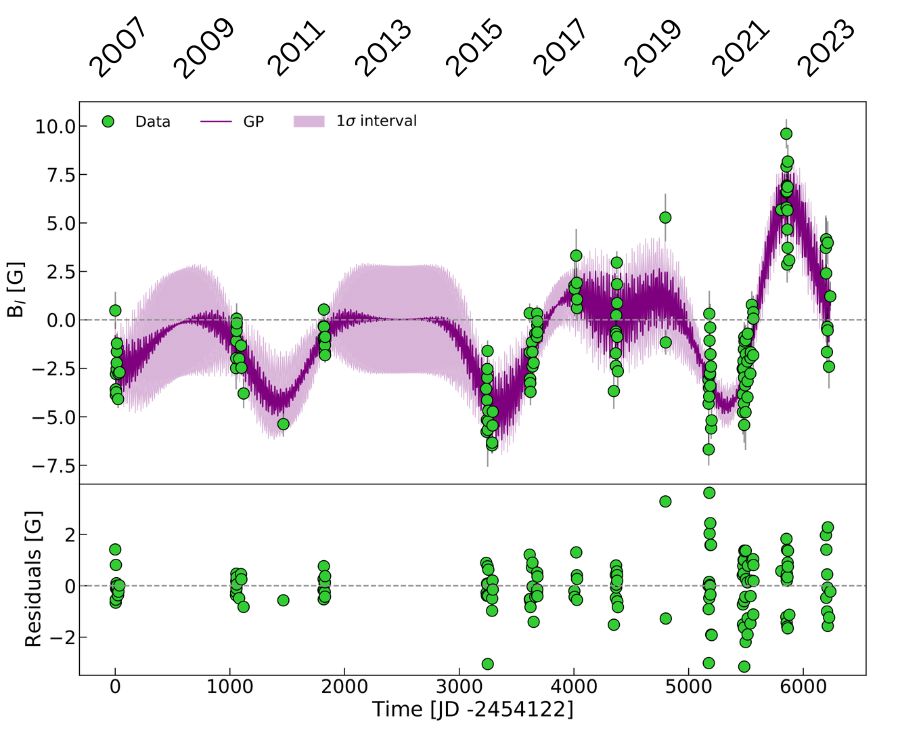}
    \includegraphics[width=\columnwidth]{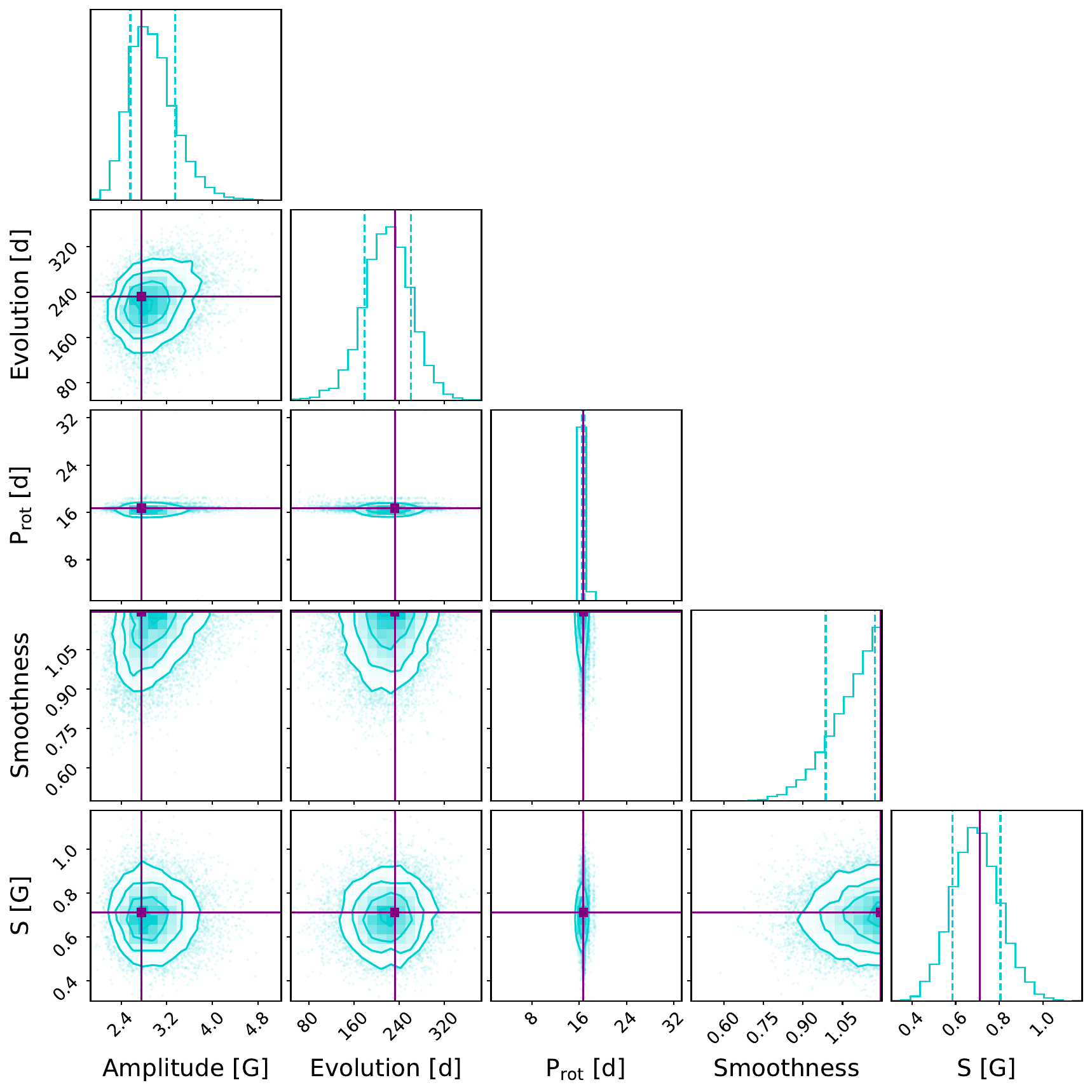}
    \caption{Time series of longitudinal magnetic field measurements for HD\,76151 and posterior distribution of the GP regression. The format is the same as Fig.~\ref{fig:Bl_hd9986}. The rotation period scale is restricted to a maximum of 35\,d for visualisation purposes, but the uniform prior encompassed the interval 1-50\,d.}
    \label{fig:Bl_hd76151}
\end{figure}

\begin{figure}[t]
    \includegraphics[width=\columnwidth]{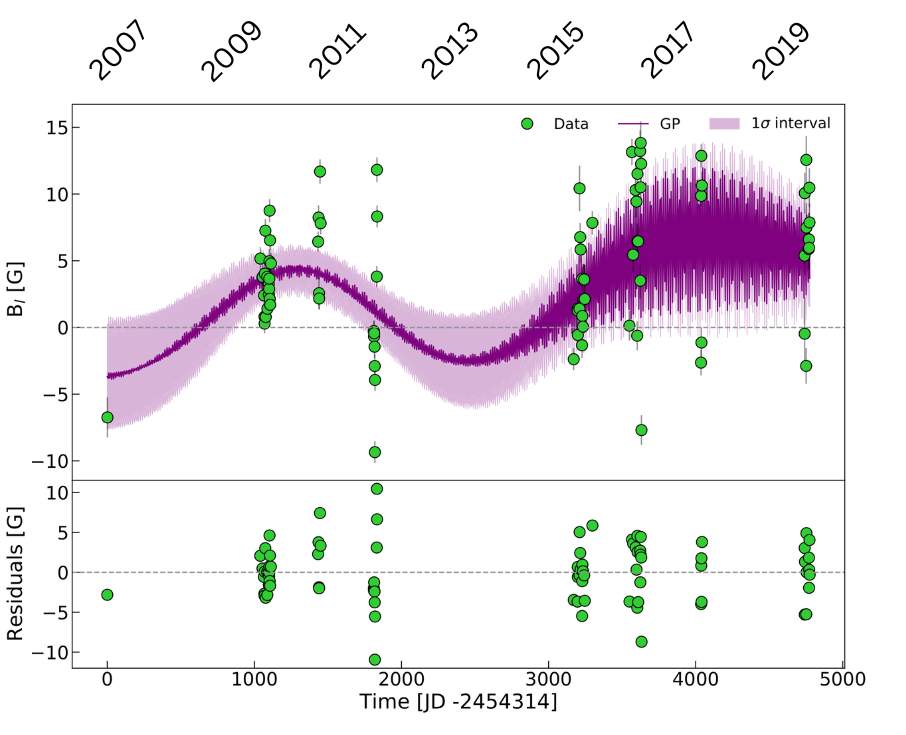}
    \includegraphics[width=\columnwidth]{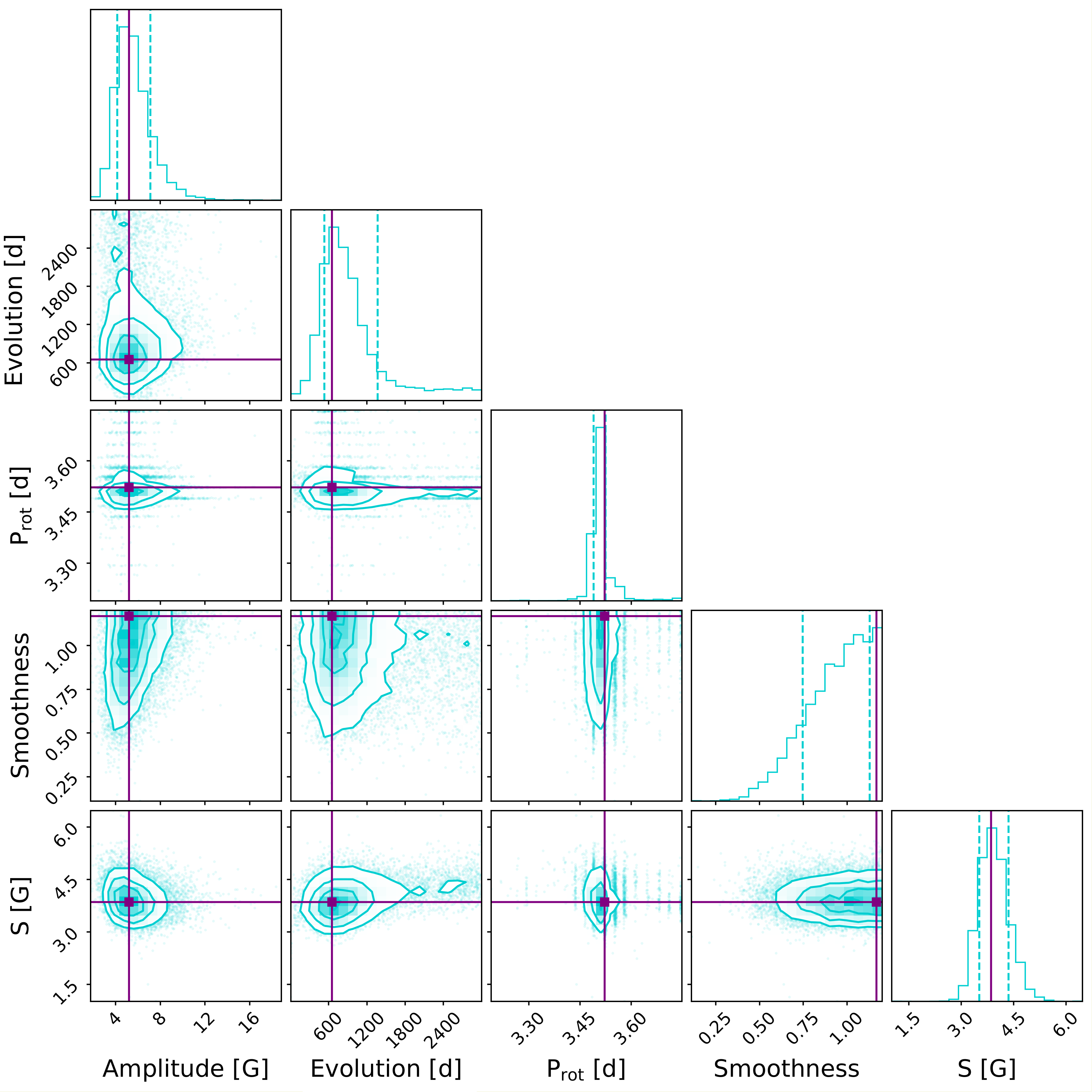}
    \caption{Time series of longitudinal magnetic field measurements for HD\,166435 and posterior distribution of the GP regression. The format is the same as Fig.~\ref{fig:Bl_hd9986}.}
    \label{fig:Bl_hd166435}
\end{figure}

\begin{figure}[t]
    \includegraphics[width=\columnwidth]{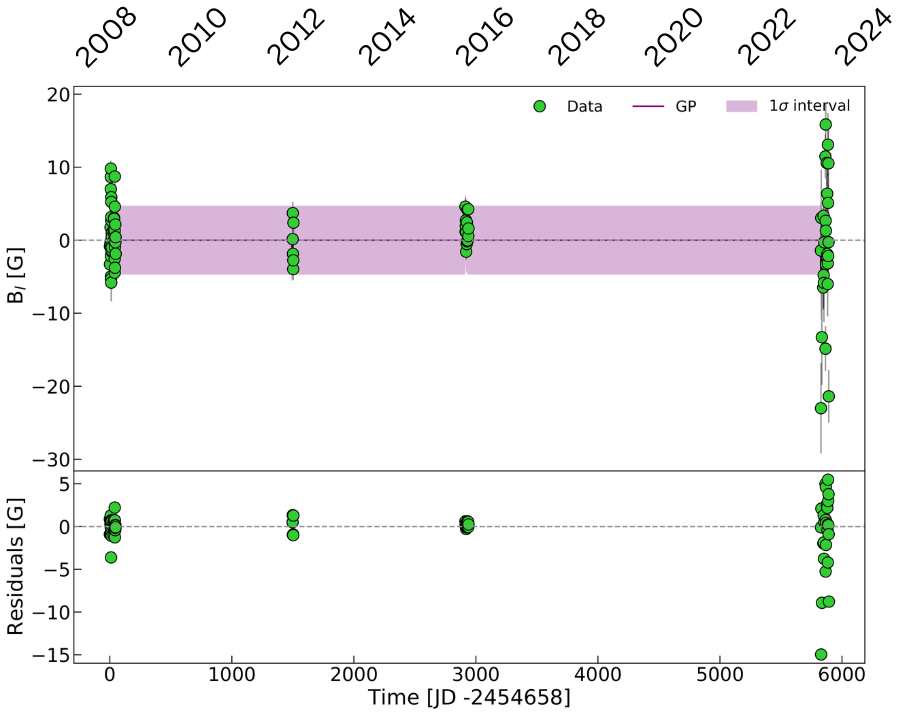}
    \includegraphics[width=\columnwidth]{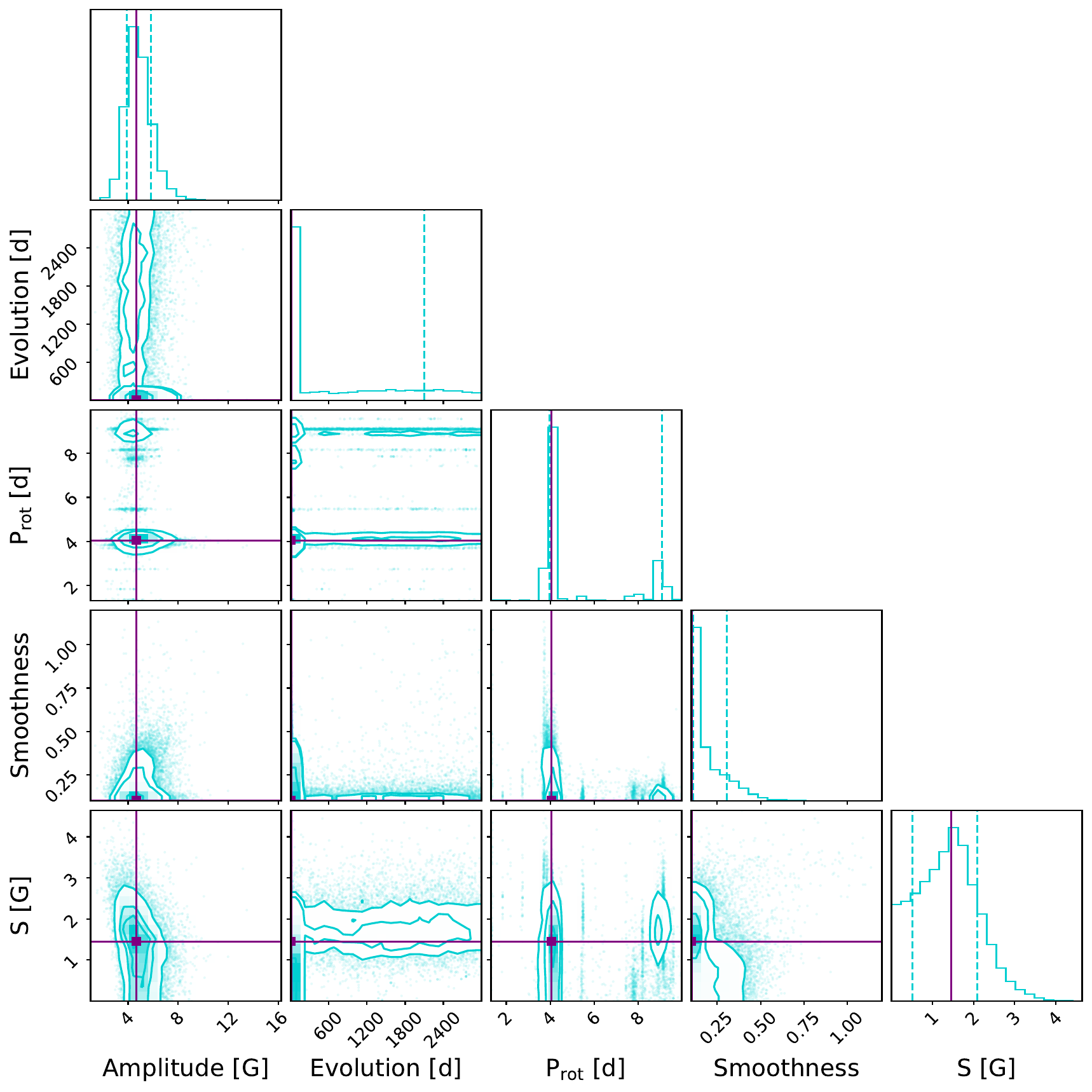}
    \caption{Time series of longitudinal magnetic field measurements for HD\,175726 and posterior distribution of the GP regression. The format is the same as Fig.~\ref{fig:Bl_hd9986}.}
    \label{fig:Bl_hd175726}
\end{figure}

\section{Additional figures Zeeman-Doppler imaging}\label{app:zdi}

We present the observed LSD Stokes~$V$ profiles together with their ZDI models. In each figure, we show the profiles for the different epochs in which we applied ZDI.

\begin{figure}[t]
    \includegraphics[width=\columnwidth]{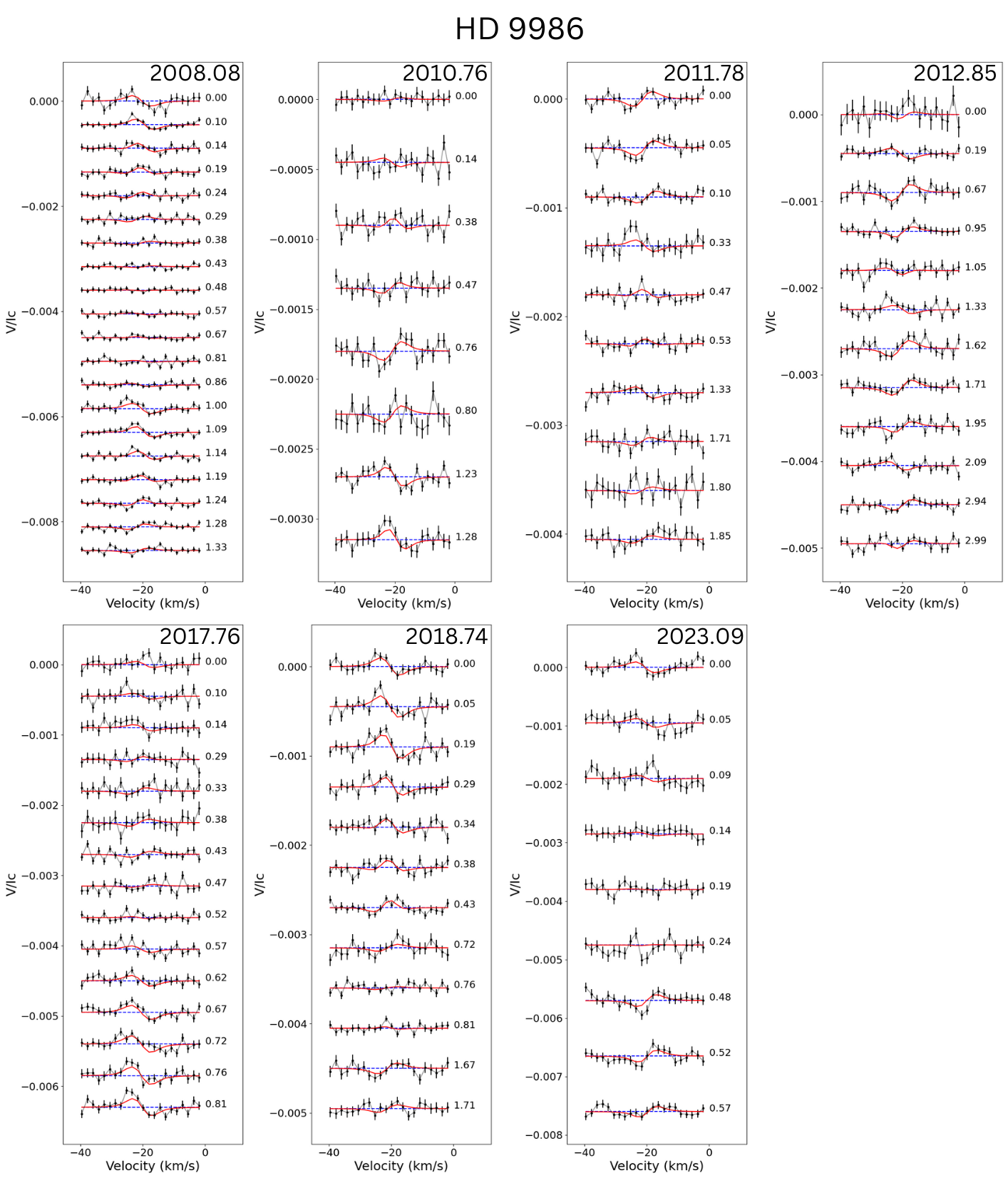}
    \caption{Time series of Stokes~$V$ LSD profiles and the ZDI models for HD\,9986. The observations are shown in grey and the models in red. The numbers on the right indicate the rotational cycle computed from Eq.~\ref{eq:ephemeris} using the first observation of an epoch as reference date. The horizontal line represents the zero point of the profiles, which are shifted vertically based on their rotational phase for visualisation purposes.}
    \label{fig:stokesV_hd9986}
\end{figure}

\begin{figure}[t]
    \includegraphics[width=\columnwidth]{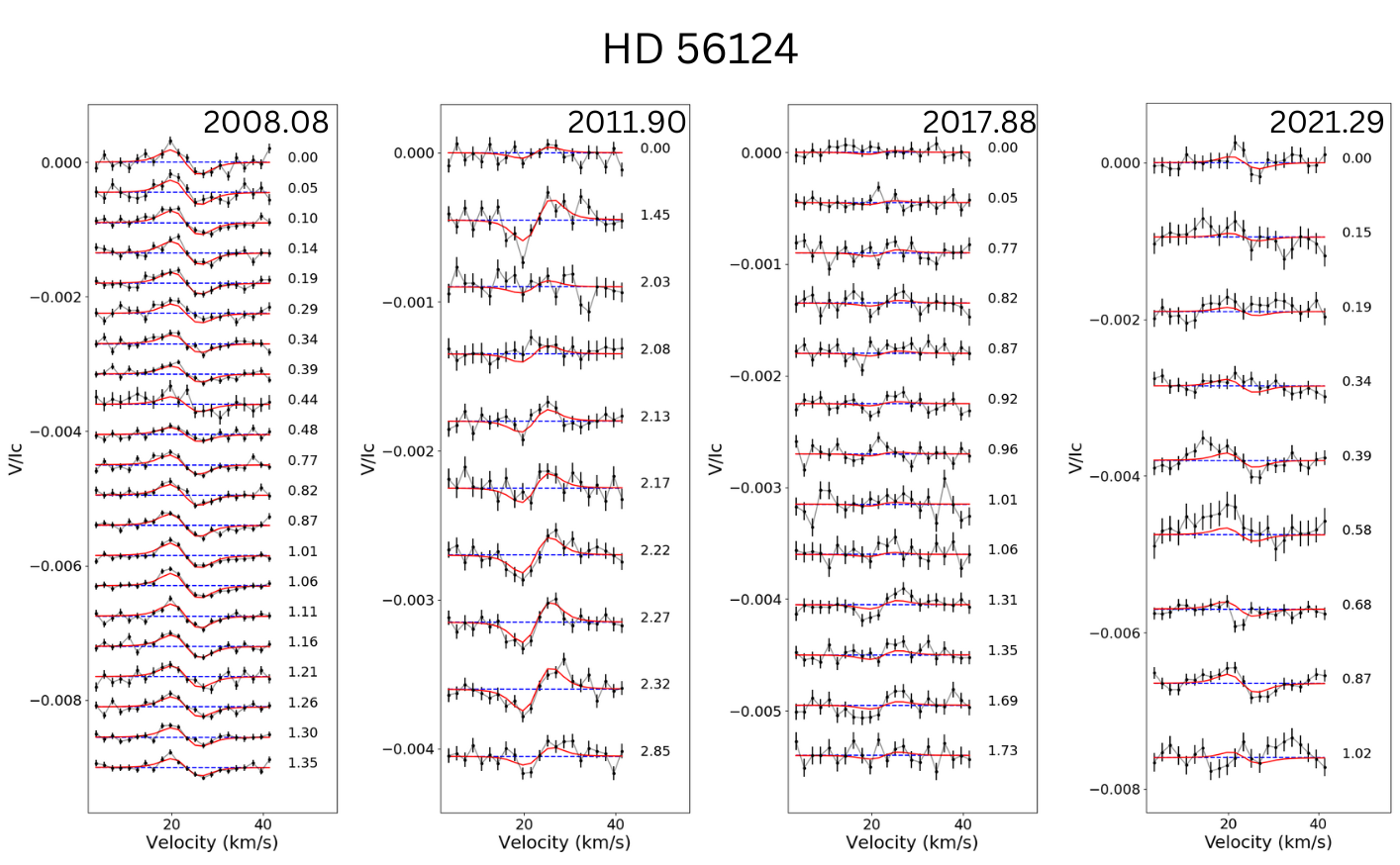}
    \caption{Time series of Stokes~$V$ LSD profiles and the ZDI models for HD\,56124. The format is the same as Fig.~\ref{fig:stokesV_hd9986}.}
    \label{fig:stokesV_hd56124}
\end{figure}

\begin{figure}[t]
    \includegraphics[width=\columnwidth]{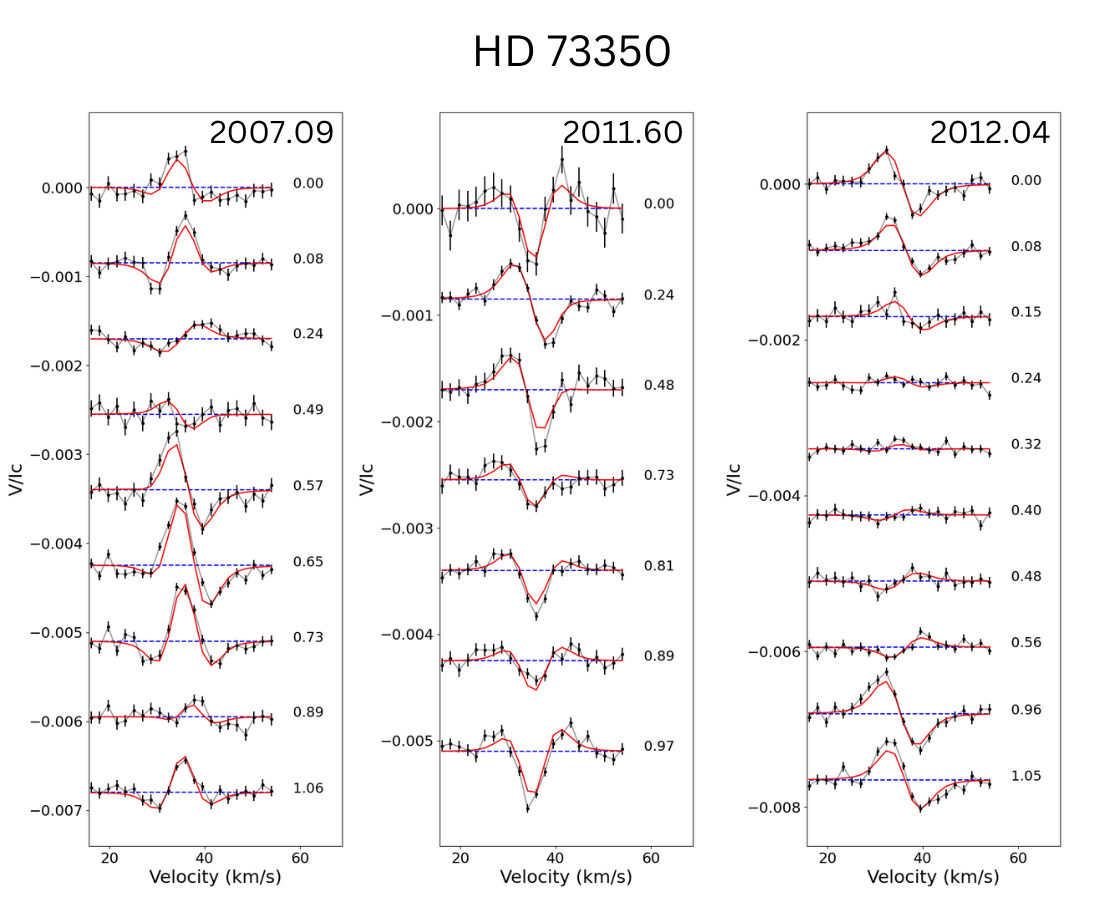}
    \caption{Time series of Stokes~$V$ LSD profiles and the ZDI models for HD\,73350. The format is the same as Fig.~\ref{fig:stokesV_hd9986}.}
    \label{fig:stokesV_hd73350}
\end{figure}

\begin{figure}[t]
    \includegraphics[width=\columnwidth]{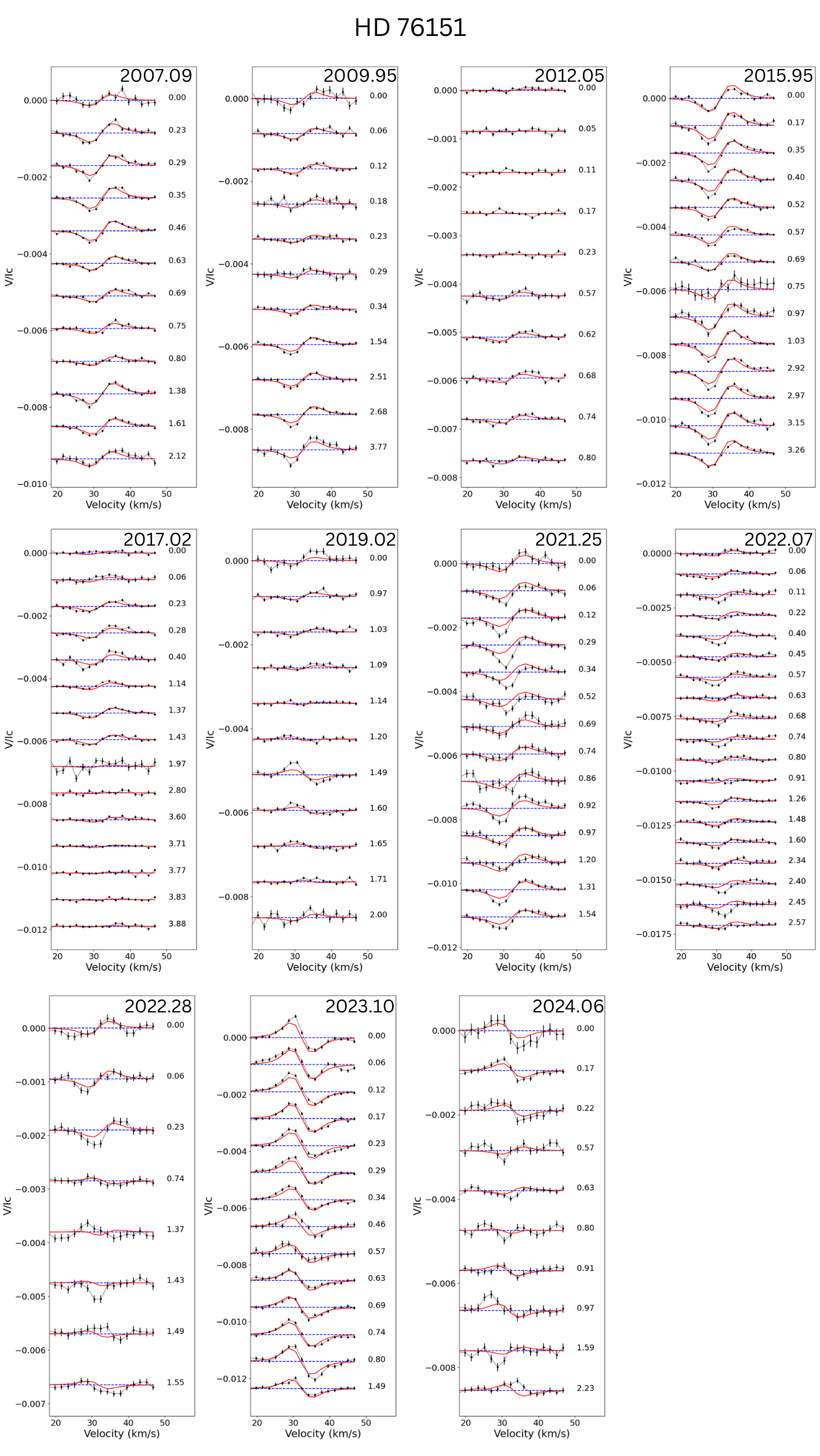}
    \caption{Time series of Stokes~$V$ LSD profiles and the ZDI models for HD\,76151. The format is the same as Fig.~\ref{fig:stokesV_hd9986}.}
    \label{fig:stokesV_hd76151}
\end{figure}

\begin{figure}[t]
    \includegraphics[width=\columnwidth]{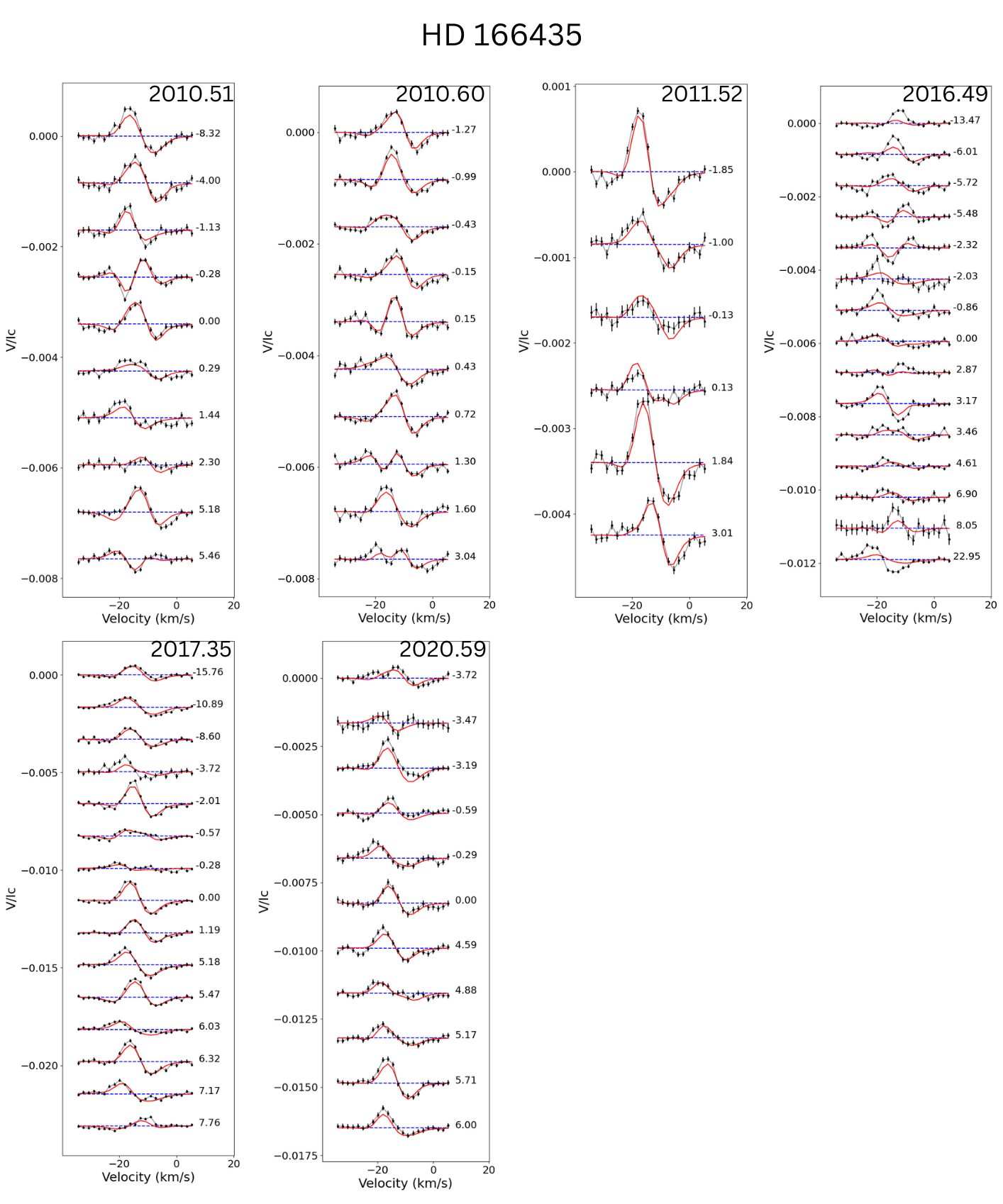}
    \caption{Time series of Stokes~$V$ LSD profiles and the ZDI models for HD\,166435. The format is the same as Fig.~\ref{fig:stokesV_hd9986}.}
    \label{fig:stokesV_hd166435}
\end{figure}

\begin{figure}[t]
    \centering
    \includegraphics[width=\columnwidth]{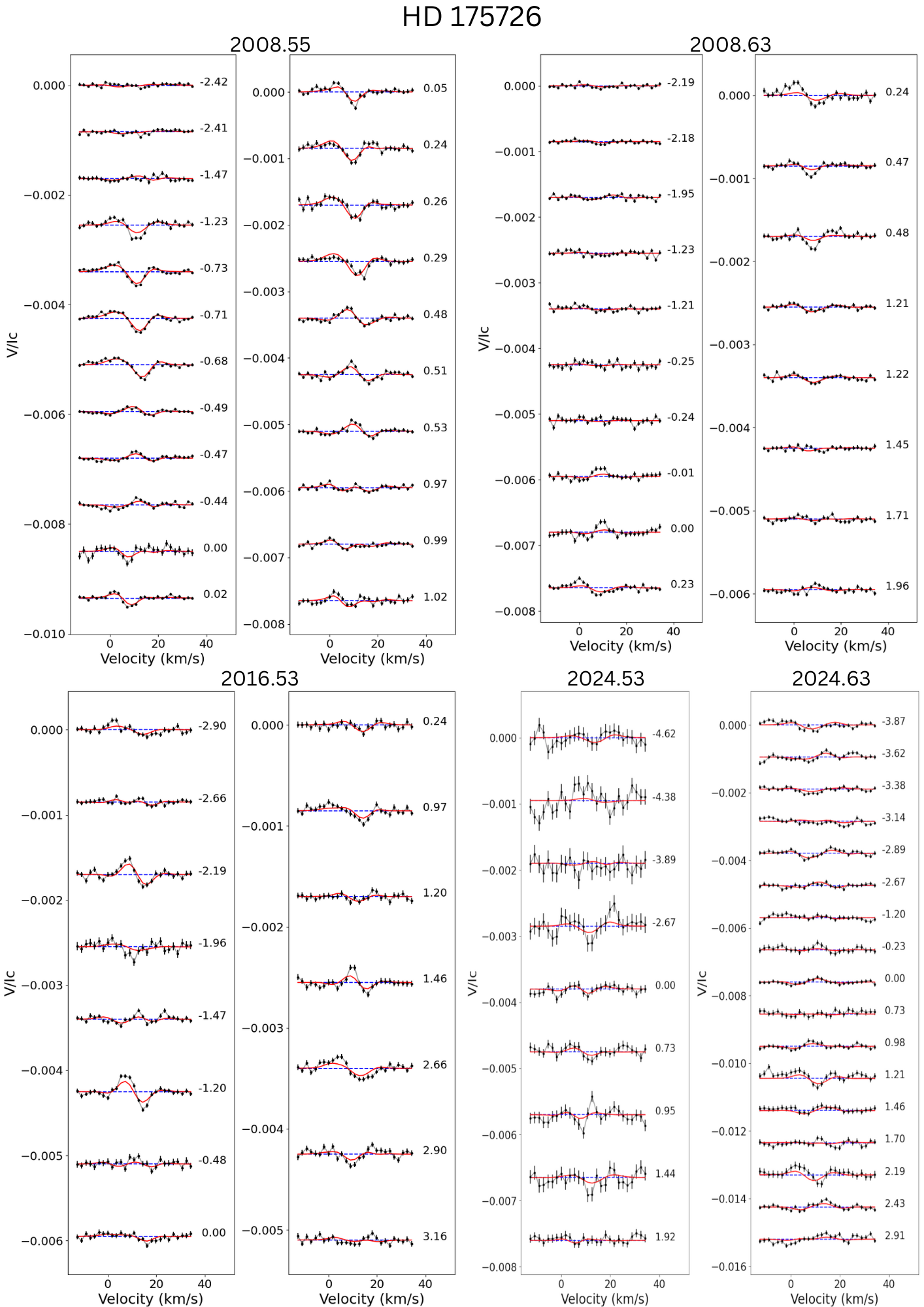}
    \caption{Time series of Stokes~$V$ LSD profiles and the ZDI models for HD\,175726. The format is the same as Fig.~\ref{fig:stokesV_hd9986}.}
    \label{fig:stokesV_hd175726}
\end{figure}

\section{Journal of observations}\label{app:log}

In this appendix we report the list of observations conducted between 2007 and 2022 with Narval, Neo-Narval, and ESPaDOnS. The observations were carried out as part of the BCool survey \citep{Marsden2014}.

\onecolumn
\twocolumn
\onecolumn

\begin{longtable}{lcrcccrc}
\caption{\label{tab:log} Observations of HD\,9986. The columns are: (1 and 2) date and universal time of the observations, (3) rotational cycle of the observations found using Eq.~\ref{eq:ephemeris}, (4) exposure time of a polarimetric sequence, (5) signal-to-noise ratio at 650 nm per polarimetric sequence, (6) RMS noise level of Stokes $V$ signal in units of unpolarised continuum, (7) longitudinal magnetic field and formal error bar, (8) $S$-index and formal error bar.} \\
\hline\hline
Date & UT & $n_\mathrm{cyc}$ & $t_{exp}$ & S/N & $\sigma_\mathrm{LSD}$ & B$_\ell$ & $S$-index\\
 & [hh:mm:ss] & & [s] & & [$10^{-5}I_c$] & [G] & \\
\hline
\multicolumn{6}{c}{2008}\\
\hline
Jan 19 & 18:05:05 & 0.00 & 4x800.0 & 364 & 9.1 & $1.20\pm1.43$ & $0.208\pm0.013$ \\
Jan 21 & 18:12:30 & 0.10 & 4x900.0 & 672 & 4.1 & $0.32\pm0.72$ & $0.208\pm0.006$ \\
Jan 22 & 18:06:46 & 0.14 & 4x900.0 & 472 & 6.8 & $0.40\pm1.04$ & $0.210\pm0.009$ \\
Jan 23 & 18:12:35 & 0.19 & 4x900.0 & 543 & 5.8 & $-0.17\pm0.94$ & $0.210\pm0.008$ \\
\end{longtable}

\begin{longtable}{lcrcccrc}
\caption{Same as Table~\ref{tab:log} for HD\,56124} \\
\hline\hline
Date & UT & $n_\mathrm{cyc}$ & $t_{exp}$ & S/N & $\sigma_\mathrm{LSD}$ & B$_\ell$ & $S$-index\\
 & [hh:mm:ss] & & [s] & & [$10^{-5}I_c$] & [G] & \\
\hline
\multicolumn{6}{c}{2008}\\
\hline
Jan 19 & 21:50:15 & 0.00 & 4x800.0 & 425 & 7.7 & $3.13\pm0.85$ & $0.206\pm0.010$ \\
Jan 21 & 00:46:42 & 0.05 & 4x600.0 & 401 & 7.6 & $1.22\pm0.95$ & $0.206\pm0.011$ \\
Jan 21 & 21:39:48 & 0.10 & 4x900.0 & 559 & 5.2 & $3.85\pm0.63$ & $0.205\pm0.007$ \\
Jan 22 & 21:24:21 & 0.14 & 4x900.0 & 520 & 5.9 & $2.12\pm0.69$ & $0.204\pm0.008$ \\
\hline
\end{longtable}

\begin{longtable}{lcrcccrc}
\caption{Same as Table~\ref{tab:log} for HD\,73350. The observations from 2017 and 2018 were taken with ESPaDOnS.} \\
\hline\hline
Date & UT & $n_\mathrm{cyc}$ & $t_{exp}$ & S/N & $\sigma_\mathrm{LSD}$ & B$_\ell$ & $S$-index\\
 & [hh:mm:ss] & & [s] & & [$10^{-5}I_c$] & [G] & \\
\hline
\multicolumn{6}{c}{2007}\\
\hline
Jan 26 & 23:22:43 & 0.00 & 4x600.0 & 410 & 6.9 & $1.76\pm1.24$ & $0.321\pm0.013$ \\
Jan 27 & 22:24:12 & 0.08 & 4x600.0 & 464 & 6.6 & $-1.01\pm1.08$ & $0.321\pm0.011$ \\
Jan 29 & 22:14:59 & 0.24 & 4x600.0 & 513 & 5.9 & $-2.47\pm0.90$ & $0.334\pm0.009$ \\
Feb 01 & 22:53:24 & 0.49 & 4x600.0 & 350 & 9.1 & $1.24\pm1.46$ & $0.340\pm0.017$ \\
\hline
\end{longtable}

\begin{longtable}{lcrcccrc}
\caption{Same as Table~\ref{tab:log} for HD\,76151. The observations from 17 January 2017, 7 January 2018, and 9 January 2018 were taken with ESPaDOnS.} \\
\hline\hline
Date & UT & $n_\mathrm{cyc}$ & $t_{exp}$ & S/N & $\sigma_\mathrm{LSD}$ & B$_\ell$ & $S$-index\\
 & [hh:mm:ss] & & [s] & & [$10^{-5}I_c$] & [G] & \\
\hline
\multicolumn{6}{c}{2007}\\
\hline
Jan 21 & 24:11:22 & 0.00 & 4x500.0 & 352 & 8.4 & $0.48\pm0.96$ & $0.257\pm0.014$ \\
Jan 26 & 01:15:05 & 0.23 & 4x800.0 & 617 & 5.0 & $-3.58\pm0.56$ & $0.249\pm0.008$ \\
Jan 27 & 01:19:19 & 0.29 & 4x800.0 & 693 & 4.5 & $-3.88\pm0.49$ & $0.252\pm0.007$ \\
Jan 28 & 01:34:57 & 0.35 & 4x800.0 & 761 & 4.2 & $-2.79\pm0.43$ & $0.253\pm0.006$ \\
\hline
\end{longtable}

\begin{longtable}{lcrcccrc}
\caption{Same as Table~\ref{tab:log} for HD\,166435} \\
\hline\hline
Date & UT & $n_\mathrm{cyc}$ & $t_{exp}$ & S/N & $\sigma_\mathrm{LSD}$ & B$_\ell$ & $S$-index\\
 & [hh:mm:ss] & & [s] & & [$10^{-5}I_c$] & [G] & \\
\hline
\multicolumn{6}{c}{2007}\\
\hline
Aug 01 & 23:15:18 & 0.00 & 4x600.0 & 354 & 9.0 & $-6.74\pm1.50$ & $0.504\pm0.014$ \\
\hline
\multicolumn{6}{c}{2010}\\
\hline
Jun 06 & 23:45:23 & 298.00 & 4x800.0 & 616 & 5.4 & $5.16\pm0.86$ & $0.473\pm0.008$ \\
Jun 22 & 00:50:59 & 302.31 & 4x800.0 & 567 & 5.8 & $3.79\pm0.98$ & $0.468\pm0.010$ \\
\hline
\end{longtable}

\begin{longtable}{lcrcccrc}
\caption{Same as Table~\ref{tab:log} for HD\,175726} \\
\hline\hline
Date & UT & $n_\mathrm{cyc}$ & $t_{exp}$ & S/N & $\sigma_\mathrm{LSD}$ & B$_\ell$ & $S$-index\\
 & [hh:mm:ss] & & [s] & & [$10^{-5}I_c$] & [G] & \\
\hline
\multicolumn{6}{c}{2008}\\
\hline
Jul 10 & 21:28:54 & -0.03 & 4x900.0 & 20 & \ldots & \ldots & \ldots \\
Jul 10 & 23:58:05 & 0.00 & 4x900.0 & 650 & 3.0 & $-0.85\pm1.15$ & $0.376\pm0.006$ \\
Jul 11 & 01:01:31 & 0.01 & 4x900.0 & 640 & 3.0 & $-3.30\pm1.18$ & $0.376\pm0.006$ \\
Jul 14 & 21:28:33 & 0.95 & 4x900.0 & 487 & 3.9 & $-0.43\pm1.62$ & $0.378\pm0.010$ \\
\hline
\end{longtable}

\end{appendix}

\end{document}